\documentclass[10pt,A4paper,amsmath,amssymb,aps,etoolbox,floatfix,
longbibliography,nofootinbib,noshowpacs,onecolumn,preprintnumbers,reprint,
showkeys,showpacs,superscript address]{revtex4}

\newcommand{\bea}{\begin{eqnarray}}
\newcommand{\beq}{\begin{equation}}

\newcommand{\eea}{\end{eqnarray}}
\newcommand{\eeq}{\end{equation}}
\newcommand{\ds}{\displaystyle}

\newcommand{\rosso}{\color{red}}

\usepackage{amsmath}
\usepackage{amsfonts}
\usepackage{amssymb}
\usepackage{bm}       
\usepackage{caption}
\usepackage{comment}
\usepackage[usenames,dvipsnames]{color}
\usepackage{dcolumn}  
\usepackage{epsfig}
\usepackage{graphics}
\usepackage{graphicx} 
\usepackage{natbib}

\usepackage{placeins} 
\usepackage{psfrag}
\usepackage{times}
\usepackage[varg]{txfonts}
\usepackage{xcolor}
\usepackage{xspace}

\begin{document}

\title{Investigating dark energy by electromagnetic frequency shifts}

\author{Alessandro D.A.M. Spallicci}
\affiliation{\mbox{Institut Denis Poisson (IDP) UMR 7013}\\
\mbox{Universit\'e d'Orl\'eans (UO) et Universit\'e de Tours (UT)}\\
\mbox{Centre National de la Recherche Scientifique (CNRS)}\\
\mbox{Parc de Grandmont, 37200 Tours, France}}

\affiliation{\mbox{Laboratoire de Physique et Chimie de l'Environnement et de l'Espace (LPC2E) UMR 7328}\\
\mbox{Centre National de la Recherche Scientifique (CNRS)}\\
\mbox{Universit\'e d'Orl\'eans (UO)}\\
\mbox{Centre National d'\'Etudes Spatiales (CNES)}\\
\mbox {3A Avenue de la Recherche Scientifique, 45071 Orl\'eans, France}}

\affiliation{\mbox{UFR Sciences et Techniques, 
Universit\'e d’Orl\'eans (UO)}\\
\mbox{Rue de Chartres, 45100 Orl\'{e}ans, France}}

\affiliation{\mbox{Observatoire des Sciences de l'Univers en region Centre (OSUC) UMS 3116,
Universit\'e d'Orl\'eans (UO)}\\
\mbox{1A rue de la F\'{e}rollerie, 45071 Orl\'{e}ans, France}}

\author{Giuseppe Sarracino}

\affiliation{{\mbox Dipartimento di Fisica E. Pancini, Universit\`{a} degli Studi di Napoli, Federico II}\\ 
{\mbox Complesso Universitario Monte S. Angelo, Via Cinthia 9 Edificio G, 80126 Napoli, Italy}}

\affiliation{{\mbox Istituto Nazionale di Fisica Nucleare (INFN), Sezione di Napoli}\\
{\mbox Complesso Universitario Monte S. Angelo, Via Cinthia 9 Edificio G, 80126 Napoli, Italy}}

\author{Salvatore Capozziello}

\affiliation{{\mbox Dipartimento di Fisica E. Pancini, Universit\`{a} degli Studi di Napoli, Federico II}\\ 
{\mbox Complesso Universitario Monte S. Angelo, Via Cinthia 9 Edificio G, 80126 Napoli, Italy}}

\affiliation{{\mbox Istituto Nazionale di Fisica Nucleare (INFN), Sezione di Napoli}\\
{\mbox Complesso Universitario Monte S. Angelo, Via Cinthia 9 Edificio G, 80126 Napoli, Italy}}

\affiliation{{\mbox Scuola Superiore Meridionale,  Universit\`{a} degli Studi di Napoli, Federico II}\\ {\mbox Largo San Marcellino 10, 80138 Napoli, Italy}}

\date{29 January 2022}

\begin{abstract}
The observed red shift $z$ might be composed by the expansion red shift $z_{\rm C}$ and an additional frequency shift $z_{\rm S}$, towards the red or the blue, by considering Extended Theories of Electromagnetism (ETE). Indeed, massive photon theories - the photon has a real mass as in the de Broglie-Proca theory or an effective mass as in the Standard-Model Extension (SME), based on Lorentz-Poincar\'e Symmetry Violation (LSV) - or Non-Linear Electro-Magnetism (NLEM) theories may induce a cosmological expansion independent frequency shift in presence of background (inter-) galactic electromagnetic fields, and where of relevance LSV fields, even when both fields are constant. We have tested this prediction considering the Pantheon Catalogue, composed by 1048 SNe Ia, and 15 BAO data, for different cosmological models characterised by the absence of a cosmological constant. From the data, we compute which values of $z_{\rm S}$ match the observations, spanning  cosmological parameters ($\Omega$ densities and Hubble-Lema\^itre constant) domains. We conclude that the frequency shift  $z_{\rm S}$ can support  an alternative to accelerated expansion, naturally accommodating each SN Ia position in the distance-modulus versus red shift diagram, due to the light-path dependency of $z_{\rm S}$. Finally, we briefly mention laboratory test approaches to investigate the additional shift from ETE predictions.    
\end{abstract}

\keywords{Photon mass; Standard-Model Extension; Non-linear electromagnetism; Cosmology; Dark universe; Red shift; SNe Ia; BAO.} 

\maketitle

\section{Introduction}
According to the widely adopted Lambda Cold Dark Matter Model ($\Lambda$CDM), the universe is filled up for its large amount by unknown fundamental constituents, namely dark matter and dark energy. The former would constitute roughly 25\% while the latter 70\% of the total density of the universe at our epoch \cite{riess-etal-1998,riess-etal-2004,perlmutter-etal-1999,bahcall-etal-1999,spergel-etal-2003,schimd-etal-2006,mcdonald-etal-2006,bamba-etal-2012,joyce-etal-2015,salucci-etal-2021}. The former was introduced to straightforwardly explain the stellar and galactic rotation curves (excess velocity as function of radial distance to the centre) in galaxies and clusters \cite{zwicky-1933}; the latter to explain the discrepancy between the luminosity distance of Supernovae type Ia (SNe Ia), considered to be well established standard candles \cite{tripp-1998}, and the associated red shift \cite{riess-etal-1998,perlmutter-etal-1999}. 

The main limit of these assumptions \cite{lopezcorredoira-2017} is the lack of direct observations despite the investments by the scientific community \cite{zyla-etal-2020}, in testing different hypothesis \cite{pigozzo-etal-2011}. 
The difficulty of finding out new fundamental particles has led to extend General
Relativity (GR) at infrared scales as possible solution of the puzzle  \cite{capozziello-delaurentis-2011}. In particular, the issue of the Hubble tension could be a manifestation of the GR inadequacy to represent the whole cosmic history involving also the possibility to
consider a sort of Heisenberg principle working at cosmological scales 
\cite{capozziello-benetti-spallicci-2020, spallicci-benetti-capozziello-2022}.
The preceding is accompanied by the difficulty of applying GR at the microscopic scale, the suspicion of deviations from GR in the strong field regime, and the 120 orders of magnitude surplus that vacuum energy possesses with respect to the dark energy fluid \cite{carroll-2001}. 

Rather than invoking dark matter or a dark fluid for the energy-momentum tensor, the Extended Theories of Gravity (ETG) \cite{capozziello-faraoni-2011} intervene on the geometrical part of the action  \cite{blspwh11}. Applications were carried at  astrophysical \cite{astashenok-capozziello-odintsov-2013, astashenok-capozziello-odintsov-2014a,astashenok-capozziello-odintsov-2014b, astashenok-capozziello-odintsov-2015a,astashenok-capozziello-odintsov-2015b, wojnar-2019,olmo-rubieragarcia-wojnar-2020,feola-etal-2020} and cosmological \cite{bahamonde-etal-2018,carloni-etal-2019,dainotti-etal-2021,saridakis-etal-2021} scales to tackle GR in extreme environments, while confirming the successes of GR at all other scales. Naturally, since the $\Lambda$CDM model leaves GR untouched,  modifications of GR will inevitably backlash on the $\Lambda$CDM.

Herein, we follow a different conceptual route. Although multi-messenger astronomy is now enriched by gravitational wave \cite{abbottetal2016} and neutrino \cite{gelmini-etal-2010} observations, the majority of the information from astrophysical sources has an electromagnetic nature. As Special Relativity (SR) and Quantum Mechanics (QM) were born by reinterpreting light \cite{capozziello-boskoff-2021}, we are tempted to do likewise by extending the theory of electromagnetism. Indeed, the Maxwellian theory could be an approximation of a broader theory, as the Newtonian physics is for GR. We refocus from the sources to the signals, that is to the photons as messengers. 

Though the Maxwell equations provide a successful description of all electromagnetic phenomena (this applies also to General Relativity, which does not impede the formulation of alternative theories of gravitation), there are reasons to reconsider this theory: i) the standard formulation of electrodynamics leads to severe problems in the self-force problem (radiation reaction) of charged point particles, as unphysical pre-acceleration and run-away solutions; ii) any modification of the phenomena of electrodynamics will also be directly related to a modified space-time geometry; iii) since according to our present understanding GR and Quantum Mechanics (QM) are not compatible, a new theory combining the geometric and quantum aspects of our world has to be different from at least one of these theories that, for consistency, would also change the Maxwell equations. Accordingly, it is of utmost importance to find out whether the equations underlying all electromagnetic phenomena are the well-known
Maxwell equations or whether modifications have to be taken into account. 

Despite its numerous successes, the Standard-Model (SM) of elementary particles has some shortcomings, as it fails to explain the unbalance of matter and anti-matter \cite{dibari-2022} and the neutrino masses \cite{mohapatra-senjanovic-1980,formaggio-etal-2021}. Thereby, the photon remains the only free particle still portrayed as massless. 
But masslessness is not provable and since the Heisenberg principle in the energy-time form forbids to determine any mass below $10^{-69}$ kg \cite{capozziello-benetti-spallicci-2020,spallicci-benetti-capozziello-2022},  there are at least fifteen orders of magnitude for experiments and observation to reach the Heisenberg limit.  

Furthermore, the SM features the absence of candidate particles for dark matter and dark energy. These and other issues have made necessary to look for physics beyond the SM: the SM Extension (SME) \cite{colladay-kostelecky-1997, colladay-kostelecky-1998}, where the photons acquire an effective mass proportional to the Lorentz-Poincaré Symmetry Violation (LSV) factors \cite{bonetti-dossantosfilho-helayelneto-spallicci-2017, bonetti-dossantosfilho-helayelneto-spallicci-2018}. 

Well before the SME, de Broglie proposed first a massive photon \cite{debroglie-1922}, estimated below $10^{-53}$ kg \cite{debroglie-1923} through dispersion analysis \cite{debroglie-1924} and wrote the modified Maxwell-Amp\`ere-Faraday-Gauss (MAFG) equations \cite{debroglie-1936,db40}. His scholar Proca wrote a Lagrangian for electrons, positrons, neutrinos and photons being convinced of the masslessness of the latter two \cite{proca-1936b,proca-1936c,proca-1936d,proca-1937}. The de Broglie-Proca (dBP) theory is not (Lorenz) gauge-invariant, but it satisfies Lorentz-Poincaré Symmetry (LoSy), for which measurements are independent of the orientation or the velocity of the observer. 
Later developments regained gauge-invariance \cite{bopp-1940,podolski-1942,stueckelberg-1957}, including the SME. Renormalisability, unitarity, origin of mass and electric charge conservation were addressed \cite{boulware-1970,guendelman-1979,nussinov-1987,itzykson-zuber-2012,scharffgoldhaber-nieto-2010}.

Many tests have been performed regarding the measurements of the upper limits of the photon mass. The strongest limit reached up to now is $10^{-54}$ kg \cite{Ryutov-1997,Ryutov-2007,Ryutov-2009} but it unfortunately arises from modelling of the solar wind rather than measurements \cite{retino-spallicci-vaivads-2016}. Surprisingly, after one century of painstaking efforts the photon mass upper limit is just one order of magnitude below the de Broglie estimate of 1923 \cite{debroglie-1923}. Limits given by direct measurements on dispersion have been computed using Fast Radio Bursts (FRBs), by this collaboration \cite{boelmasasgsp2016,boelmasasgsp2017}. The best result was achieved $3.9 \times 10^{-51}$ kg at 95\% confidence level \cite{wang-miao-shao-2021}. Lunar nano-satellites have been proposed to open a new window at very-low radio-frequencies \cite{bebosp2017}. 
The best laboratory test verified Coulomb's law, setting the limit at $2\times 10^{-50}$~kg \cite{wifahi71}.  

Non-Linear Electromagnetism (NLEM) theories describe the interactions between electromagnetic fields in vacuum. Born and Infeld solved the problem of the divergence of a point charge by relating the maximum of the electric field with the size and rest mass of the electron  \cite{born-infeld-1934a,born-infeld-1935}. Heisenberg and Euler theory is adopted  \cite{heisenberg-euler-1936} for strong magnetic fields and in Quantum Electrodynamics (QED). 

Both dBP and SME photons exchange energy with the either galactic or intergalactic background electromagnetic fields (The SME photon exchanges also with the LSV background), thereby undergoing frequency shifts towards the red or the blue \cite{helayelneto-spallicci-2019,spallicci-etal-2021}. Also for NLEM theories, there is such an exchange and frequency shift  \cite{helayelneto-spallicci-2022}. It must be added that a frequency shift occurs also in the Maxwellian theory, but it requires a variable background.

We here investigate the consequences of adding a frequency shift $z_{\rm S}$, coming from one of the Extended Theories of Electromagnetism (ETE) towards the red or the blue due to non-classical electromagnetic effects to the expansion red shift $z_{\rm C}$. 
The size of the shift $z_{\rm S}$ may span from a negligible to a large fraction of the observed $z$. 
From this view point, it is a tool that can allow to reinterpret and correct astrophysical observations under a different standard.

Herein, through recasting the observed $z$, we avoid introducing the accelerated expansion. To this end, we use the publicly available Pantheon Sample of SNe Ia \cite{scolnic-etal-2018}, which is a catalogue of 1048 objects collected from various observational programs, and data constrains using the Baryon Acoustic Oscillations (BAOs) data \cite{beutler-etal-2011,blake-etal-2011,ross-etal-2015, dumasdesbourboux-etal-2020,alam-etal-2021}. Most of the BAO data considered herein have been published in the latest Sloan Digital Sky Survey (SDSS) release, while others in the  6dF Galaxy and the WiggleZ Dark Energy surveys. 

The paper is organised as follows. In Sect. 2, we briefly recall how the frequency shift is produced in ETE framework. The interested reader may refer to the cited publications. Our intent is simply to alert the astrophysics oriented reader of the literature, but we ought to stress that we do not stick to a specific mechanism underlying this static and non-classical frequency shift. We show that a frequency shift $z_{\rm S}$ naturally emerges in the framework of these theories, and in, specific circumstances, also in the Maxwellian theory. In Sect. 3, we discuss the approach of implementing $z_{\rm S}$ in SN Ia and BAO data, as well for mock red shift $z$ values going beyond the furthest SN Ia. In Sect. 4, we show the results of our computations. Finally, discussions, conclusions and perspectives are reported in Sect. 5.

\section{Frequency shift from Extended Theories of Electromagnetism}

In this section, we sketch how it is possible to infer a further component in the total detected red shift of astrophysical observations. We present cases that cover a large class of ETE, without any pretension of exhaustivity.  

In SI units, the photon energy-momentum tensor component has dimensions of Jm$^{-3}$, where $\theta^{0}_{\ 0}$ is the energy density, $\theta^{0}_{\ k}$ is the energy flux divided by $c$ along the $k$ direction, $\theta^{k}_{\ 0}$ is the momentum density through the orthogonal surface to $k$, multiplied by c. The photon energy-momentum non-conservation implies that the photon exchanges energy with the background. 

The energy-momentum density tensor $\theta_{\ \tau}^{\alpha}$ variation is expressed by a derivative, that is $\partial_\alpha \theta_{\ \tau}^{\alpha}$ [Jm$^{-4}$] and the wave-particle correspondence, down to a single photon \cite{aspect-grangier-1987}, determines that the light-wave energy non-conservation corresponds to photon energy variation and thereby to a red or a blue shift

\beq
\partial_\alpha \theta^\alpha_{~\tau} \longrightarrow \Delta \nu~.
\eeq

For later use, the expression of the scalar fields ${\cal F}$ and its dual ${\cal G}$ are

\beq
{\cal F} = -\frac{1}{4\mu_0} {F}^2 = -\frac{1}{4\mu_0}{F}_{\sigma\tau}{F}^{\sigma\tau} = \frac{1}{2\mu_0}\left(\frac{{\vec { E}^2}}{c^2} - \vec { B}^2\right)~, 
\label{FF}
\eeq
and 

\beq
{\cal G} = -\frac{1}{4\mu_0} {F}{G} = -\frac{1}{4\mu_0} {F}_{\sigma\tau} G^{\sigma\tau} = \frac{1}{\mu_0}\frac {{\vec {E}}}{c}\cdot{\vec {B}}~,
\label{FG}
\eeq
where ${F}_{\sigma\tau}$ is the electromagnetic field tensor and ${G}^{\sigma\tau} = {\ds \frac{1}{2}}\epsilon^{\sigma\tau\alpha\beta} F_{\alpha\beta}$ is its dual; $\mu_0= 4\pi \times 10^{-7} \approx 1.256 $ H m$^{-1}$ or V s A$^{-1}$ m$^{-1}$ is the vacuum permeability, and $c = 2.998 \times 10^8$ m s$^{-1}$ is the speed of light. 

We now associate the fields above to a background field and imagine a photon crossing such a background. Thereby, we split the total (T) electromagnetic tensor field $F_{\rm T}$ and the total (T) electromagnetic 4-potential $A_{\rm T}$ in the background (capital letters) and photon (small letters) 

\beq
A^\beta_{\rm T} = A^\beta + a^\beta~, ~~~~~~~~~~~~~~~~~F^{\alpha\beta}_{\rm T} = F^{\alpha\beta} + f^{\alpha\beta}~,~~~~~~~~~~~~~~~~~G^{\alpha\beta}_{\rm T} = G^{\alpha\beta} + g^{\alpha\beta}~. 
\eeq 

\subsection{Maxwellian theory}

It is known, although seldom mentioned, that a photon may exchange energy-momentum (density), represented by the tensor $\theta^\alpha_{~\tau}$, with the background, even in the Maxwellian theory, if the background field is space-time dependent. The energy-momentum density tensor variation is

\beq
\partial_\alpha \theta^\alpha_{~\tau} = 
\underbrace{
j^\alpha f_{\alpha\tau}- {\ds \frac{1}{\mu_0}}(\partial_\alpha F^{\alpha\beta})f_{\beta\tau}
}_{\text{Maxwellian terms}}~, 
\label{Mnoncon}
\eeq 
being $j^\beta$ a possibly existing external 4-current. In conclusion, a frequency shift may exist even in the framework of the standard electromagnetic theory.

\subsection{de Broglie-Proca theory}

Stepping into the dBP formalism, the photon therein interacts with the background through the potential even when the background field is constant. Indeed, if a field is constant, its associated potential is not. For ${\cal M} = m_\gamma c /\hbar$, being $m_\gamma$ the photon mass, 
 $\hbar = 1.055 \times 10^{-34}$ kg m$^2$ s$^{-1}$, the energy-momentum density tensor variation becomes \cite{spallicci-etal-2021}

\beq
\partial_\alpha \theta^\alpha_{~\tau} = 
\underbrace{
j^\alpha f_{\alpha\tau}- {\ds \frac{1}{\mu_0}}(\partial_\alpha F^{\alpha\beta})f_{\beta\tau}
}_{\text{Maxwellian terms}} 
\underbrace{ +
{\ds \frac{1}{\mu_0}} {\cal M}^2 (\partial_\tau A^\beta)a_\beta
}_{\text{de Broglie-Proca term}}~.
\label{dBnoncon}
\eeq

Incidentally, the dBP photon does not display energy changes in absence of a background, unless invoking imaginary masses and frequencies \cite{thiounn-1960,yourgrau-woodward-1974}.

\subsection{Standard-Model Extension}

The SME-LSV factors are represented by a $k^{\rm AF}_{\alpha}$ [meter $^{-1}$] 4-vector when the handedness of the 
Charge conjugation-Parity-Time reversal (CPT) symmetry is odd and by a $k_{\rm F}^{\alpha\nu\rho\sigma}$ [dimensionless] 
tensor when even. The $k^{\rm AF}_{\alpha}$ vector, coming from the Carroll-Field-Jackiw Lagrangian \cite{carroll-field-jackiw-1990}, induces always a mass, while the $k_{\rm F}^{\alpha\nu\rho\sigma}$ tensor only in a supersymmetrised context after photino integration  \cite{bonetti-dossantosfilho-helayelneto-spallicci-2017, bonetti-dossantosfilho-helayelneto-spallicci-2018}.

We consider the LSV present in the photonic sector only. The space-time metric, spin connection 
and curvature are unaffected by the LSV, and we stick to the Minkowski space-time with the anisotropies parameterised by $k^{\rm AF}_{\alpha}$ and $k_{\rm F}^{\alpha\nu\rho\sigma}$.  

In contrast to the LSV tensor, the LSV vector does violate the CPT theorem. The frequency shift is thereby an observable of CPT violation, since it depends also on the LSV vector. 

Indicating by $G$ or $g$ the dual field of the background and of the photon, respectively, the photon energy-momentum density tensor variation $\partial_{\alpha} \theta_{\ \tau}^{\alpha}$ [Jm$^{-4}$] is given by 


\begin{eqnarray}  
&\partial_{\alpha}\theta_{\ \tau}^{\alpha} = 
\underbrace{
j^{\nu}f_{\nu \tau} - {\ds \frac{1}{\mu_0}} \left(\partial_{\alpha}F^{\alpha \nu}\right)f_{\nu \tau}
}
_{\text{Maxwellian terms}} -
& {\ds \frac{1}{\mu_0}} \left [ 
\underbrace{ 
 \frac{1}{2}
\left(\partial_{\alpha}k^{\rm AF}_{\tau}\right) g^{\alpha\nu}a_{\nu} - \frac{1}{4}\left(\partial_{\tau}k_{\rm F}^{\alpha \nu\kappa\lambda}\right)f_{\alpha \nu}f_{\kappa\lambda} 
}_{\text{EM background independent terms}} + 
\underbrace{\partial_{\alpha}\left(k_{\rm F}^{\alpha \nu\kappa\lambda}F_{\kappa\lambda}\right)f_{\nu \tau}
}
_{\text{non-constant term}} + 
\underbrace{
k^{\rm AF}_{\alpha}G^{\alpha \nu}f_{\nu \tau}
}
_{\text{constant term}}
\right]~. 
\label{pemt-nc-0}
\end{eqnarray}

The LSV 4-vector $k_{\rm AF}$ and the rank-4 tensor 
$k_{\rm F}$ are the vacuum condensation of a vector and a tensor field in the 
context of string models \cite{kosteleckysamuel1989b}. They describe part of the vacuum structure, which appears in the form of space-time
 anisotropies. Therefore, their presence on the right-hand side of Eq. 
(\ref{pemt-nc-0}) reveals that vacuum effects are responsible for the energy variation of light-waves. Anisotropies are under considerations in cosmology \cite{migkas-etal-2021}.

Last but not least, we have shown that in the context of SME, photons are dressed of an effective mass proportional to the LSV vector for CPT-odd and related to the LSV tensor for CPT-even  \cite{bonetti-dossantosfilho-helayelneto-spallicci-2017,bonetti-dossantosfilho-helayelneto-spallicci-2018}.  

\subsection{Non-linear electromagnetism}

For NLEM, we have set a generalised Lagrangian, encompassing the formalisms of Born-Infeld and Euler-Heisenberg
\cite{born-infeld-1934a,born-infeld-1935,heisenberg-euler-1936}, as a polynomial, function of integer powers of the field and its dual 
\cite{helayelneto-spallicci-2022}. Indeed, the generalised Lagrangian is written as 

\beq
{\cal L} = {\cal L}\left(\cal F, \cal G \right)~.
\eeq

The photon energy-momentum density tensor variation $\partial_{\alpha} \theta_{\tau}^{\alpha}$ [Jm$^{-4}$] is given by 
\begin{eqnarray}
\partial_{\alpha} \theta_{\tau}^{\alpha} = 
- \partial_\alpha \left(C_1 F^{\alpha\nu}+ C_2 G^{\alpha\nu}\right) f_{\nu\tau} + 
\frac{1}{4} \left( \partial_\tau C_1 \right) f^2 + \frac{1}{4} \left( \partial_\tau C_2 \right) gf - \frac{1}{8} \left( \partial_\tau s^{\nu\alpha\kappa\lambda} \right ) f_{\nu\alpha}f_{\kappa\lambda} 
- \frac{1}{4} \left( \partial_\tau t^{\nu\alpha\rho\sigma} \right ) f_{\nu\alpha}f_{\rho\sigma}~, 
\end{eqnarray}

where the coefficients are computed on the background and are

\beq
\left. \frac{\partial{\cal L}}{\partial {\cal F}}  \right|_{\rm B} = C_1~~~~
\left. \frac{\partial{\cal L}}{\partial {\cal G}}  \right|_{\rm B} = C_2~~~~
\left. \frac{\partial^2{\cal L}}{\partial {\cal F}^2}  \right|_{\rm B} = D_1~~~~ \left. \frac{\partial^2{\cal L}}{\partial {\cal G}^2}  \right|_{\rm B} = D_2~~~~ \left. \frac{\partial^2{\cal L}}{\partial {\cal F}\partial {\cal G}}  \right|_{\rm B} = D_3~,
\eeq

\beq
s^{\mu\nu\kappa\lambda} = D_1 F^{\mu\nu} F^{\kappa\lambda}
+ D_2 G^{\mu\nu} G^{\kappa\lambda} ~~~~~~~~~~~~~~~~~~~~~~~~~ 
t^{\mu\nu\kappa\lambda} = D_3 F^{\mu\nu} F^{\kappa\lambda}~,
\eeq

In this context, it is natural to obtain a frequency shift also in the NLEM framework. Whether this shift is accompanied necessarily by a photon with an effective mass, it is subject of on-going investigations.  

\section{Models of the ETE shifts and their possible contributions to the observations.}

We propose four functions for the ETE frequency shifts and discuss their possible contributions to astrophysical observations and how they
could affect the interpretation of cosmological models.

\begin{itemize}
\item{An estimate of the frequency change that light would undergo was given \cite{helayelneto-spallicci-2019,spallicci-etal-2021}. In terms of magnitude $z_{\rm S}$ can range from a negligible quantity to the totality of the red shift. Such a large range is explained by}:
\begin{itemize}
\item{the uncertainties of the values of the background magnetic fields (galactic and inter-galactic fields, roughly $10^{-10}$ - $10^{-9}$ T) \cite{alves-ferriere-2018} and where applicable, of the values of the LSV fields (there are about twenty orders of magnitude difference between the terrestrial and astrophysical estimates \cite{gomesmalta2016,kosteleckyrussell2011}).}
\item{the lack of information on the alignments of the multiple fields crossed by the photon; differently oriented magnetic fields may provide opposite effects to the photon frequency.}
\item{possibly the space-time dependency of the LSV and electromagnetic fields.}
\end{itemize}
\item{The frequency shift $z_{\rm S}$ is very much light-path dependent. A remarkable feature of this additional frequency shift is its natural capability of explaining different red shifts of equally distant sources.}
\item{The frequency shift can be towards the red or the blue.}
\end{itemize}

To any SN Ia a specific $z_{\rm S}$ can be associated, towards the red or the blue, big or small. Independently of
any model, the individual effects, as well the general trend emerging from the Pantheon sample, complemented by the BAO data, can be both checked.

We recall the definition of $z = \Delta \nu/\nu_o$, where $\Delta \nu = \nu_{\rm e} - \nu_o$ is the difference between the observed $\nu_o$ and emitted 
$\nu_{\rm e}$ frequencies, or else $z = \Delta \lambda/\lambda_{\rm e}$ for the wavelengths. Expansion obliges $\lambda_{\rm e}$ to stretch to $\lambda_{\rm C}$; that is, $\lambda_{\rm C} = (1+z_{\rm C})\lambda_{\rm e}$. The quantity $z_{\rm C}$ refers only to expansion. 

The wavelength $\lambda_{\rm C}$ could be further stretched or conversely shrunk for the ETE shift $z_{\rm S}$ to $\lambda_{\rm o} = 
(1+z_{\rm S})\lambda_{\rm C}  = (1+z_{\rm S}) (1+z_{\rm C}) \lambda_{\rm e}$. 
But since $\lambda_{\rm o} = (1+z)\lambda_{\rm e}$, we have 
$1 +  z = (1+z_{\rm C}) (1+z_{\rm S})$; thus
\vspace*{-0.2cm} 

\begin{eqnarray}
z = z_{\rm C} + z_{\rm S}  + z_{\rm C}z_{\rm S} \coloneqq z_{\rm o}~. 
\label{newz}
\end{eqnarray}
where $z_{\rm o}$ is the spectroscopically or photometrically observed $z$. The second order is non-negligible for larger $z_{\rm C}$. 

The procedure for computing $z_{\rm S}$ considers that the expansion of the universe is solely represented by $z_{\rm C}$, with respect to which we derive the cosmological distances (and not with respect to the observed red shift $z$). From Eq. (\ref{newz}), $z_{\rm C}$ is given by 

\beq 
    z_{\rm C}=\frac{z-z_{\rm S}}{1+z_{\rm S}}~.
\label{zc}
\eeq

We note that if $z_{\rm S}$ is blue (negative), the photon gains energy in the path to us due to one of the ETE processes. This in turn means that the expansion red shift $z_{\rm C}$ is bigger than the observed $z$, which in turn means that the astrophysical object is actually further than what we could have detected in the $\Lambda$CDM model. If instead $z_{\rm S}$ is red (positive), the photon loses energy in its path, implying that $z_{\rm C}$ is smaller than $z$, meaning that the astrophysical object is closer to us than what it would been dictated by the same $\Lambda$CDM model.

Modelling the behaviour of $z_{\rm S}$ with respect to the distance and the frequency, we suppose four different behaviours of the frequency variation that can be proportional to:

\begin{itemize}
    \item the instantaneous frequency and the distance;
    \item the emitted frequency and the distance;
    \item only the distance;
    \item the observed frequency and the distance.
\end{itemize}

The relations between $z_{\rm S}$, the distance $r$ and the frequency $\nu$ that are shown in Tab. \ref{tabdeltanu}. Two comments related to this table are in order: first, the distance $r$ appearing in the $z_{\rm S}$ formulas is the light-travel distance, defined as

\begin{equation} \label{light travel}
    r=\frac{c}{H_0} \int_0^z \frac{dz'}{(1+z')E(z')}~,
\end{equation}
which is the actual distance travelled by the photon to reach us in an expanding universe. The function $E(z)$ is defined in the most general form by

\begin{equation} \label{E(z)}
    E(z)=\frac{H(z)}{H_0}=\sqrt{\Omega_r(1+z)^4+\Omega_M(1+z)^3+\Omega_k(1+z)^2+\Omega_{\Lambda}}~,
\end{equation}
where the different density contributions are reported: radiation $\Omega_r$, matter $\Omega_M$, curvature $\Omega_k$ and dark energy (or cosmological constant) $\Omega_{\Lambda}$. Second, referring to the Pantheon database, the coefficients $k_{i}$ can either be sought as depending on the specific SN Ia or else as an unique value valid for all SNe Ia. Both roads are viable, and we are going to follow both in our computations.

\begin{table}
    \centering
    \begin{tabular}{|c|c|c|c|c|}
    \hline
       Type  &  1 & 2 & 3 & 4\\[6pt]\hline
         $d\nu$ & $k_{1}\nu dr$ &$k_{2}\nu_e dr$ &$k_{3} dr$ &$k_{4}\nu_o dr$ \\[6pt]\hline
         $\nu_o$ & $\nu_e e^{k_1 r}$ & $\nu_e (1+k_2 r)$ & $\nu_e + k_3 r$ & ${\ds \frac{\nu_e}{1-k_4 r}}$ \\[6pt]\hline
         $z_{\rm S}$ & $e^{-k_1 r}-1$ & ${\ds -\frac{k_2 r}{1+k_2 r}}$ & ${\ds -\frac{k_3 r}{\nu_e+k_3 r}}$ & $-k_4 r$ \\[6pt]\hline
           $k_i$ & $-{\ds \frac{\ln(1+z_{\rm S})}{r}}$  
         & $-{\ds \frac {z_{\rm S}}{r(1 + z_{\rm S})}}$ 
         & $-{\ds \frac {\nu_{\rm e} z_{\rm S}}{r(1 + z_{\rm S})}}$ 
         & $-{\ds \frac {z_{\rm S}}{r}}$ \\[6pt]\hline
         $r$ & $-{\ds \frac{\ln(1+z_{\rm S})}{k_1}}$  
         & $-{\ds \frac {z_{\rm S}}{k_2(1 + z_{\rm S})}}$ 
         & $-{\ds \frac {\nu_{\rm e} z_{\rm S}}{k_3(1 + z_{\rm S})}}$ 
         & $-{\ds \frac {z_{\rm S}}{k_4}}$ \\[6pt]\hline
         \end{tabular}
    \caption{The different variations of the frequency $\nu$ can be summarised by four different cases of proportionality: 1. to the instantaneous frequency and the distance; 2. to the emitted frequency and the distance; 3. to the distance only; 4. to the observed frequency and the distance. These variations determine the frequency observed $\nu_o$, the shift $z_{\rm S}$, the parameters $k_i$ and the distance $r$. The positiveness of the distance $r$ constraints $z_{\rm S} > 0 $ for $k_1<0$, and $- 1 < z_{\rm S} < 0 $ for $k_1 >0$; 
    $ z_{\rm S}  > 0$ for $k_2<0$, and $- 1 < z_{\rm S} < 0 $ for $k_2 >0$; 
    $ z_{\rm S}  > 0$ for $k_3<0$, and $- 1 < z_{\rm S} < 0 $ for $k_3 >0$; 
    $ z_{\rm S}  > 0$ for $k_4<0$, and $z_{\rm S} < 0 $ for $k_4 >0$. }
    \label{tabdeltanu}
\end{table}

For deriving the values of $z_{\rm S}$ necessary to replace the effects due to dark energy, we assume cosmological models where $\Omega_{\Lambda}=0$. Considering also the radiation contribution as negligible, we follow three main models:

\begin{itemize}
    \item {Cosmology model A: we set $\Omega_{M}=0.3$ and consider $\Omega_{K}=0$, implying a flat universe where the "cosmic triangle" relation $\Omega_{M}+\Omega_{K}+\Omega_{\Lambda}=1$, is not satisfied {\it ab initio}. Nevertheless, the dark energy effect could be replaced {\it a posteriori} by the effect of $z_{\rm S}$. This approach supposes that $z_{\rm S}$ is a manifestation of the LSV vacuum energy in string models, in case of the SME \cite{kosteleckysamuel1989b}}.
    
    \item {Cosmology model B: we take into account an open universe model, where $\Omega_{M}=0.3$ and $\Omega_{K}=0.7$, so that $\Omega_{K}+\Omega_{M}=1$.}
    
    \item {Cosmology model C: we return to the Einstein-de Sitter conception \cite{einstein-desitter-1932}, that is a flat, matter dominated universe with $\Omega_{M}=1$. This was one of the most popular cosmological model before the advent of the dark energy hypothesis \cite{perlmutter-etal-1997}.}
    
    \end{itemize}
    
    We run over three different values of the Hubble-Lema\^itre parameter $H_0$ km s$^{-1}$ per Mpc: 67 (consistent with the result given by the Planck data coming from the Cosmic Microwave Background (CMB) radiation \cite{aghanimetal2020}), 74 (consistent with the cosmological ladder measurements \cite{riess-etal-2019}), and 70 which is an average and rounded intermediate value.  
    
In the next section, we explore first mock red shifts at regular intervals that do not correspond to any observation, and for which we have computed the correction $z_{\rm S}$ to mimic the effects of dark energy without recurring to the accelerated expansion in all the aforementioned cosmological models. 

Later, we explore real data related to the SNe Ia from the Pantheon Sample as well as the BAO constraints. In both cases, we have used the luminosity distance \cite{hogg-1999}
\begin{equation} \label{luminosity distance}
    d_{\rm L}(z)=(1+z)d_{\rm M}(z)~,
\end{equation}
where $d_{\rm M}(z)$ is the transverse comoving distance
\begin{equation} \label{comoving flat}
    d_{\rm M}(z)=\frac{c}{H_0} \int_0^z \frac{dz'}{E(z')}~,
\end{equation}
for $\Omega_{K}=0$, and
\begin{equation} \label{comoving open}
    d_{\rm M}(z)=\frac{c}{H_0 \sqrt{\Omega_{K}}} \sinh \biggl(\frac{H_0 \sqrt{\Omega_{K}}}{c} \int_0^z \frac{dz'}{E(z')} \biggr)~,
\end{equation}
for $\Omega_{K}>0$. 

Applying Eq. (\ref{luminosity distance}) to SN Ia data, the factor $(1+z)$ refers to the heliocentric red shift, that is the apparent red shift affected by the relative motion of the Sun, and not to the red shift observed in a cosmological rest frame \cite{kenworthy-scolnic-riess-2019},  Instead, the latter appears in the integrals of Eqs. (\ref{comoving flat}, \ref{comoving open}).  Both these quantities are provided by the Pantheon catalogue. We underline that these quantities have been computed in our approach considering the expansion red shift $z_{\rm C}$ for all the probes we have used.

For the study of the Pantheon sample, as mentioned before, we have chosen to compute the best fit values of the $k_i$ parameters and of $z_{\rm S}$ both considering each SN Ia individually or finding a general value valid for all SNe Ia, through a best fit, accounting for the covariance matrix of the systematic and statistical errors. For both these cases, we have used the distance-modulus, defined as (length units in Mpc)

   \begin{equation} \label{distance modulus}
        \mu=m-M=5\log_{10}\left [d_{\rm L}(z_{\rm C})\right]+25~.
    \end{equation}
    
    Incidentally, we emphasise once more that the distance is computed with respect to the expansion red shift $z_{\rm C}$. We have fixed the value for the absolute magnitude $M$ for the SNe Ia to -19.35 \cite{dainotti-etal-2021}, to avoid the degeneracy with $H_0$, thus observing how our results change when varying the Hubble-Lema\^itre constant while keeping constant the intrinsic luminosity for the SNe Ia sample. 
    
    When we evaluate the effect for each SN Ia individually, we minimise the following quantity for each SN Ia
    \begin{equation} \label{min each SN}
        \chi_1^2=\frac{(\mu_{obs}-\mu_{th})^2}{(\mu_{obs, err})^2}~,
    \end{equation}
    where $\mu_{obs}$ is the distance-modulus given by the Pantheon Sample with the correspondent error $\mu_{obs, err}$, while $\mu_{th}$ is the theoretical distance-modulus computed according to our cosmological models including  $z_{\rm S}$. We emphasise that this is not the $\chi^2$ of the entire Pantheon Sample: the values derived by these computations are individually belonging to each SN Ia. This approach allows to explain each SN Ia location in the ($\mu$, $z$) diagram, due to different 
    SN Ia host environments, light paths, intervening electromagnetic (and LSV if applicable) fields and their alignments, and obviously distance.
    
    For the second approach, for which we consider all the SN Ia data and find a best fit, we use the following $\chi^2$ function
    \begin{equation} \label{eq_chi2_SNe}
    \chi^2= (\mu_{th}-\mu_{obs})^T\times \mathcal{C}^{-1} \times (\mu_{th}-\mu_{obs})~,   
    \end{equation}
    where $\mathcal{C}^{-1}$ is the inverse of the covariance matrix of the Pantheon Sample \cite{scolnic-etal-2018, dainotti-etal-2021}. For this case, a single $k_{i}$ value is computed as a best fit parameter for all SNe Ia, which implies that $z_{\rm S}$ depends only on the light-travel distance between a specific SN Ia and us, see Tab. \ref{tabdeltanu}. The advantages of this approach are the characterisation of the three cosmological models through single $k_i$ values, but even more the possibility of studying the impact of the variation of other cosmological parameters, as the $\Omega$ densities and $H_0$, and the $k_i$ parameters.
    
    The BAO data are accounted only in this second approach. Although the BAO data points are only 15, behind each of those there are thousands of observations of galaxies, quasars and Lyman $\alpha$ forests \cite{alam-etal-2021}. Given their "mean value" nature, we have included them in the second general approach and not in the first individual one (in which they would have added 15 points to 1048 ones provided by the Pantheon Sample). 
    
    Furthermore, while the SNe Ia detections are all associated to the distance-modulus and the luminosity distance, this is not the case for the BAOs. Indeed, the quantities measured using BAOs are related through the following equations
    
    \begin{equation}
        d_V(z)=\biggr[{d_M}^2(z)\frac{cz}{H(z)} \biggr]^\frac{1}{3},
    \label{eq_dilationscale}
    \end{equation}
    \begin{equation}
        A(z)=\frac{100 d_V(z)\sqrt{\Omega_M h^2}}{cz},
        \label{Aparameter}
    \end{equation}
    \begin{equation}
        d_H(z)=\frac{c}{H(z)},
        \label{dH}
    \end{equation}
    and the comoving distance defined in Eqs. (\ref{comoving flat}, \ref{comoving open}). Here, $h=H_0/100$ km s$^{-1}$ per Mpc. Also in \cite{alam-etal-2021}, the results reported of these detected quantities are rescaled by the sound horizon $r_d$, for which, in our computations, we have used the following numerical approximation \cite{aubourg-etal-2015,sharov-2016}
    
    \begin{equation}
    r_d=\frac{55.154 \cdot e^{[-72.3(\Omega_{\nu}h^{2}+0.0006)^2]}}{(\Omega_{M}h^{2})^{0.25351}(\Omega_{b}h^{2})^{0.12807}}Mpc~,
    \label{eq_rsfiducialtrue}
    \end{equation}
    where $\Omega_{b}$ is the baryonic density in the universe, set to the value $\Omega_{b}\cdot h^2=0.02237$ \cite{aghanimetal2020}, while $\Omega_{\nu}$ is the neutrino density in the universe, fixed to the value provided by the $\Lambda$CDM model $\Omega_{\nu} \cdot h^2=0.00064$.
    
    Regarding the $\chi^2$ associated to the BAOs, we considered each subset of the total BAOs sample, given the different quantities measured and the possible covariance terms \cite{beutler-etal-2011, blake-etal-2011,ross-etal-2015,dumasdesbourboux-etal-2020, alam-etal-2021}. The forms of these functions are the same of Eqs. (\ref{min each SN}, \ref{eq_chi2_SNe}).
    
   We use a Bayesian approach, assigning flat priors (specified in the next section) to the variables under consideration. The computations rely on \textit{Cobaya} \cite{torrado-lewis-2021}, a code for Bayesian analysis in Python, that adopts a Markow-Chain Monte-Carlo (MCMC) method.
 
\section{Results}

\subsection{Mock red shifts}

\begin{figure}
    \centering
    \includegraphics[width=0.33\hsize,height=0.3\textwidth,angle=0,clip]{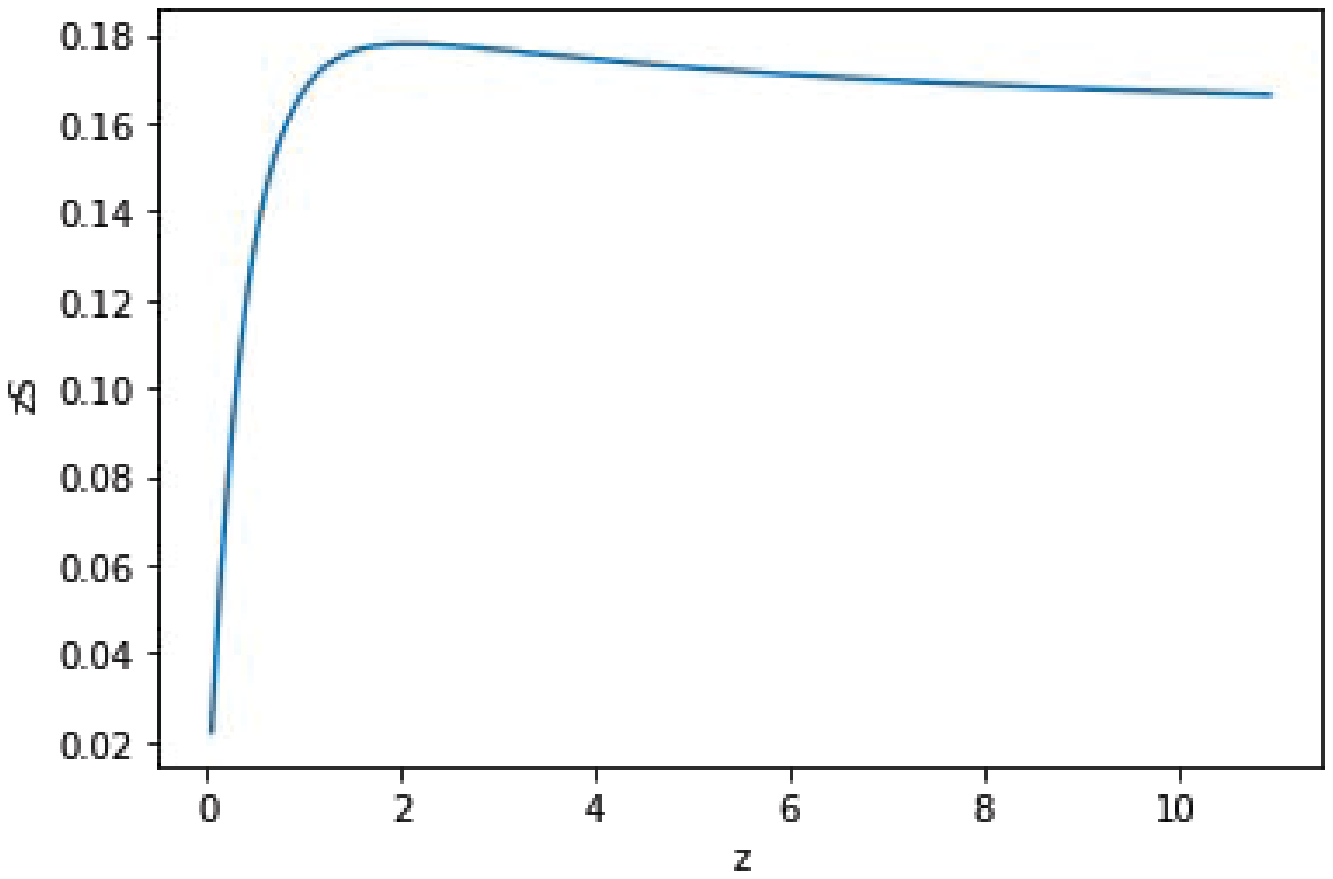}
    \includegraphics[width=0.33\hsize,height=0.3\textwidth,angle=0,clip]{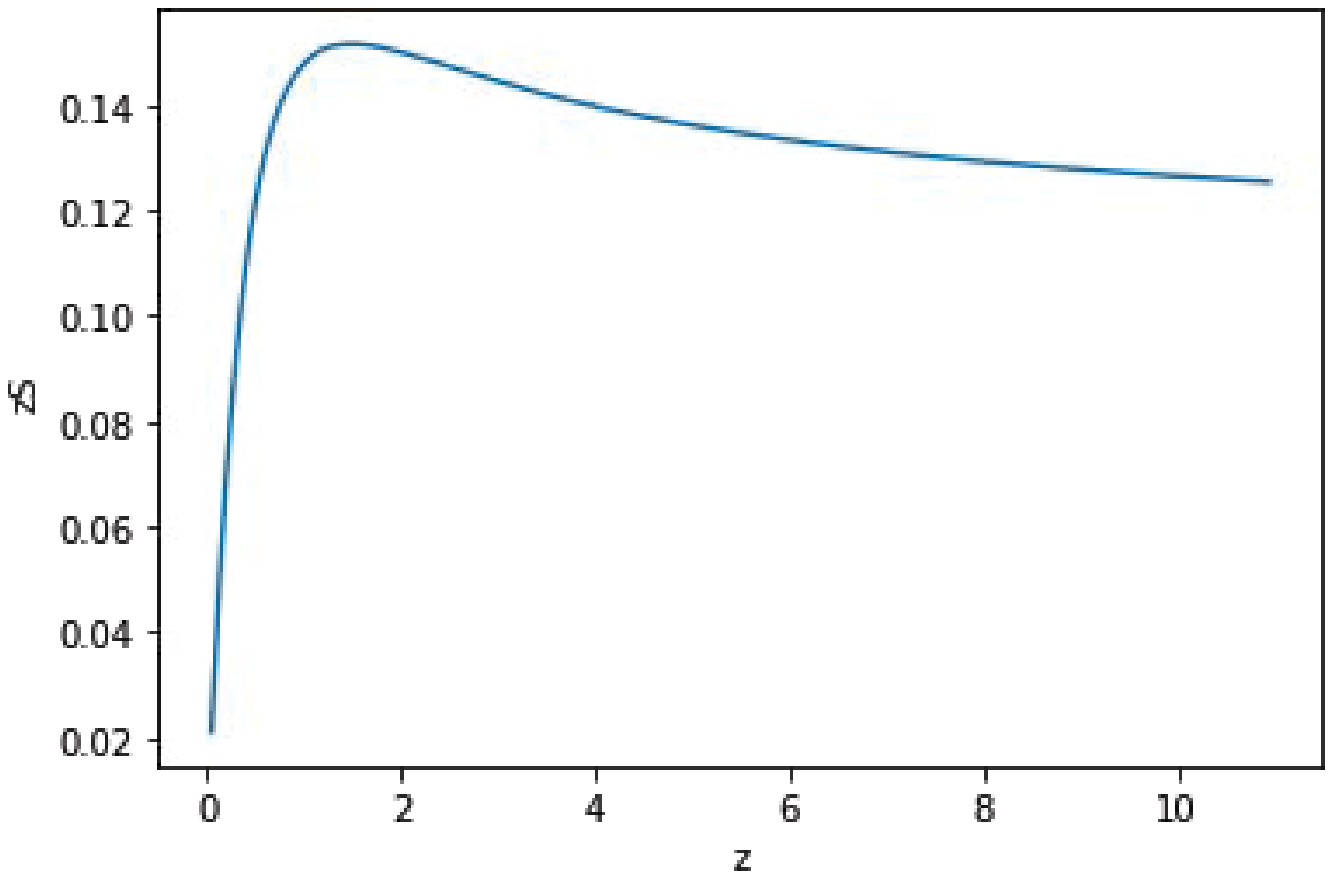}
    \includegraphics[width=0.33\hsize,height=0.3\textwidth,angle=0,clip]{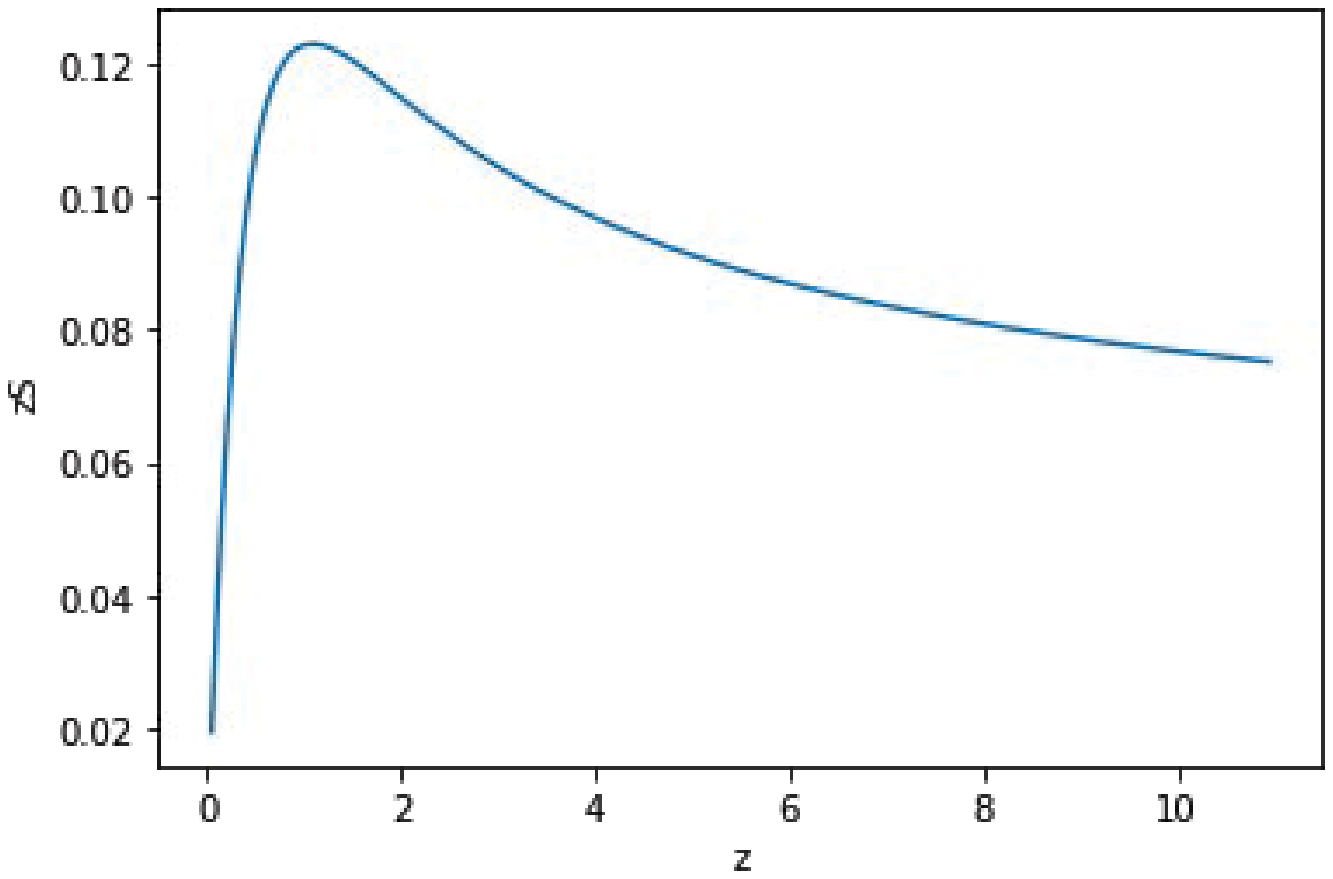}
    \caption{Plots of $z_{\rm S}$ versus the mock $z$ for the Cosmology model A, where $\Omega_{M}=0.3$, $\Omega_{k}=\Omega_{\Lambda}=0$. We have considered $H_0=67, 70, 74$ for the left, central and right panels, respectively. The values for $H_0$ are in km s$^{-1}$ per Mpc. The ratio $z_{\rm s}/z$ goes from $1.5 \%$ (for $z\approx 11$) to $44.7\%$ (for $z\approx 0$), left panel; from $1.1\%$ ($z\approx 11$) to $42.7\%$ ($z\approx 0$), central panel; from $0.7\%$ (for $z\approx 11$) to $39.1\%$ (for $z\approx 0$), right panel. The peaks for the absolute values of $z_{\rm s}$ have been reached for $z=2.1$ (left panel), $z=1.5$ (central panel), and $z=1.1$ (right panel). The shifts in all cases are towards the red and thus dissipative.}
    \label{fig:firstthreecasesbuiltredshift}
\end{figure}

\begin{figure}
    \centering
    \includegraphics[width=0.33\hsize,height=0.3\textwidth,angle=0,clip]{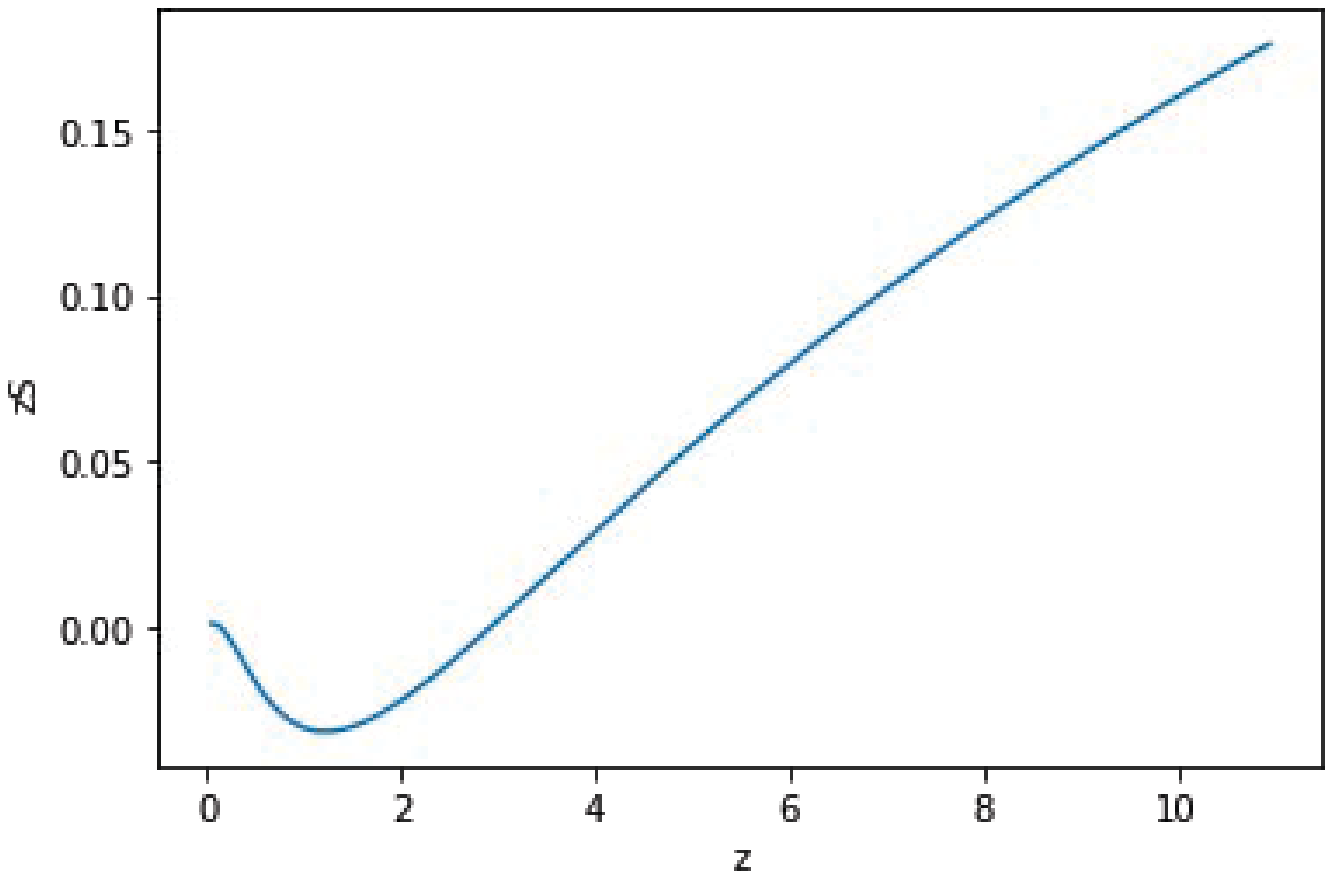}
    \includegraphics[width=0.33\hsize,height=0.3\textwidth,angle=0,clip]{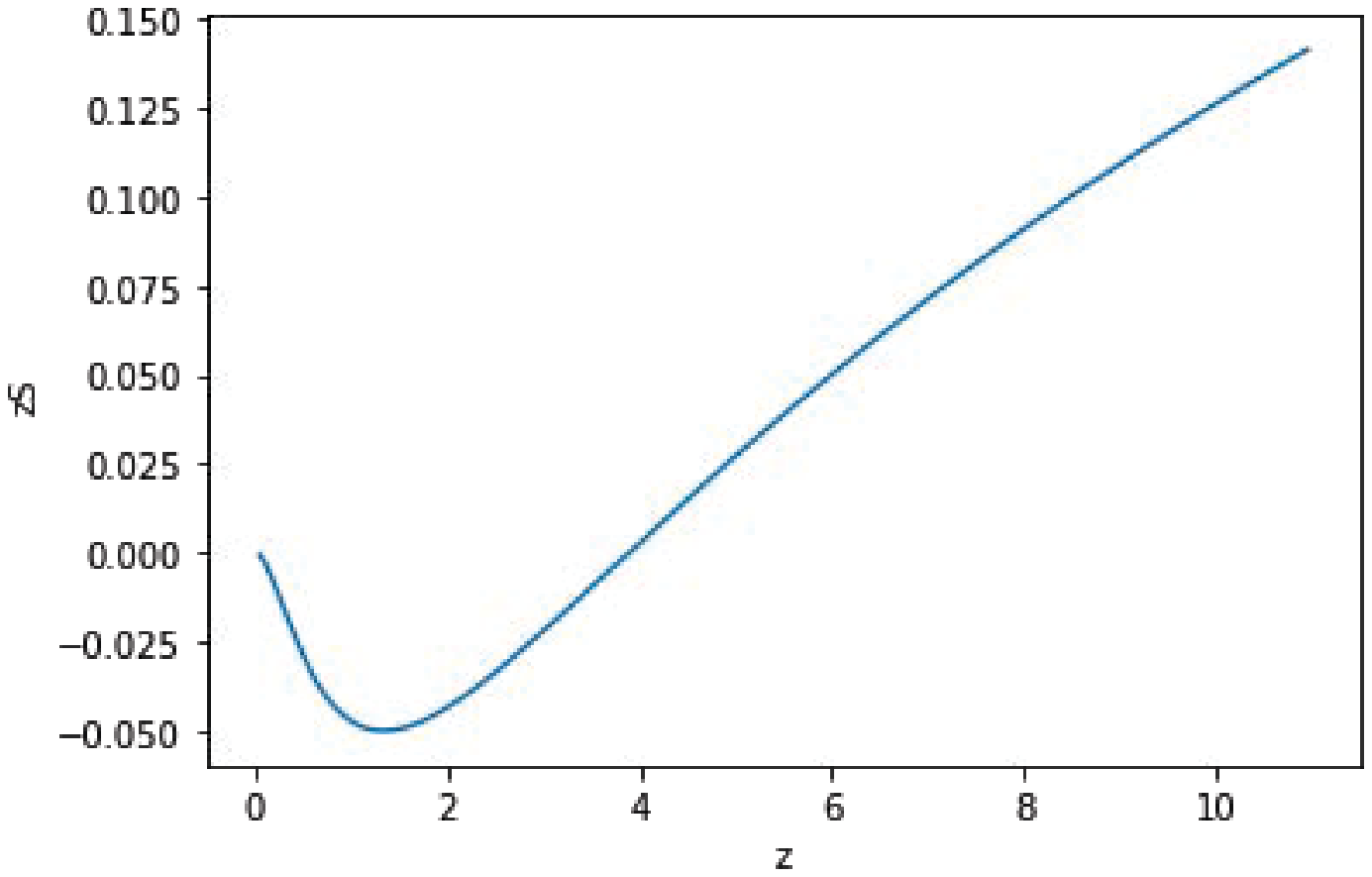}
    \includegraphics[width=0.33\hsize,height=0.3\textwidth,angle=0,clip]{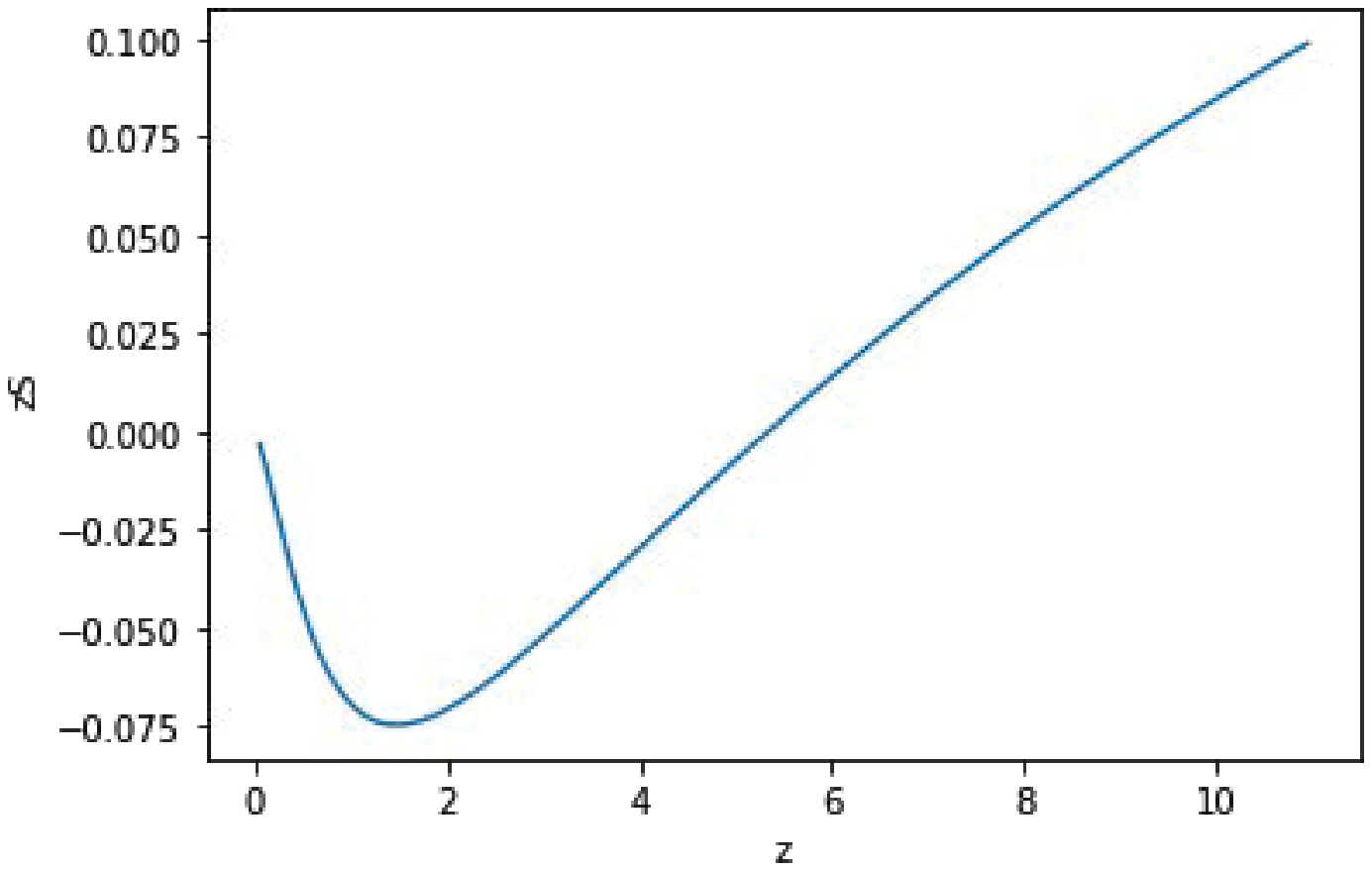}
    \caption{Plots of $z_{\rm S}$ versus the mock $z$ for the Cosmology model B, where $\Omega_{M}=0.3$, $\Omega_{k}=0.7$, and $\Omega_{\Lambda}=0$. We have considered $H_0=67, 70, 74$ for the left, central and right panels, respectively. The values for $H_0$ are in km s$^{-1}$ per Mpc. The ratio $z_{\rm s}/z$ goes from $-3.4 \%$ (for $z\approx 0.65$) to $2.5\%$ (for $z\approx 0$), left panel; from $-5.8\%$ (for $z\approx 0$) to $2,5\%$ (for $z\approx 11$), central panel; from $-9.3\%$ (for $z\approx 0.35$) to $0.9\%$ (for $z\approx 11$), right panel. The minima for the values of $z_{\rm s}$ correspond to  $z=1.2$ (left panel), $z=1.35$ (central panel), and $z=1.45$ (right panel). The shifts are towards the red or the blue, depending on the value of $z$.}
    \label{fig:secondthreecasesbuiltredshift}
\end{figure}

\begin{figure}
    \centering
    \includegraphics[width=0.33\hsize,height=0.3\textwidth,angle=0,clip]{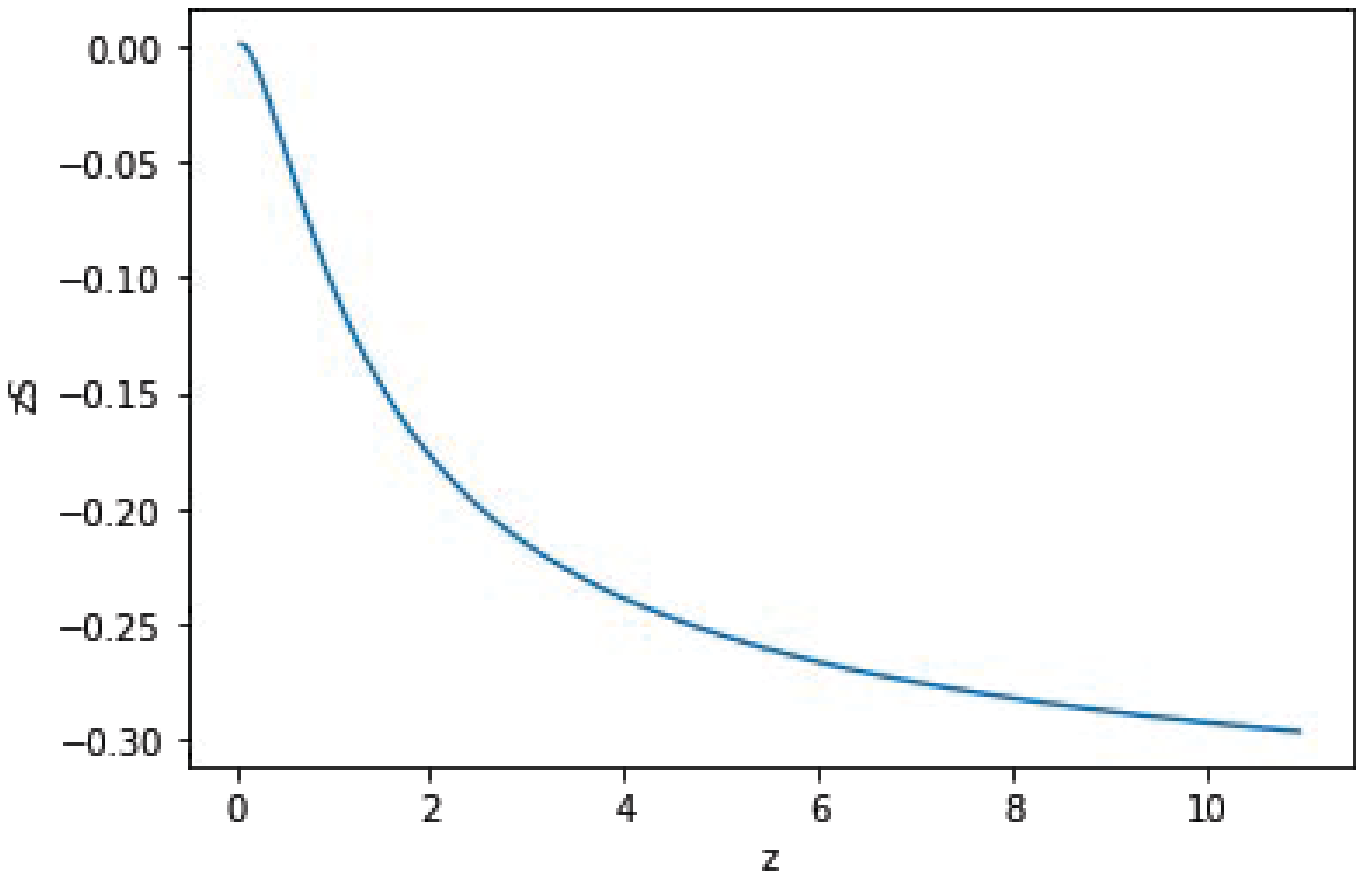}
    \includegraphics[width=0.33\hsize,height=0.3\textwidth,angle=0,clip]{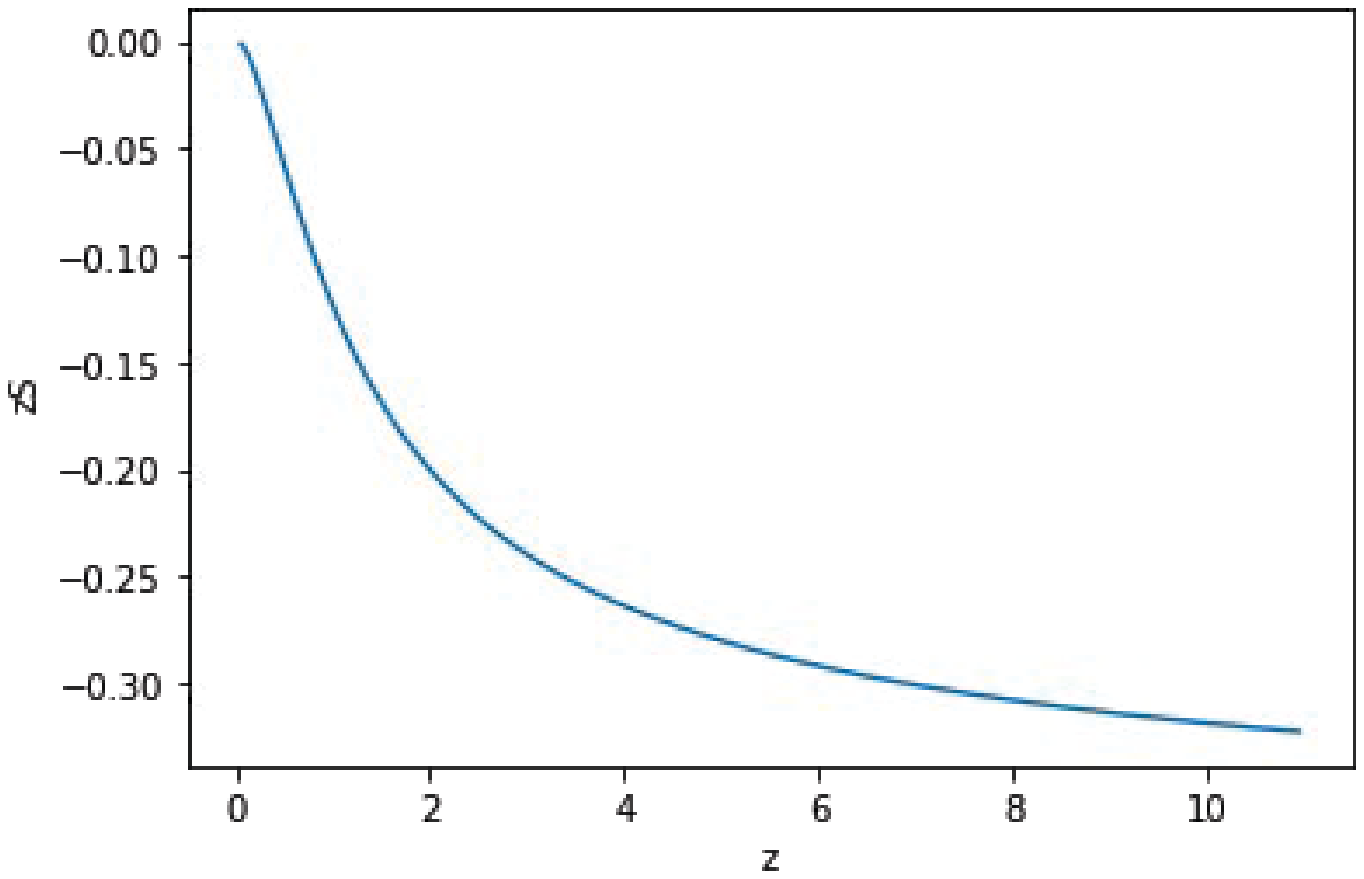}
    \includegraphics[width=0.33\hsize,height=0.3\textwidth,angle=0,clip]{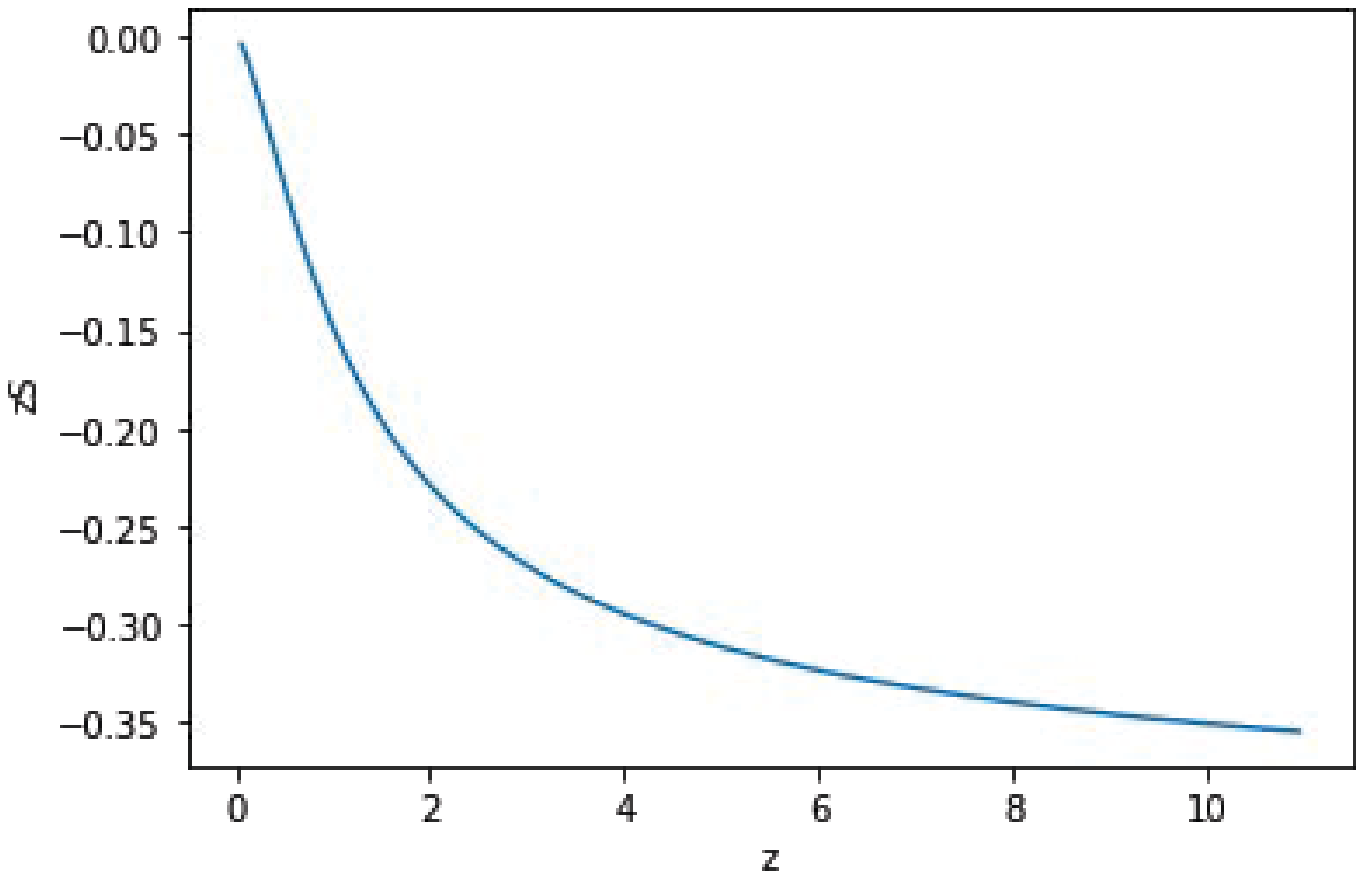}
    \caption{Plots of $z_{\rm S}$ versus the mock $z$ for the Cosmology model C, where $\Omega_{M}=1$, $\Omega_{k}=\Omega_{\Lambda}=0$. We have considered $H_0=67, 70, 74$ for the left, central and right panels, respectively. The values for $H_0$ are in km s$^{-1}$ per Mpc. The ratio $z_{\rm s}/z$ goes from $-10.4 \%$ (for $z\approx 1$) to $1.7\%$ (for $z\approx 0$) and the shifts are towards the red and the blue (left panel); from $-12.5\%$ (for $z\approx 0.8$) to $-2,4\%$ (for $z\approx 0$) and the shifts are only towards the blue (central panel); from $-15.5\%$ (for $z\approx 0.65$) to $-3.2\%$ (for $z\approx 11$) and the shifts are only towards the blue (right panel).}
    \label{fig:thirdthreecasesbuiltredshift}
\end{figure}

We start dealing with mock red shifts and later we will deal with red shifts issued from the catalogues. The range of red shifts spans a large
interval from $z=0$ up to $z=11$ to encompass the furthest galaxy detected at  $z=11.09$ \cite{oesch-etal-2016}. Once more, we search the values of $z_{\rm S}$ that, inserted in Eqs. (\ref{newz}, \ref{luminosity distance}), would display the same effects of dark energy in the three Cosmology models under considerations. 
The results are shown in Figs. \ref{fig:firstthreecasesbuiltredshift}, \ref{fig:secondthreecasesbuiltredshift},
\ref{fig:thirdthreecasesbuiltredshift}. 

The first set of three plots, Fig. \ref{fig:firstthreecasesbuiltredshift}, shows the behaviour of $z_{\rm S}$ versus the mock red shift $z$ for Cosmology model A, where $\Omega_{M}=0.3$, $\Omega_{k}=\Omega_{\Lambda}=0$. We observe that $z_{\rm S}$ is always positive, thus red, which means that in this model the non-standard effects are dissipative for the photons. This occurs for all three values of $H_0$ considered (67, 70, 74 km s$^{-1}$ per Mpc). We also note a peak for $z_{\rm S}$ for the values $z$ shown in the caption of Fig. \ref{fig:firstthreecasesbuiltredshift}, after which we see a decrease of the value, with a variable steepness depending on the value of $H_0$. The steepness increases, while the peak values decreases with $H_0$.

The second set of three plots, Fig. \ref{fig:secondthreecasesbuiltredshift}, shows the behaviour of $z_{\rm S}$ versus the mock red shift $z$ for Cosmology model B, where $\Omega_{M}=0.3$, $\Omega_{k}=0.7$, and $\Omega_{\Lambda}=0$. We note a very different behaviour with respect to Cosmology model A: $z_{\rm S}$ has a negative peak for $z$ in the region around $1.3$ - the precise values are stated in the caption of  Fig. \ref{fig:secondthreecasesbuiltredshift} - then increases monotonically, becoming positive. The value of $H_0$ affects the size of the negative pick as well as the steepness during the subsequent increase, but it does not affect the overall behaviour. We recall that a negative $z_{\rm S}$ indicates an increase of the energy of the photons. These results are in agreement with the trend observed in \cite{spallicci-etal-2021}. The behaviour of $z_{\rm S}$ for $0 \leq z \leq 2.5$ echoes the behaviour of dark energy which effects of are mainly relevant in the same range of $z$.  The rising behaviour of $z_{\rm S}$ at large $z$ seems to indicate that the expansion decelerates far away. The curve moves downward with increasing $H_0$.

The third set of three plots, Fig. \ref{fig:thirdthreecasesbuiltredshift}, shows the behaviour of $z_{\rm S}$ versus the mock red shift $z$ for Cosmology model C, where $\Omega_{M}=1$, $\Omega_{k}=\Omega_{\Lambda}=0$. We observe that $z_{\rm S}$ is always negative, and decreases with $z$. Once more increasing $H_0$ shifts the curve downward without changing the overall behaviour. 

This test has demonstrated how different cosmological models influence the results and tell us which trends we should expect when using the Pantheon Sample in the red shift region covered by the real data. 

\subsection{SNe Ia catalogue data}

\subsubsection{Histograms and scatter plots of $z_{\rm S}$}

\begin{figure}
    \centering
    \includegraphics[width=0.33\hsize,height=0.3\textwidth,angle=0,clip]{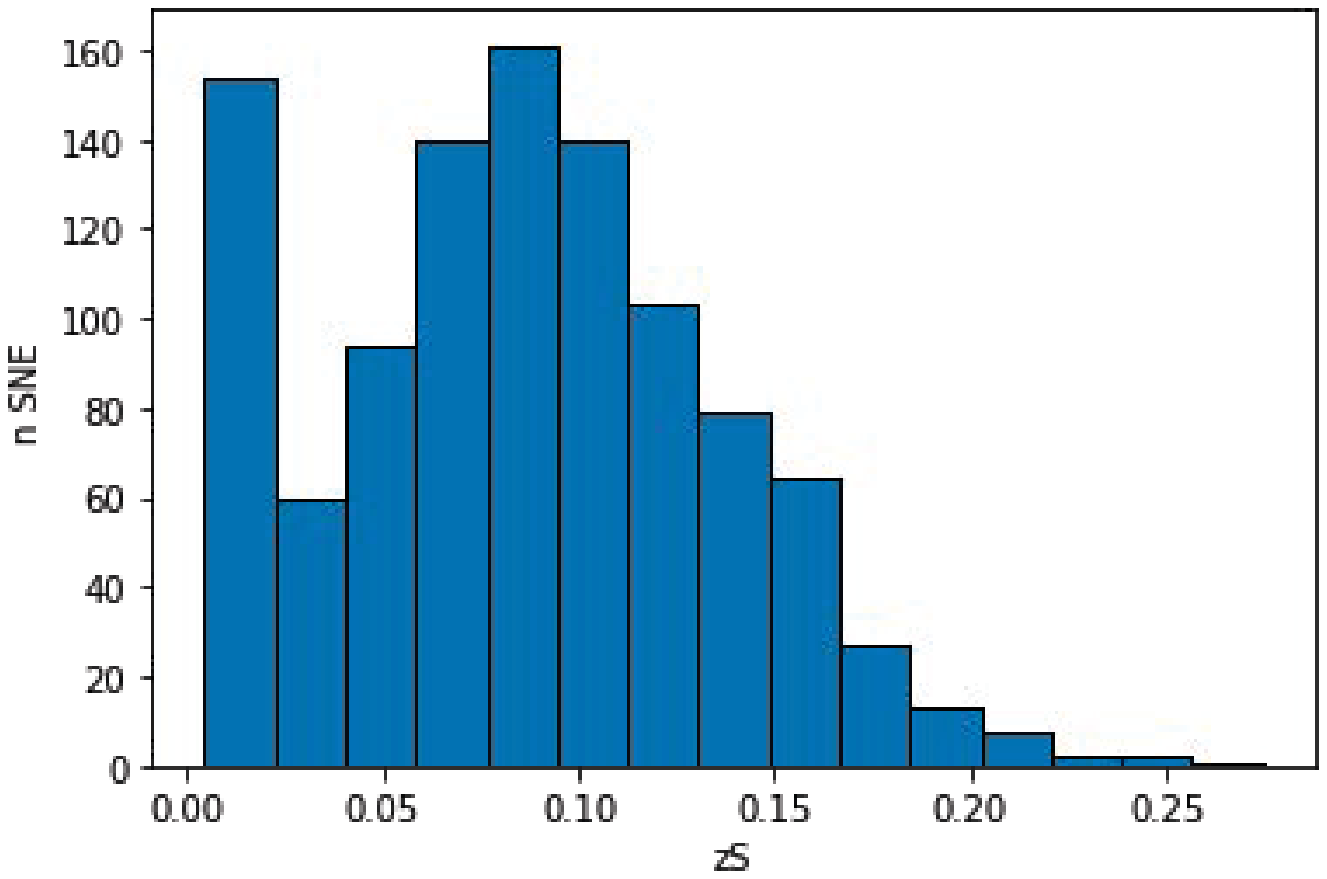}
    \includegraphics[width=0.33\hsize,height=0.3\textwidth,angle=0,clip]{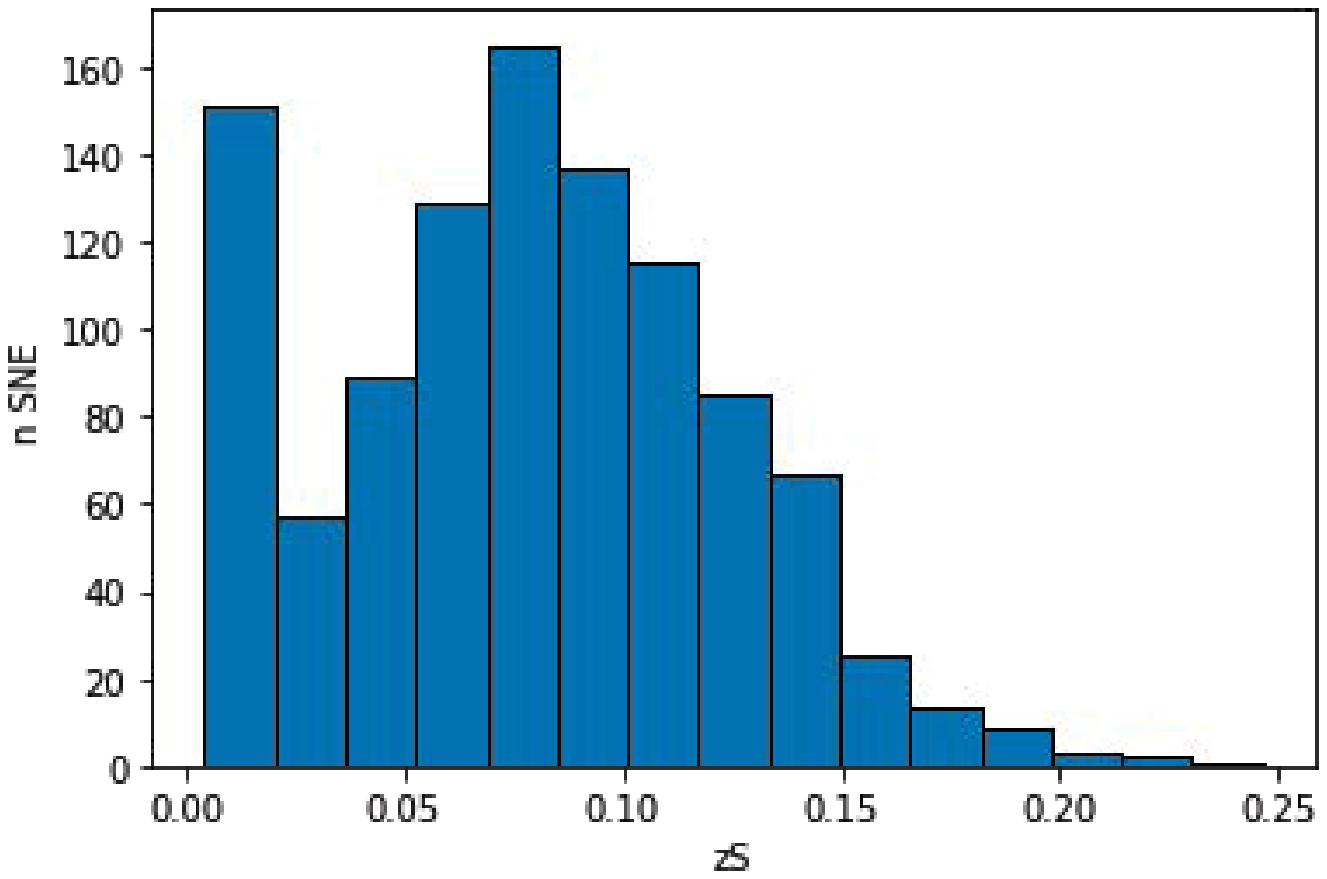}
    \includegraphics[width=0.33\hsize,height=0.3\textwidth,angle=0,clip]{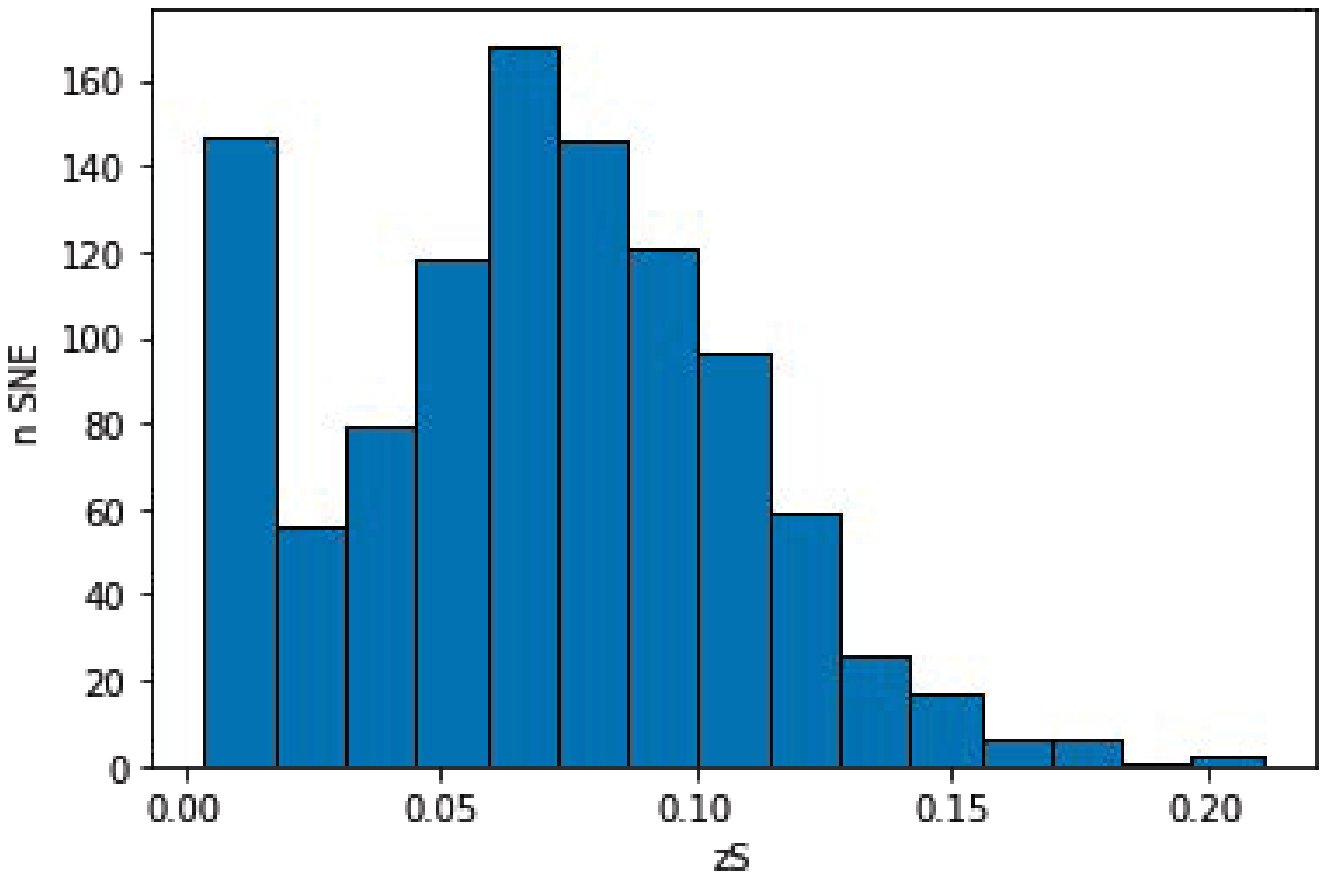}
    \includegraphics[width=0.33\hsize,height=0.3\textwidth,angle=0,clip]{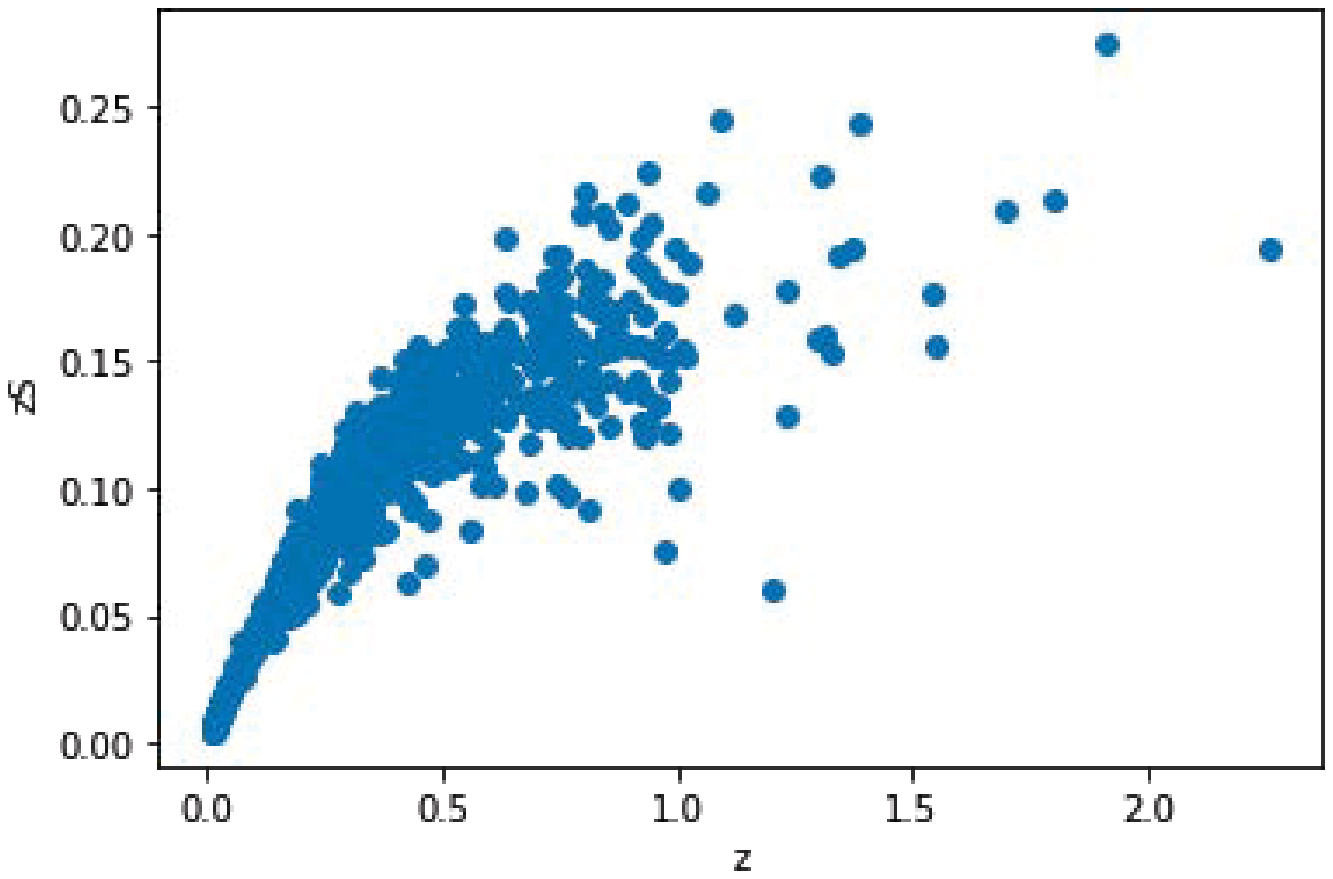}
    \includegraphics[width=0.33\hsize,height=0.3\textwidth,angle=0,clip]{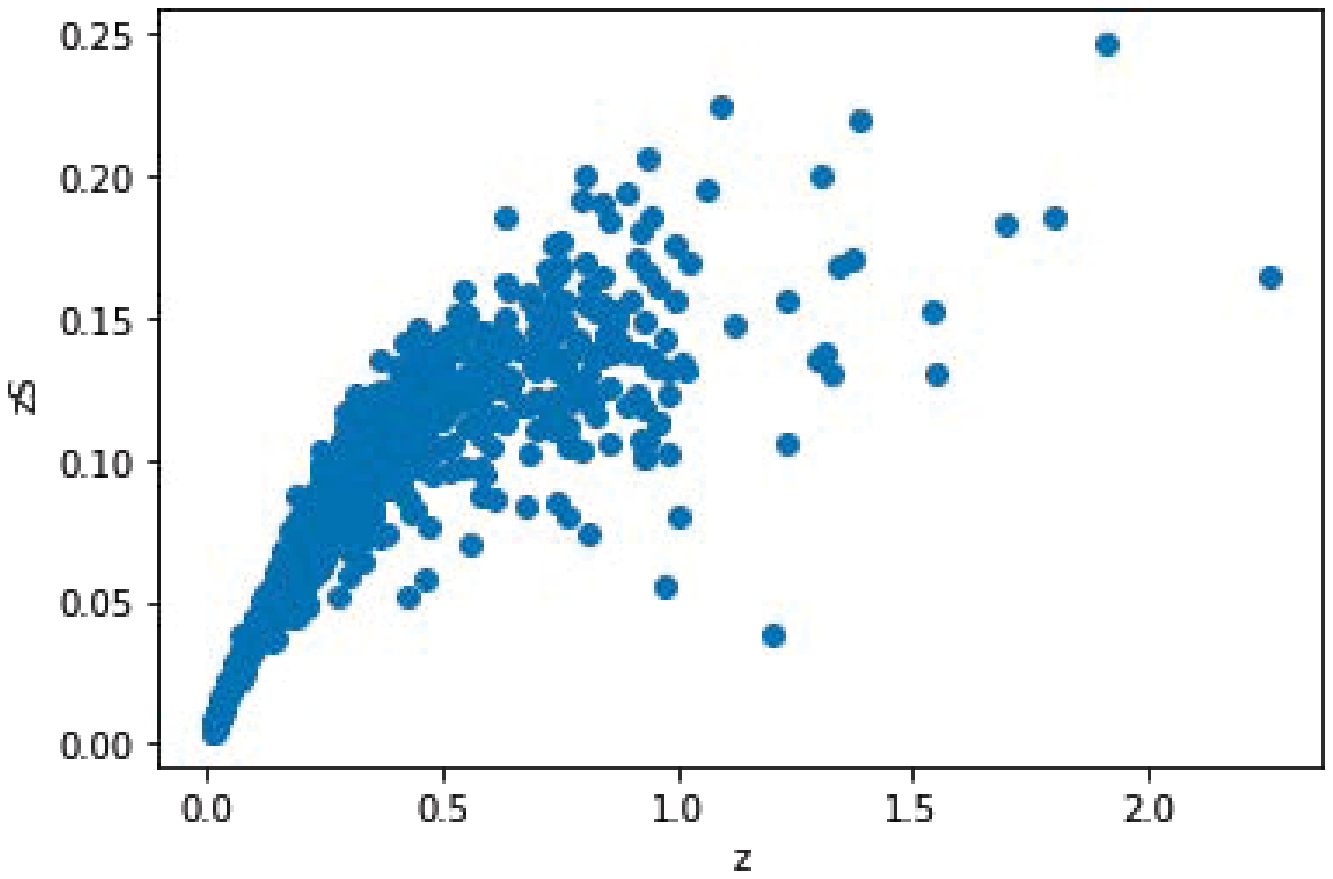}
    \includegraphics[width=0.33\hsize,height=0.3\textwidth,angle=0,clip]{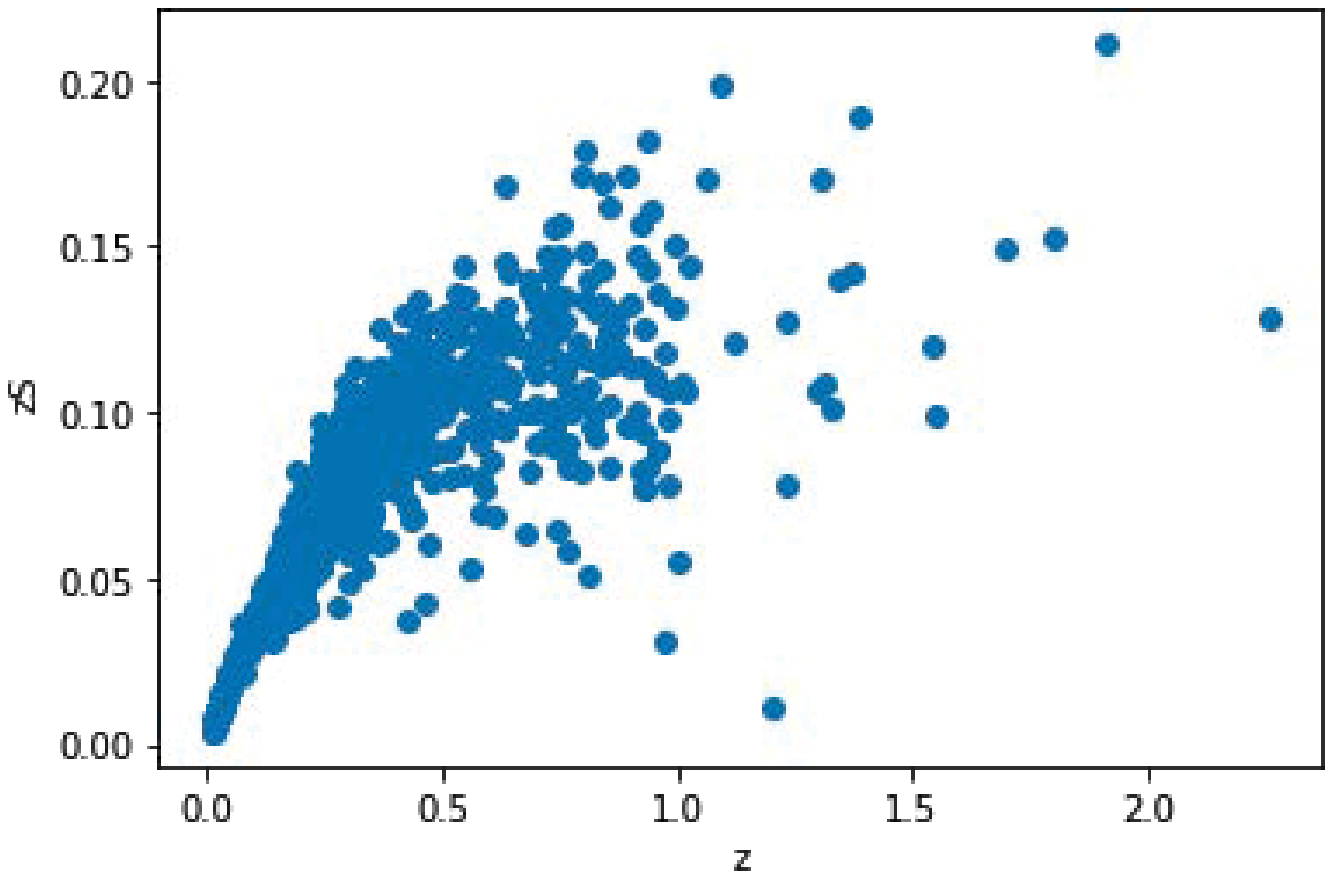}
    \caption{The first row shows the histograms of $z_{\rm S}$ for the Cosmology model A, where $\Omega_{M}=0.3$, $\Omega_{k}=\Omega_{\Lambda}=0$, related to the Pantheon Sample. The second row shows the scatter plot $z_{\rm S}$ versus $z$. $H_0$ assumes the values 67 (first and fourth panels), 70 (second and fifth panels), 74 (third and sixth panels), km s$^{-1}$ per Mpc.}
    \label{fig:firstthreecasessingular}
\end{figure}

\begin{figure}
    \centering
    \includegraphics[width=0.33\hsize,height=0.3\textwidth,angle=0,clip]{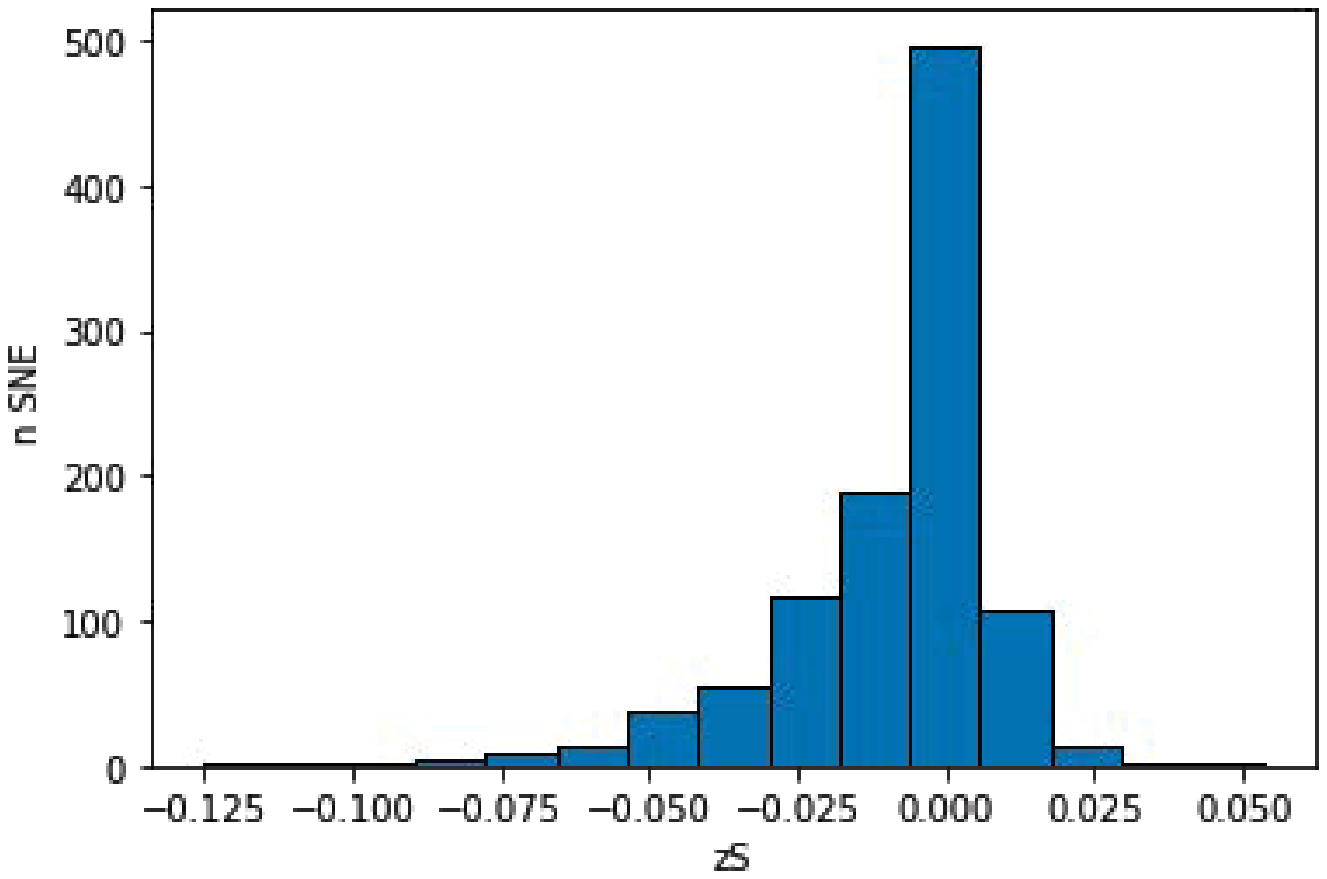}
    \includegraphics[width=0.33\hsize,height=0.3\textwidth,angle=0,clip]{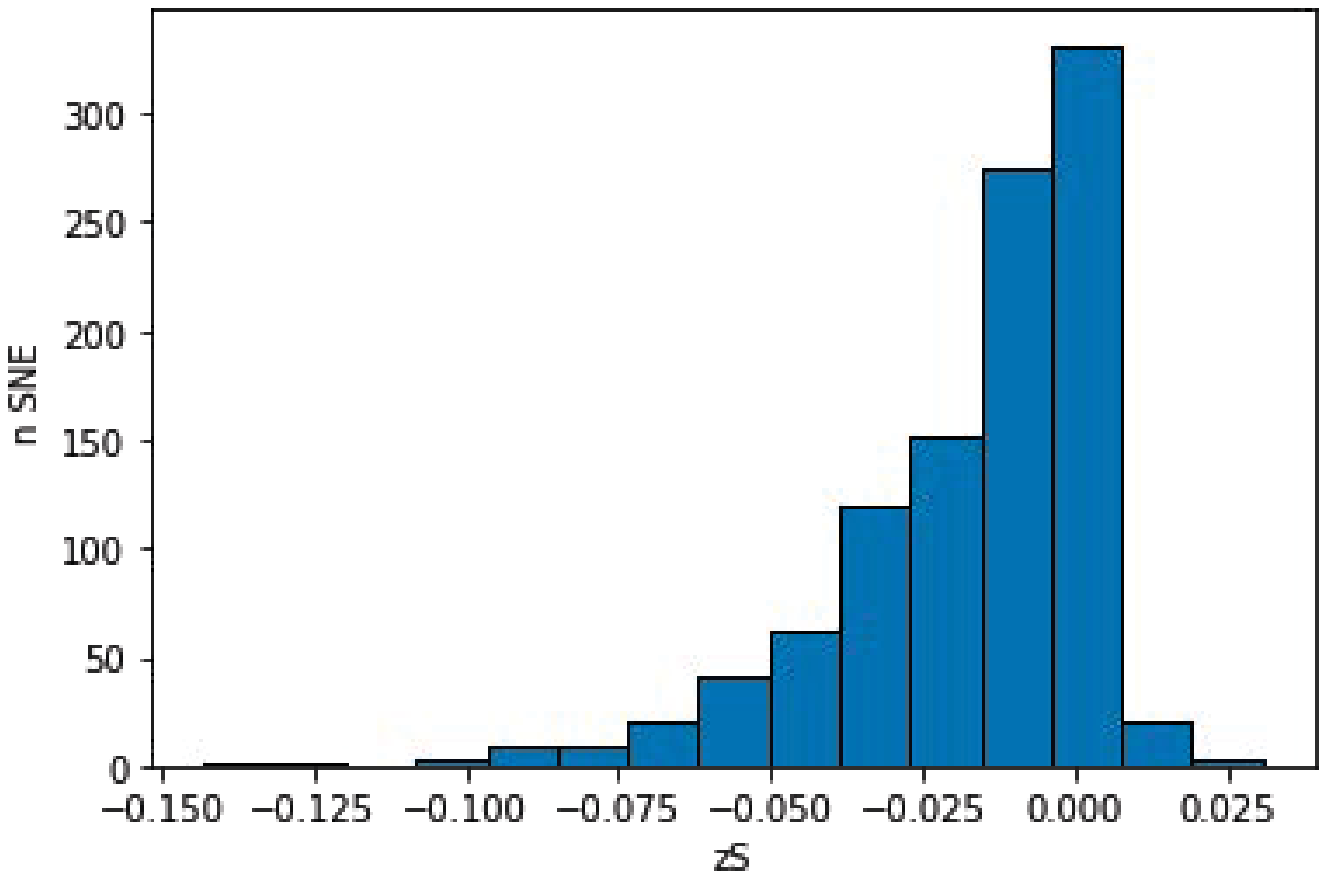}
    \includegraphics[width=0.33\hsize,height=0.3\textwidth,angle=0,clip]{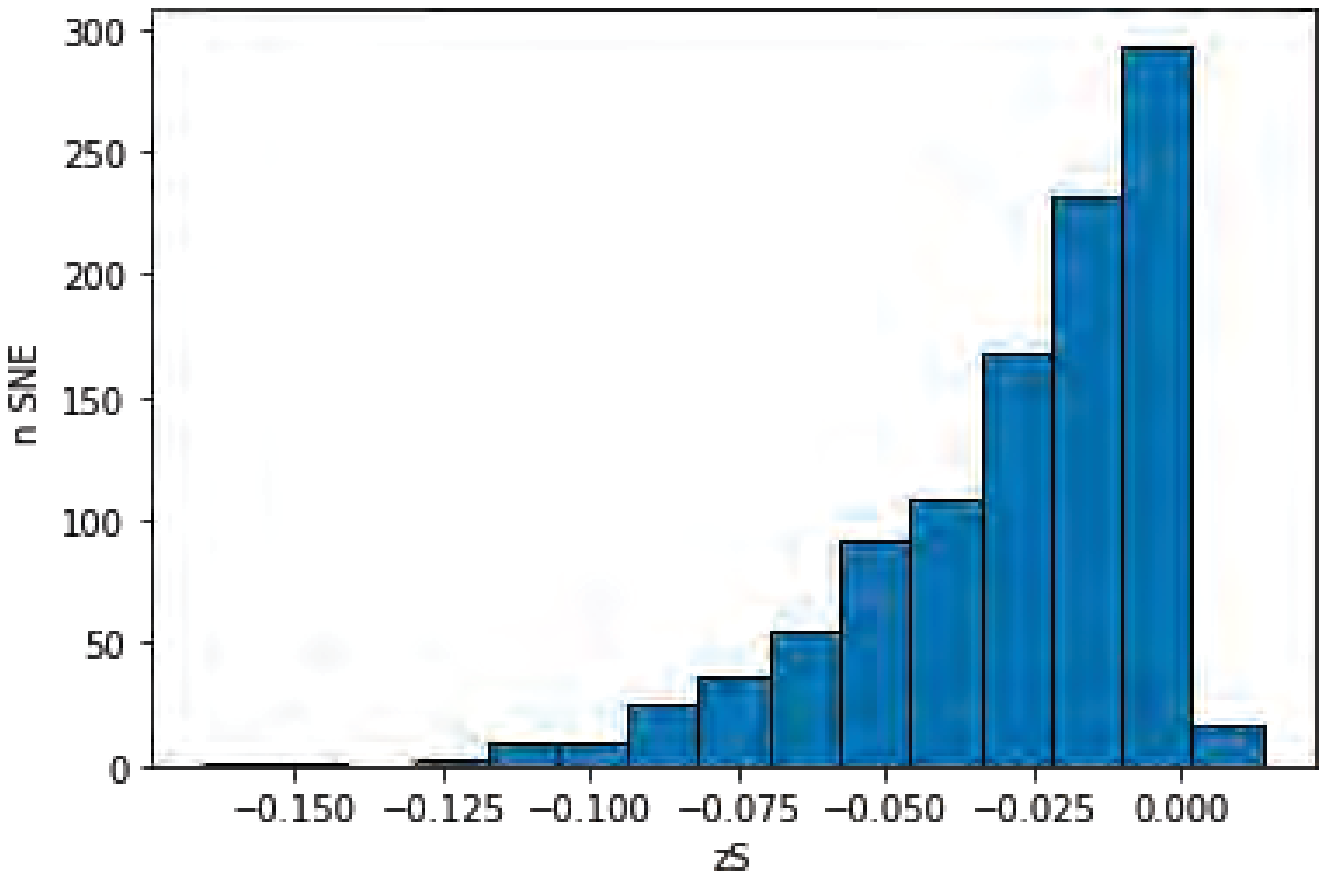}
    \includegraphics[width=0.33\hsize,height=0.3\textwidth,angle=0,clip]{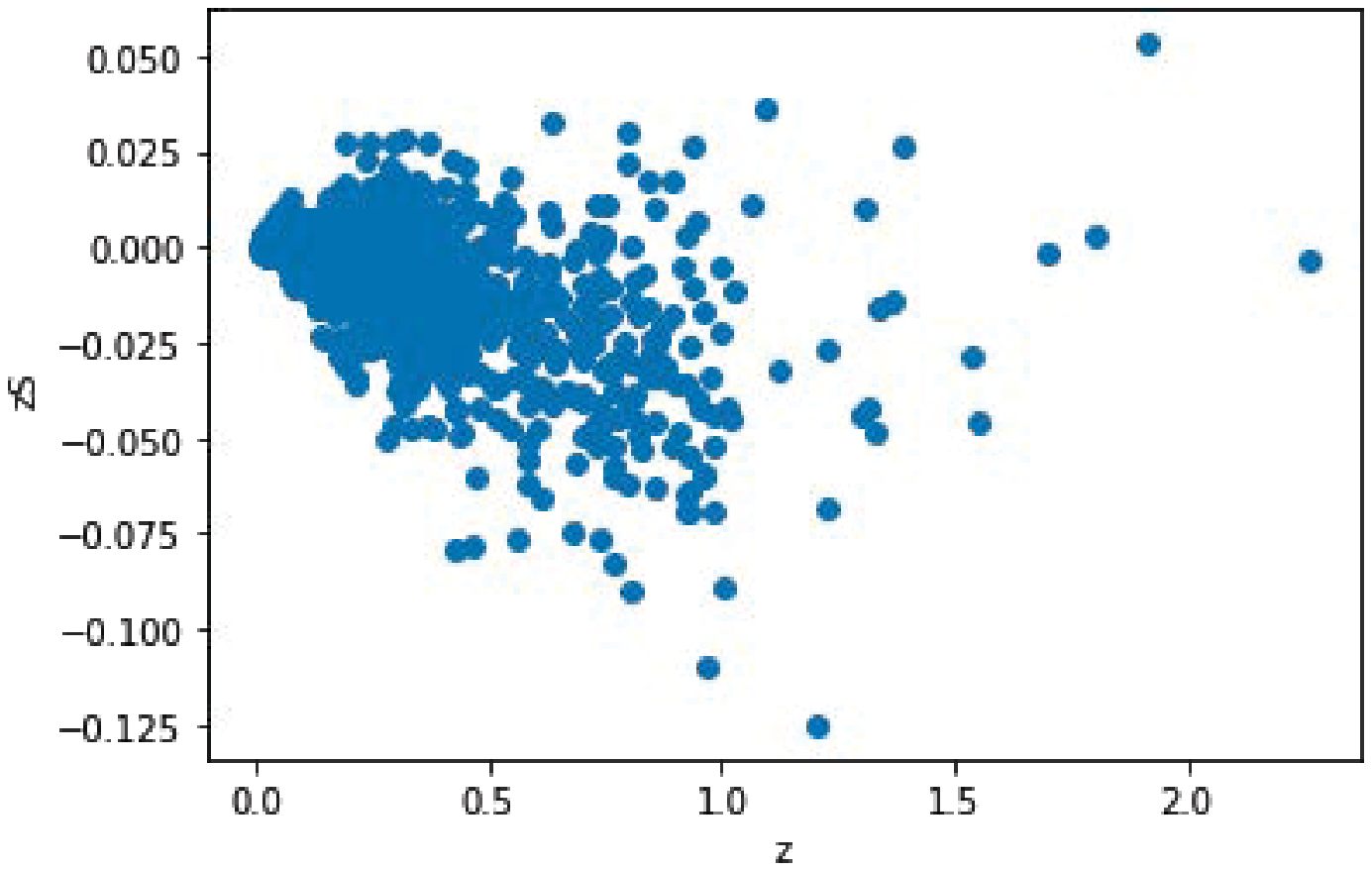}
    \includegraphics[width=0.33\hsize,height=0.3\textwidth,angle=0,clip]{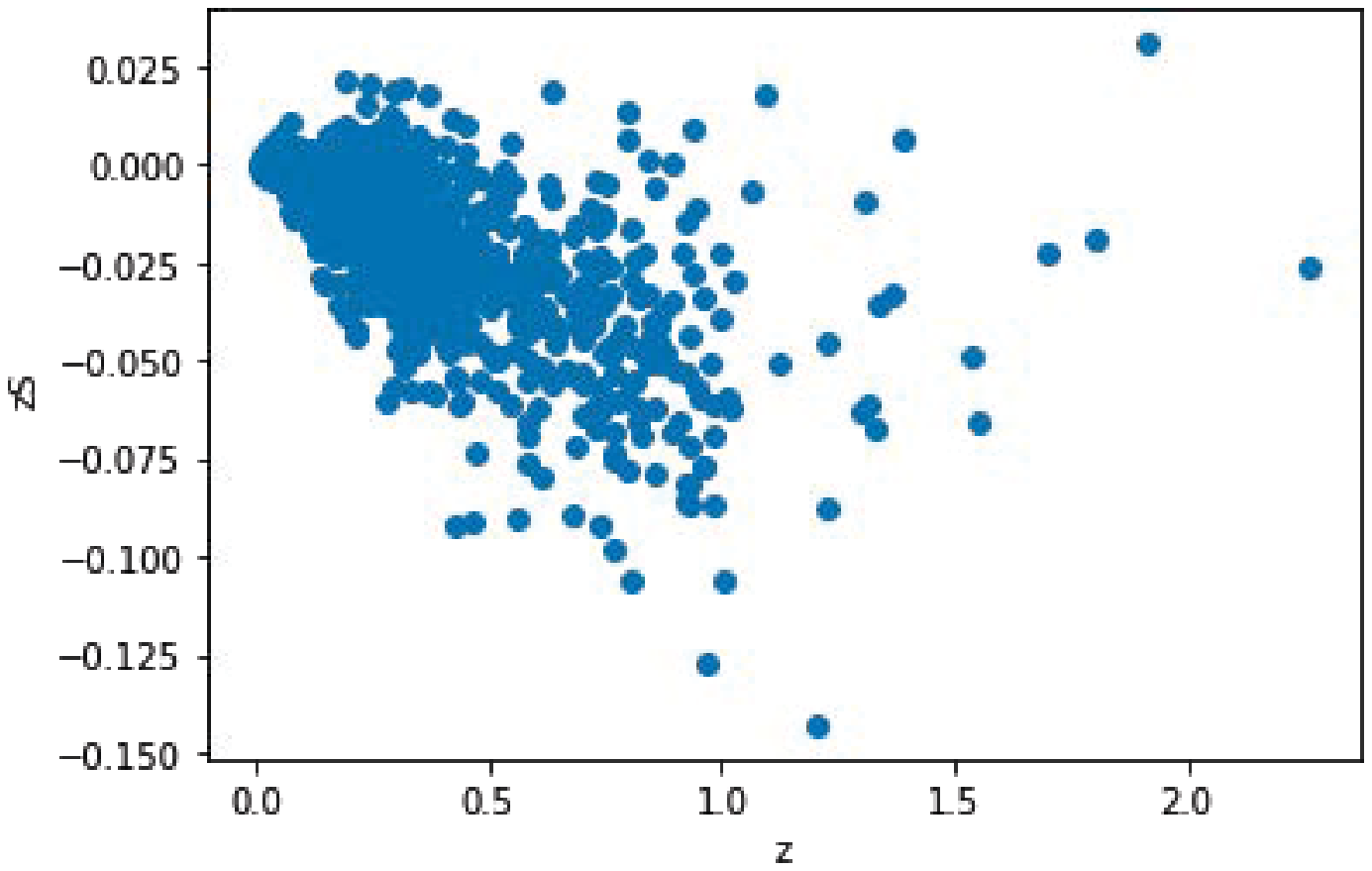}
    \includegraphics[width=0.33\hsize,height=0.3\textwidth,angle=0,clip]{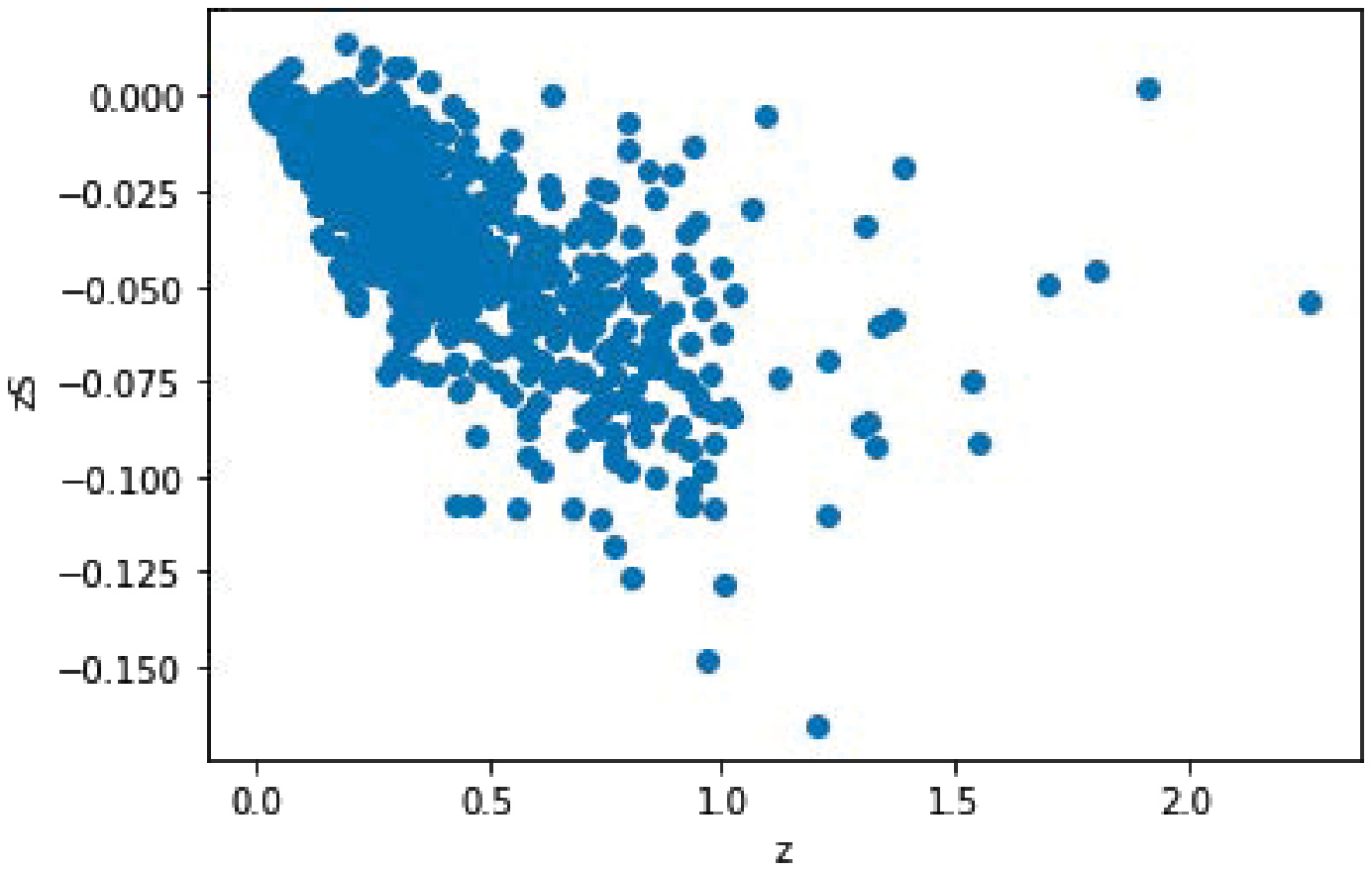}
    \caption{The first row shows the histograms of $z_{\rm S}$ for the Cosmology model B, where $\Omega_{M}=0.3$, $\Omega_{k}=0.7$, and $\Omega_{\Lambda}=0$, related to the Pantheon Sample. The second row shows the scatter plot $z_{\rm S}$ versus $z$. $H_0$ assumes the values 67 (first and fourth panels), 70 (second and fifth panels), 74 (third and sixth panels), km s$^{-1}$ per Mpc.}
    \label{fig:secondthreecasessingular}
\end{figure}

\begin{figure}
    \centering
    \includegraphics[width=0.33\hsize,height=0.3\textwidth,angle=0,clip]{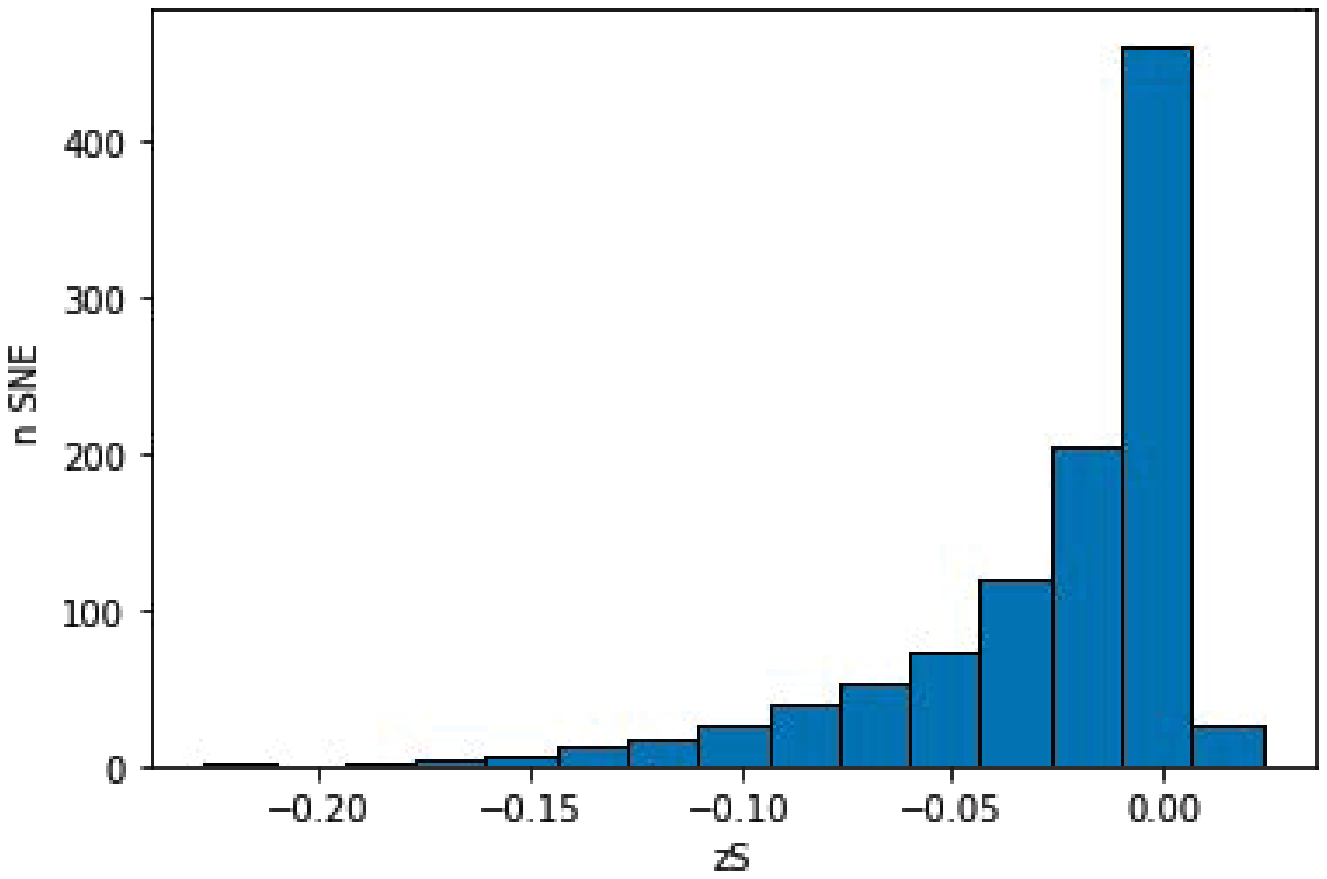}
    \includegraphics[width=0.33\hsize,height=0.3\textwidth,angle=0,clip]{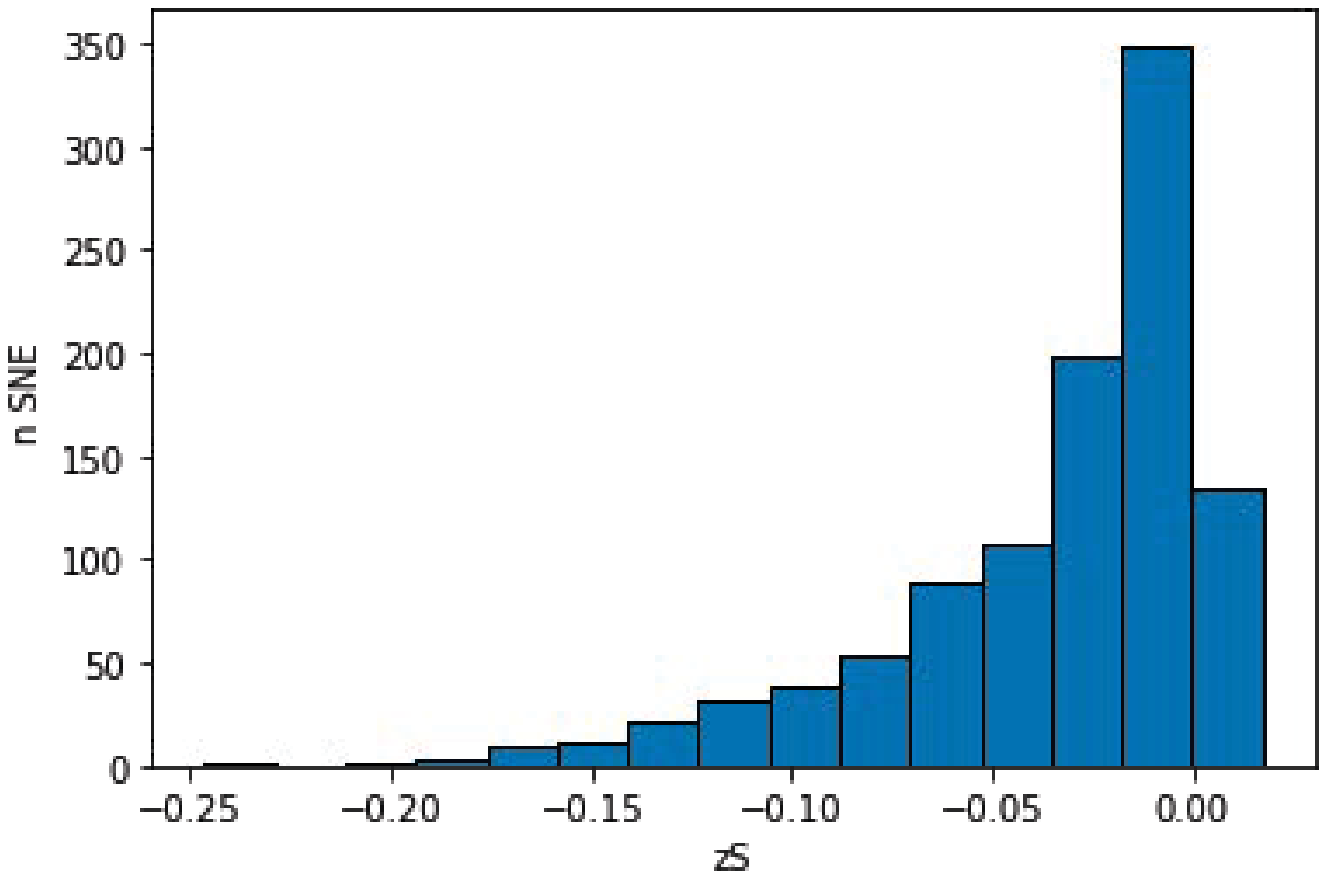}
    \includegraphics[width=0.33\hsize,height=0.3\textwidth,angle=0,clip]{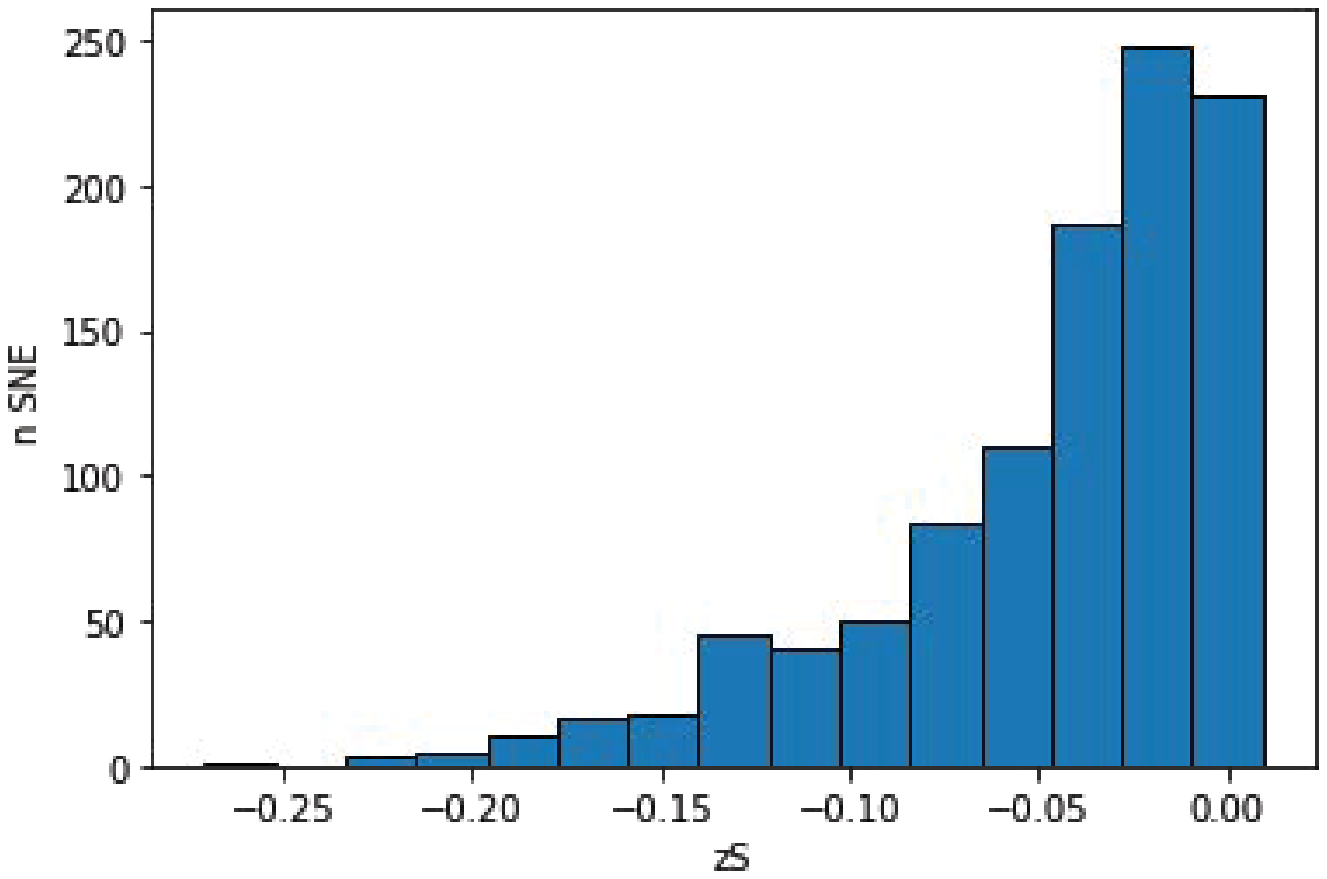}
    \includegraphics[width=0.33\hsize,height=0.3\textwidth,angle=0,clip]{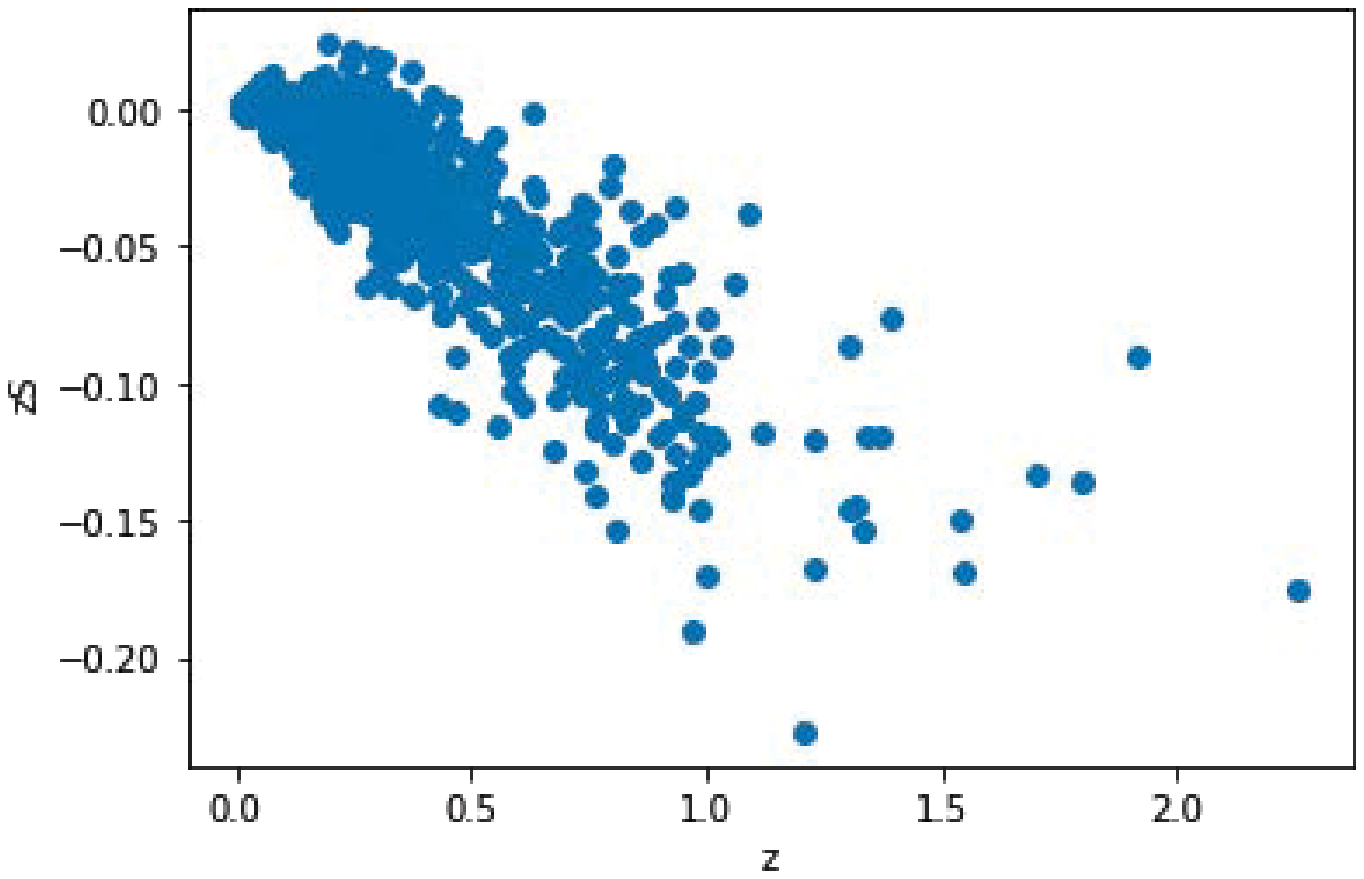}
    \includegraphics[width=0.33\hsize,height=0.3\textwidth,angle=0,clip]{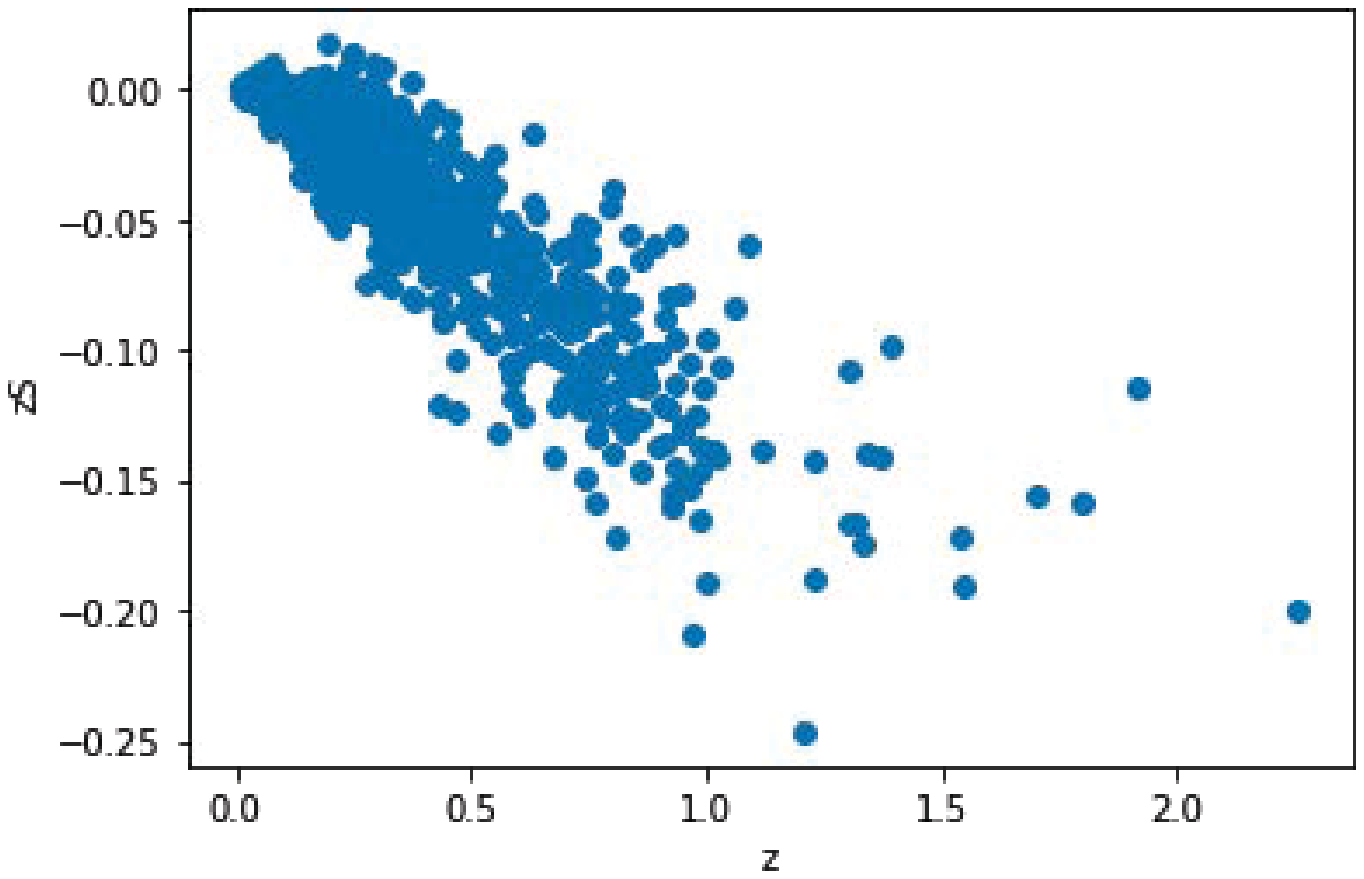}
    \includegraphics[width=0.33\hsize,height=0.3\textwidth,angle=0,clip]{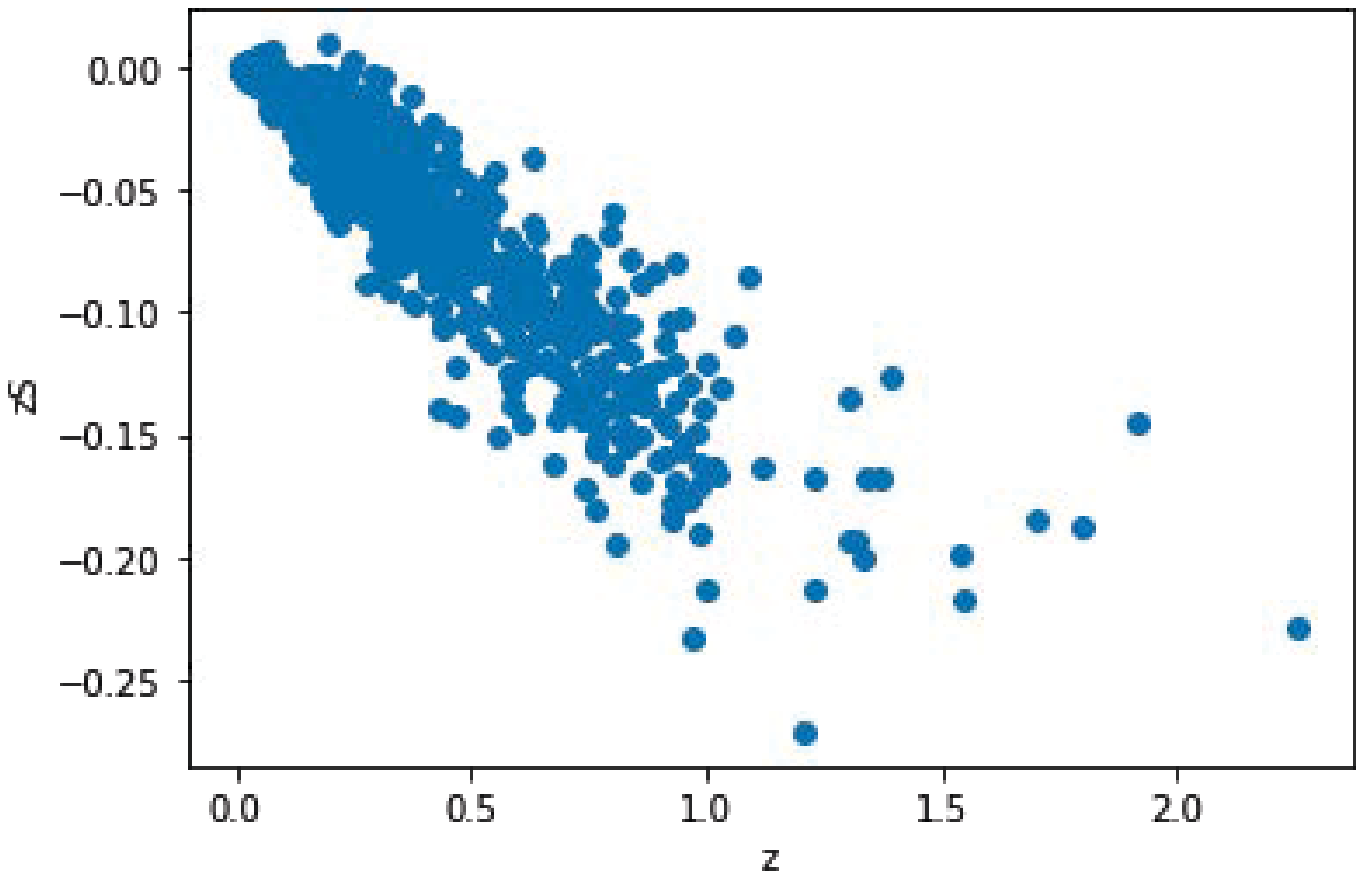}
    \caption{The first row shows the histograms of $z_{\rm S}$ for the Cosmology model C, where $\Omega_{M}=1$, $\Omega_{k}=\Omega_{\Lambda}=0$, related to the Pantheon Sample. The second row shows the scatter plot $z_{\rm S}$vs$z$. $H_0$ assumes the values 67 (first and fourth panesl), 70 (second and fifth panels), 74 (third and sixth panels), 
    km s$^{-1}$ per Mpc.}
    \label{fig:thirdthreecasessingular}
\end{figure}

We now show the results of our computations based on the Pantheon Sample, picking individual SN Ia, Figs. \ref{fig:firstthreecasessingular}, \ref{fig:secondthreecasessingular}, \ref{fig:thirdthreecasessingular}. In each of these figures, we show the histograms of $z_{\rm S}$ for the three Cosmology models and the plots of $z_{\rm S}$ versus $z$. 

In Fig. \ref{fig:firstthreecasessingular}, we are considering the Cosmology model A, where $\Omega_{M}=0.3$, $\Omega_{k}=\Omega_{\Lambda}=0$. From the histograms displayed in the first row, we note that the peak number of SNe Ia is corresponded by $z_{\rm S}$ below 0.1, especially for increasing $H_0$. Also, we note that $z_{\rm S}$ is always positive, confirming the mock red shift values, for which the non-standard electromagnetic effects are dissipative.
In the second row, we visualise the behaviour of $z_{\rm S}$ with respect to the detected $z$. We see a clear dispersion of the points obtained by our computations, probably due to the uncertainties on the real measurements, especially on the distance-modulus. 
Comparing these last plots with those obtained with the mock red shifts, Fig. \ref{fig:firstthreecasesbuiltredshift}, we find a similar trend where the domains of $z$ overlap. Also, we note the same decrease in the peak value of $z_{\rm S}$ both in the real data computations and in the mock $z$ test. Finally, we note that higher values of $z_{\rm S}$ are reached by the real data with respect to the test with mock red shift. This could be an effect due to the already mentioned dispersion. 
Finally, the $z_{\rm S}$ values get lower for increasing $H_0$ as in the mock red shift test.  

 In Fig. \ref{fig:secondthreecasessingular}, we take into account the Cosmology model B, where $\Omega_{M}=0.3$, $\Omega_{k}=0.7$, and $\Omega_{\Lambda}=0$. From the histograms, we note that the peak of $z_{\rm S}$ is close to $0$. Most of the $z_{\rm S}$ are negative. Increasing $H_0$ shifts the majority of SNe Ia to more negative values of $z_{\rm S}$. Looking at the second row, we see the general trend of $z_{\rm S}$ and comparing with the test results displayed in Fig. \ref{fig:secondthreecasesbuiltredshift}, we see that the negative peak predicted with the mock red shifts is more visible. We also note that raising $H_0$ increases the absolute value of the negative peak reached by $z_{\rm S}$, as expected from the mock red shifts plots. We also notice that the positive values of $z_{\rm S}$ can be found at both low and high $z$. This could be due to data dispersion or to the changing sign of the non-standard electromagnetic contribution, effect unobserved for the mock shifts. In passing, we observe that the scattering seems to decrease in the third plot, for the highest value for $H_0$.
 
 In Fig. \ref{fig:thirdthreecasessingular}, we consider the Cosmology model C, where $\Omega_{M}=1$, $\Omega_{k}=\Omega_{\Lambda}=0$. Looking at the first row, we see in the histograms that the great majority of $z_{\rm S}$ are negative, especially for larger $H_0$. This trend can be noted also in the second row, where we notice that the positive values of $z_{\rm S}$ are found at low $z$, from which a monotonic decrease is observed. This behaviour confirms the mock red shifts test, Fig. \ref{fig:thirdthreecasesbuiltredshift}.
 
 Generally, the behaviour observed for the mock red shifts is confirmed by real data in the common $z$ range. 
 Thereby, we are induced to assume that the predicted behaviour of $z_{\rm S}$ for large  red shifts beyond the Pantheon catalogue has some reliability. The discrepancies, namely the dispersion and the values of $z_{\rm S}$ themselves, can be ascribed to the observational errors. 
 
 \subsubsection{The $k_i$ parameters issued from individual computations}
 
 The analysis based on individual SN Ia elaborates for each data point a specific value of the parameters $k_i$, Tab. \ref{tabdeltanu}. This has been carried out for all values of $\Omega$ and $H_0$. The results (mean and standard deviation of the $k_i$ distributions are shown in Tab. \ref{tab:resultskiSingular}. From Tab. \ref{tabdeltanu}, we note that a negative value of $k_i$ for $i=1,4$ means a positive value for $z_{\rm S}$ and vice versa, while for $i=2,3$ the sign depends also on the denominator. For our computations, we have chosen for the emitted frequency $k_{3}$, $\nu_{e}=6.74 \times 10^{14} s^{-1}$, which is an optical frequency in the B-Band. 
 
  \begin{table}
    \centering
    \begin{tabular}{c|c|c|c|c}
    \hline
       Cosmology  &  $k_1$ & $k_2$ & $k_3$ & $k_4$\\\hline
         Cosmology 1, $H_0=67$ & $(-8.18 \pm 2.15) \times 10^{-5}$ &$(-7.89 \pm 2.20) \times 10^{-5}$ &$(-5.28 \pm 1.47) \times 10^{10}$ &$(-8.49 \pm 2.10) \times 10^{-5}$ \\\hline
         Cosmology 1, $H_0=70$ & $(-7.69 \pm 2.13 )\times 10^{-5}$ &$(-7.44 \pm 2.16 )\times 10^{-5}$ &$(-4.98 \pm 1.45 )\times 10^{10}$ &$(-7.96 \pm 2.10 )\times 10^{-5}$ \\\hline
         Cosmology 1, $H_0=74$ & $(-7.03 \pm 2.11 )\times 10^{-5}$ &$(-6.82 \pm 2.13) \times 10^{-5}$ &$(-4.57 \pm 1.42) \times 10^{10}$ &$(-7.24 \pm 2.10 )\times 10^{-5}$ \\\hline
         Cosmology 2, $H_0=67$ & $(0.39 \pm 1.53 )\times 10^{-5}$ &$(0.40 \pm .54)\times 10^{-5}$ &$(0.27 \pm 1.07) \times 10^{10}$ &$(0.42 \pm 1.67 )\times 10^{-5}$ \\\hline
         Cosmology 2, $H_0=70$ & $(1.34 \pm 1.57 )\times 10^{-5}$ &$(1.37 \pm 1.61)\times 10^{-5}$ &($0.92 \pm 1.08) \times 10^{10}$ &($1.32 \pm 1.54) \times 10^{-5}$ \\\hline
         Cosmology 2, $H_0=74$ & $(2.72 \pm 1.63) \times 10^{-5}$ &$(2.77 \pm 1.69)\times 10^{-5}$ &$(1.79 \pm 1.10) \times 10^{10}$ &($2.66 \pm 1.57 )\times 10^{-5}$ \\\hline
         Cosmology 3, $H_0=67$ & $(1.66 \pm 2.31 )\times 10^{-5}$ &$(1.73 \pm 2.41)\times 10^{-5}$ &$(1.16 \pm 1.62) \times 10^{10}$ &($1.61 \pm 2.23 )\times 10^{-5}$ \\\hline
         Cosmology 3, $H_0=70$ & $(2.73 \pm 2.38 )\times 10^{-5}$ &$(2.83 \pm 2.53)\times 10^{-5}$ &($1.89 \pm 1.69) \times 10^{10}$ &($2.63 \pm 2.25) \times 10^{-5}$ \\\hline
          Cosmology 3, $H_0=74$ & ($4.17 \pm 2.49) \times 10^{-5}$ &($4.33 \pm 2.71)\times 10^{-5}$ &($2.90 \pm 1.81) \times 10^{10}$ &($4.01 \pm 2.30 )\times 10^{-5}$ \\\hline
    \end{tabular}
    \caption{Mean and standard deviation values of the $k_i$ parameters issued from the individual computations of $z_{\rm S}$, for the three cosmological models. The computations have been performed considering the distances in Mpc, which means that the $k_i$ parameters are in Mpc$^{-1}$ for $i=1,2,4$ and Mpc$^{-1}$ s$^{-1}$ for $k=3$, while the values for $H_0$ are in km s$^{-1}$ per Mpc.} 
    \label{tab:resultskiSingular}
\end{table}

The absolute values of the $k_i$ parameters are in the range $4 \times 10^{-6} - 8.5 \times 10^{-5}$ per Mpc for $i=1,2,4$ and of $10^{10}$ per Mpc s$^{-1}$ for $i=3$: indeed, a given $\delta \nu$, is achieved by $k_3 dr$ without the support of a frequency as multiplying factor, see the first row in Tab. \ref{tabdeltanu}. Thereby, it appears a similar order of magnitude of the $k_i$, with the exception of $k_3$. We note a relatively high standard deviation with respect to the mean due to the dispersion that already appeared in the previous plots. These results do not assume the meaning of best fit of $k_i$ for all the SNe Ia, conversely to the next section, but allow us to appreciate the expected order of magnitude. 

\subsubsection{The $k_i$ parameters computed as best fit}
 
We now search the $k_i$ parameters valid for every SN Ia. We will first analyse the results considering uniquely the Pantheon Sample, and later we will add the BAO data. 

An iterative procedure has been performed using Eqs. (\ref{zc}, \ref{light travel} , \ref{luminosity distance}) and Tab. \ref{tabdeltanu} in the following way

\[ z\rightarrow r (z)\rightarrow k_i (z) \rightarrow z_{\rm S} (z) \rightarrow \\
z_{\rm C} (z) \rightarrow r (z_{\rm C}) \rightarrow k_i (z_{\rm C}) \rightarrow {\rm and \: back \: again}~.
\]

Starting from the Catalogue observed red shift $z$, we compute the light-travel distance and, by our Bayesian procedure, we derive the best fit values for the $k_i$ parameters via Eq. (\ref{eq_chi2_SNe}). These parameters have been used to derive $z_{\rm S}$ and $z_{\rm C}$. Once we know $z_{\rm C}$, we go back to the best fit derivation of the $k_i$ parameters using this new value for the expansion red shift instead of $z$, to compute a new light-travel distance and repeat the procedure. We stop the computations once the difference between the light-travel distances computed in two subsequent steps becomes small enough (in particular, a mean difference between these two values of $\sim 1 Mpc$ considering all SNe Ia). The same has been done for the BAOs. 

Fixing the $\Omega$ densities and the $H_0$ parameter, we gather nine cases as before. Flat priors have been chosen for these parameters, while the intervals have been chosen case by case, keeping in mind the values obtained in the previous computation for individual SNe Ia. The results obtained for the $k_{i}$ parameters are shown in Tab. \ref{tab:resultskigeneral}. 

\begin{table}
    \centering
    \begin{tabular}{c|c|c|c|c}
    \hline
       Cosmology  &  $k_1$ & $k_2$ & $k_3$ & $k_4$\\\hline
         Cosmology 1, $H_0=67$ & $(-8.19\pm 0.02)  \times 10^{-5}$ &$(-7.74 \pm 0.02) \times 10^{-5}$ &$(-5.17\pm 0.01 ) \times 10^{10}$ &$(-8.53\pm 0.03)  \times 10^{-5}$ \\\hline
         Cosmology 1, $H_0=70$ & $(-7.66\pm 0.03)  \times 10^{-5}$ &$(-7.30 \pm 0.02) \times 10^{-5}$ &$(-4.89\pm 0.02 ) \times 10^{10}$ &$(-8.00\pm 0.03)  \times 10^{-5}$ \\\hline
         Cosmology 1, $H_0=74$ & $(-7.01\pm 0.03)  \times 10^{-5}$ &$(-6.72 \pm 0.02) \times 10^{-5}$ &$(-4.51\pm 0.01 ) \times 10^{10}$ &$(-7.29\pm 0.02)  \times 10^{-5}$ \\\hline
        Cosmology 2, $H_0=67$ & $(0.39\pm 0.04)  \times 10^{-5}$ &$(0.39 \pm 0.04) \times 10^{-5}$ &$(0.26\pm 0.03 ) \times 10^{10}$ &$(0.39\pm 0.04)  \times 10^{-5}$ \\\hline
         Cosmology 2, $H_0=70$ & $(1.34\pm 0.04)  \times 10^{-5}$ &$(1.34 \pm 0.05) \times 10^{-5}$ &$(0.90\pm 0.03 ) \times 10^{10}$ &$(1.32\pm 0.04)  \times 10^{-5}$ \\\hline
         Cosmology 2, $H_0=74$ & $(2.60\pm 0.04)  \times 10^{-5}$ &$(2.63 \pm 0.04) \times 10^{-5}$ &$(1.76\pm 0.03 ) \times 10^{10}$ &$(2.58\pm 0.04)  \times 10^{-5}$ \\\hline
         Cosmology 3, $H_0=67$ & $(1.52\pm 0.04)  \times 10^{-5}$ &$(1.54 \pm 0.04) \times 10^{-5}$ &$(1.02\pm 0.03 ) \times 10^{10}$ &$(1.51\pm 0.04)  \times 10^{-5}$ \\\hline
        Cosmology 3, $H_0=70$ & $(2.56\pm 0.04)  \times 10^{-5}$ &$(2.58 \pm 0.04) \times 10^{-5}$ &$(1.73\pm 0.03 ) \times 10^{10}$ &$(2.54\pm 0.04)  \times 10^{-5}$ \\\hline
          Cosmology 3, $H_0=74$ & $(3.96\pm 0.04)  \times 10^{-5}$ &$(4.02 \pm 0.05) \times 10^{-5}$ &$(2.69\pm 0.04 ) \times 10^{10}$ &$(3.91\pm 0.05)  \times 10^{-5}$ \\\hline
    \end{tabular}
    \caption{
   The best fit values with the relative errors for the $k_i$ parameters, considering the general best fit for all the SNe Ia belonging to the Pantheon Sample, for the three cosmological models. The computations have been performed considering the distances in Mpc, which means that the $k_i$ parameters are in Mpc$^{-1}$ for $i=1,2,4$ and Mpc$^{-1}$ s$^{-1}$ for $k=3$, while the values for $H_0$ are in km s$^{-1}$ per Mpc.} 
    \label{tab:resultskigeneral}
\end{table}

Targeting an unique value for the $k_i$ parameters for every SN Ia, we assume that $z_{\rm S}$ depends only on the distance, Tab. \ref{tabdeltanu}, and not on other effects, which are forcefully neglected by a single $k_i$. Nevertheless, we observe how the results in Tabs. \ref{tab:resultskiSingular} and \ref{tab:resultskigeneral} are very similar. Indeed, they are consistent at 1 $\sigma$ (even if this is not very informative given the high standard deviations found in Tab. \ref{tab:resultskiSingular}), and in some cases we get identical mean values, at the level of the chosen precision. Contrarily, the computed uncertainties for the general parameters are two orders of magnitude smaller than for those obtained in the previous computations, although we are compelled to comment on the different meanings of the results presented. In Tab. \ref{tab:resultskiSingular}, we read the mean and the standard deviation for a distribution of $k_i$ while in Tab. \ref{tab:resultskigeneral}, we read the single best fit value with the relative $\sigma$ error valid for all SNe Ia.  

\begin{figure}
    \centering
    \includegraphics[width=0.33\hsize,height=0.3\textwidth,angle=0,clip]{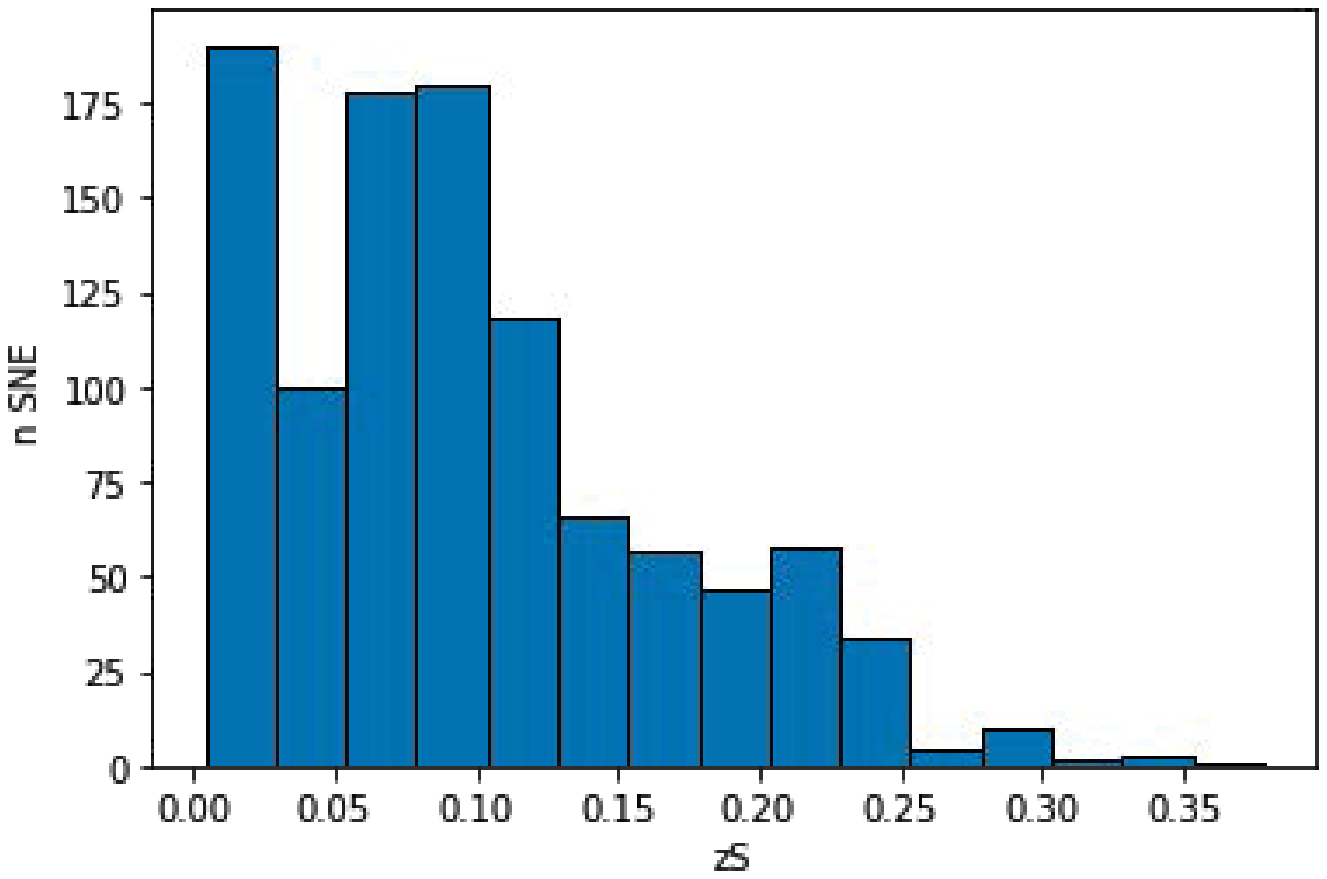}
    \includegraphics[width=0.33\hsize,height=0.3\textwidth,angle=0,clip]{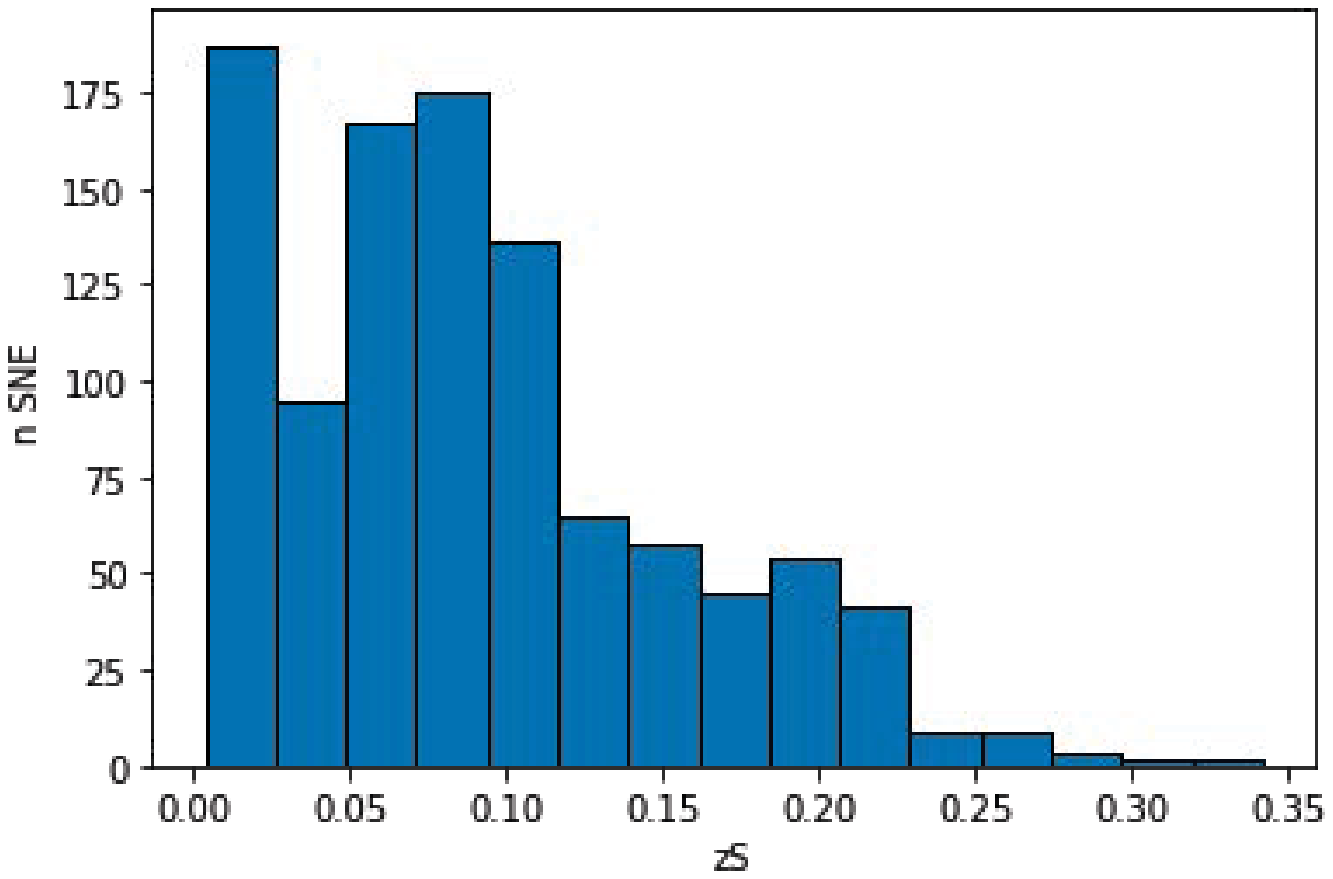}
    \includegraphics[width=0.33\hsize,height=0.3\textwidth,angle=0,clip]{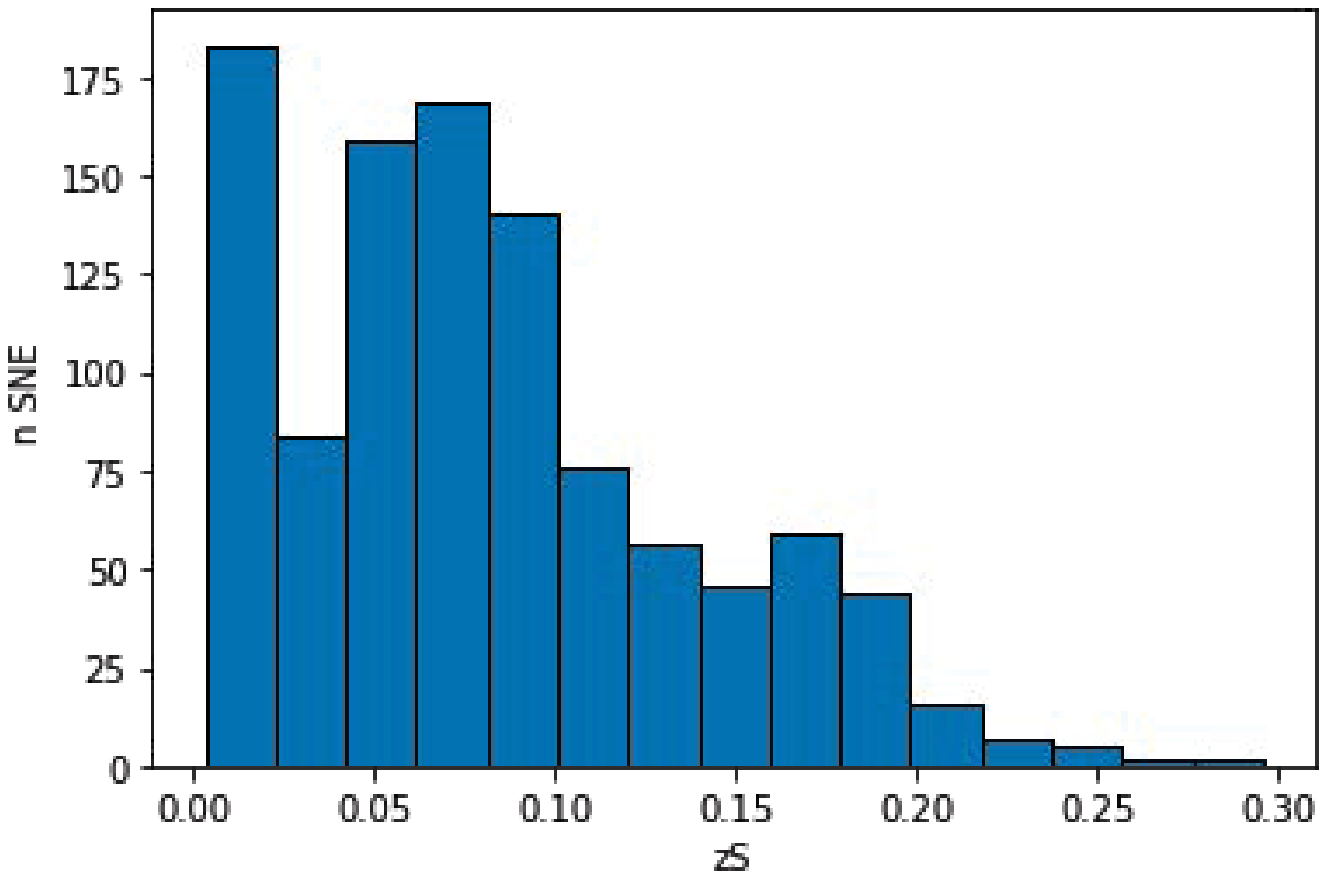}
    \caption{Histograms of the computed $z_{\rm S}$ from the $k_1$ parameter considering the Cosmology model A, where  $\Omega_{M}=0.3$, $\Omega_{k}=\Omega_{\Lambda}=0$ ($H_0=67, 70, 74$ for the left, central and right panels. The values for $H_0$ are in km s$^{-1}$ per Mpc).}
    \label{fig:firstthreecasesgeneralredshift}
\end{figure}

\begin{figure}
    \centering
    \includegraphics[width=0.33\hsize,height=0.3\textwidth,angle=0,clip]{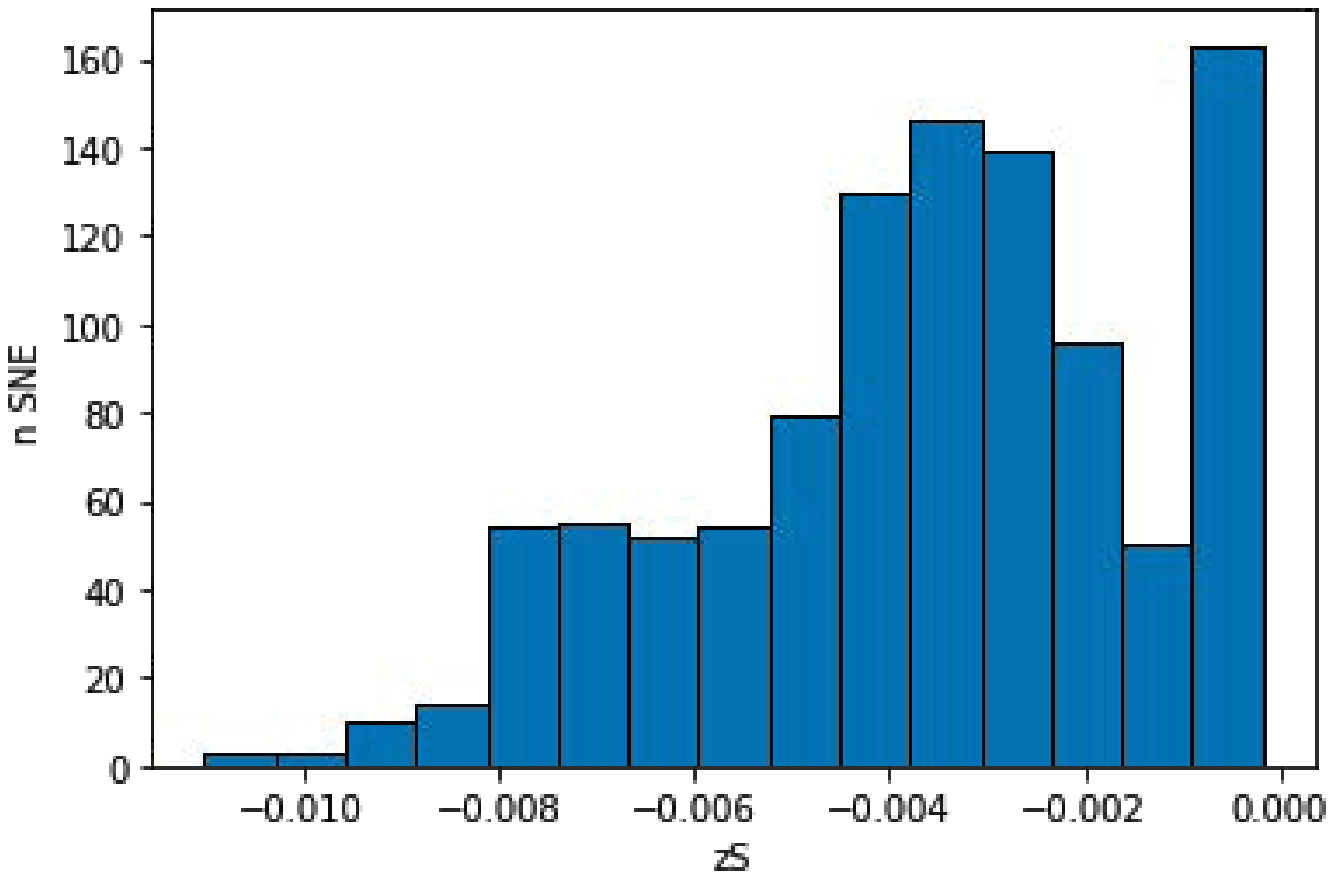}
    \includegraphics[width=0.33\hsize,height=0.3\textwidth,angle=0,clip]{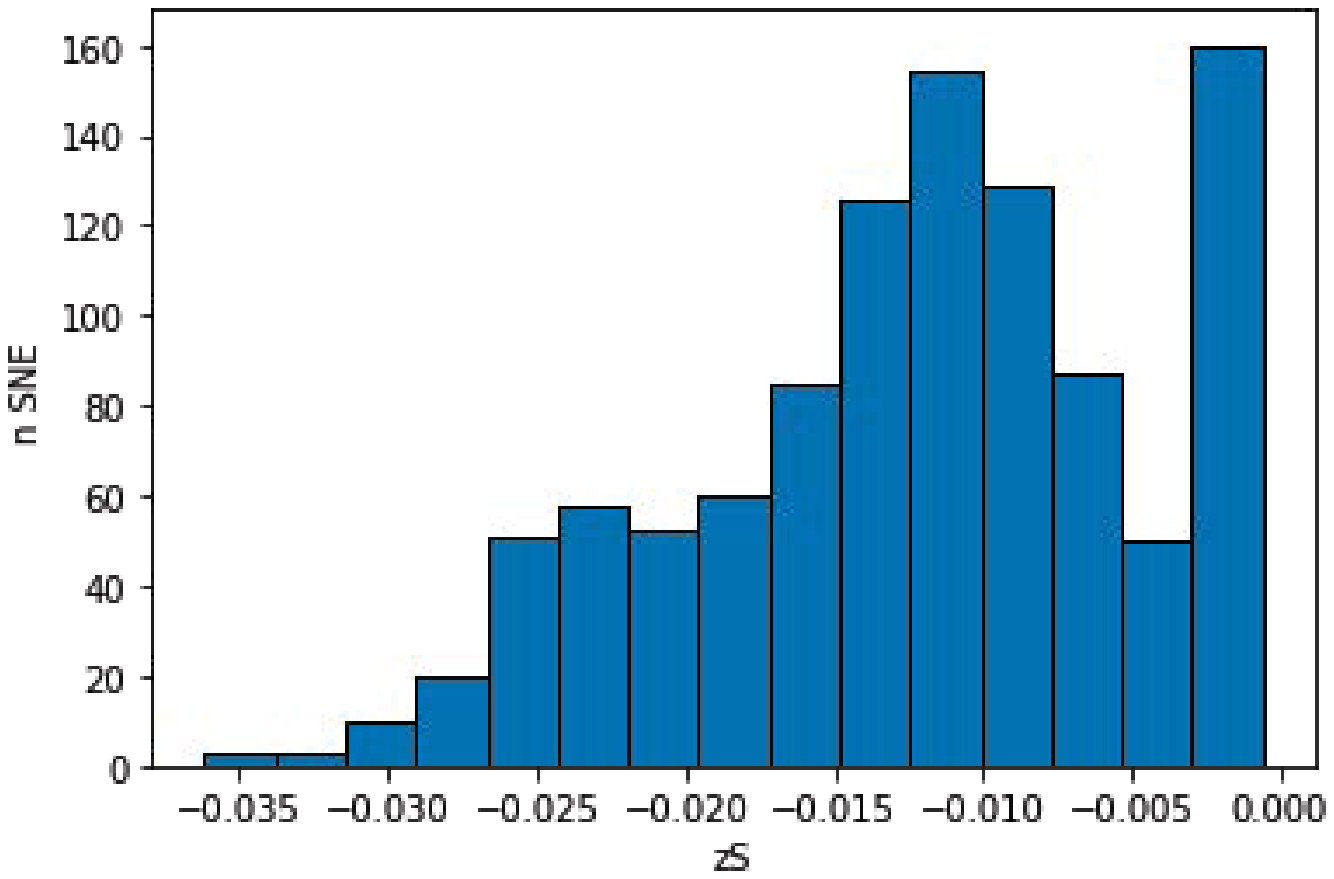}
    \includegraphics[width=0.33\hsize,height=0.3\textwidth,angle=0,clip]{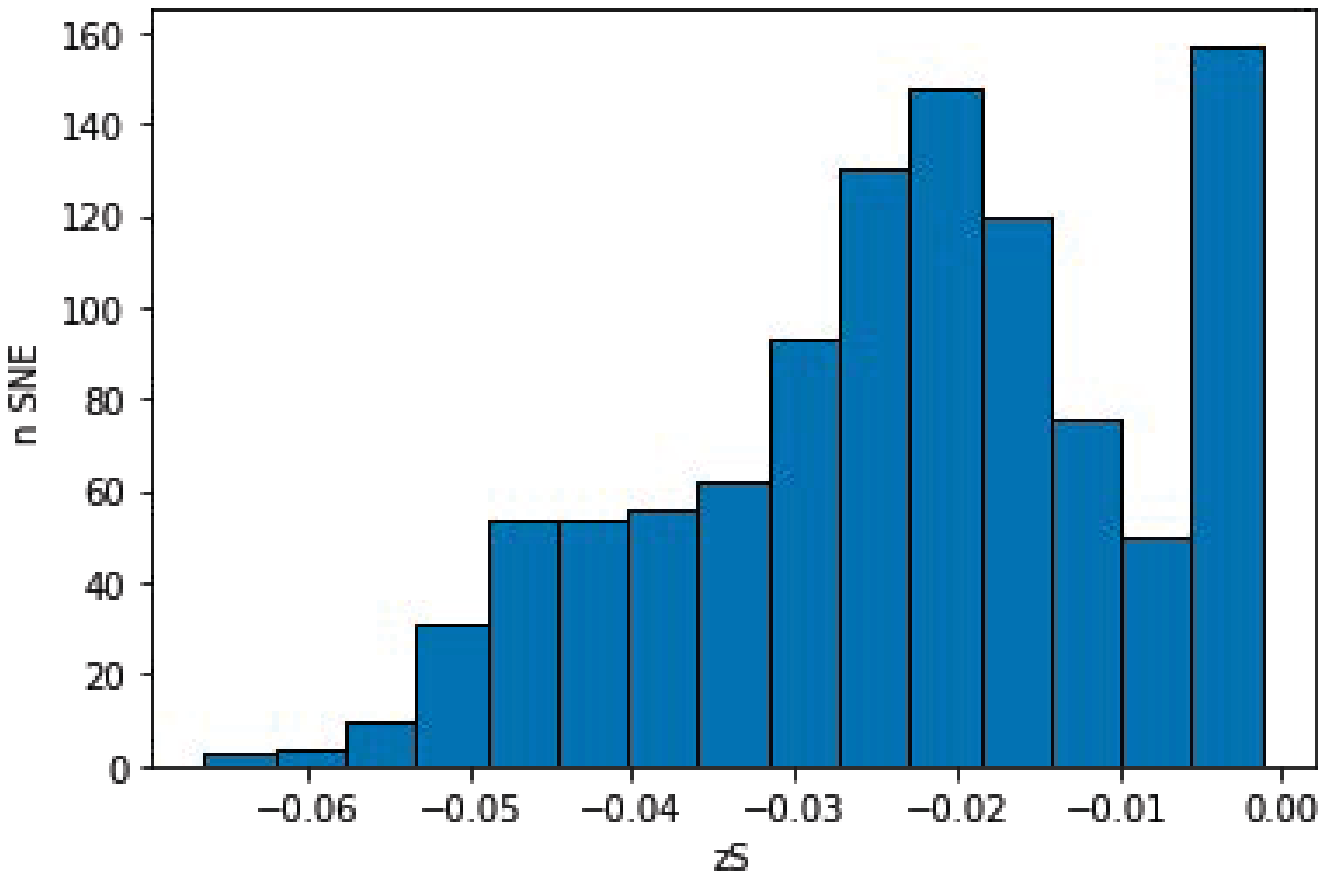}
    \caption{Histograms of the computed $z_{\rm S}$ from the $k_1$ parameter considering the Cosmology model B, where  $\Omega_{M}=0.3$, $\Omega_{k}=0.7$, $\Omega_{\Lambda}=0$ ($H_0=67, 70, 74$ for the left, central and right panels. The values for $H_0$ are in km s$^{-1}$ per Mpc).}
    \label{fig:secondthreecasesgeneralredshift}
\end{figure}

\begin{figure}
    \centering
    \includegraphics[width=0.33\hsize,height=0.3\textwidth,angle=0,clip]{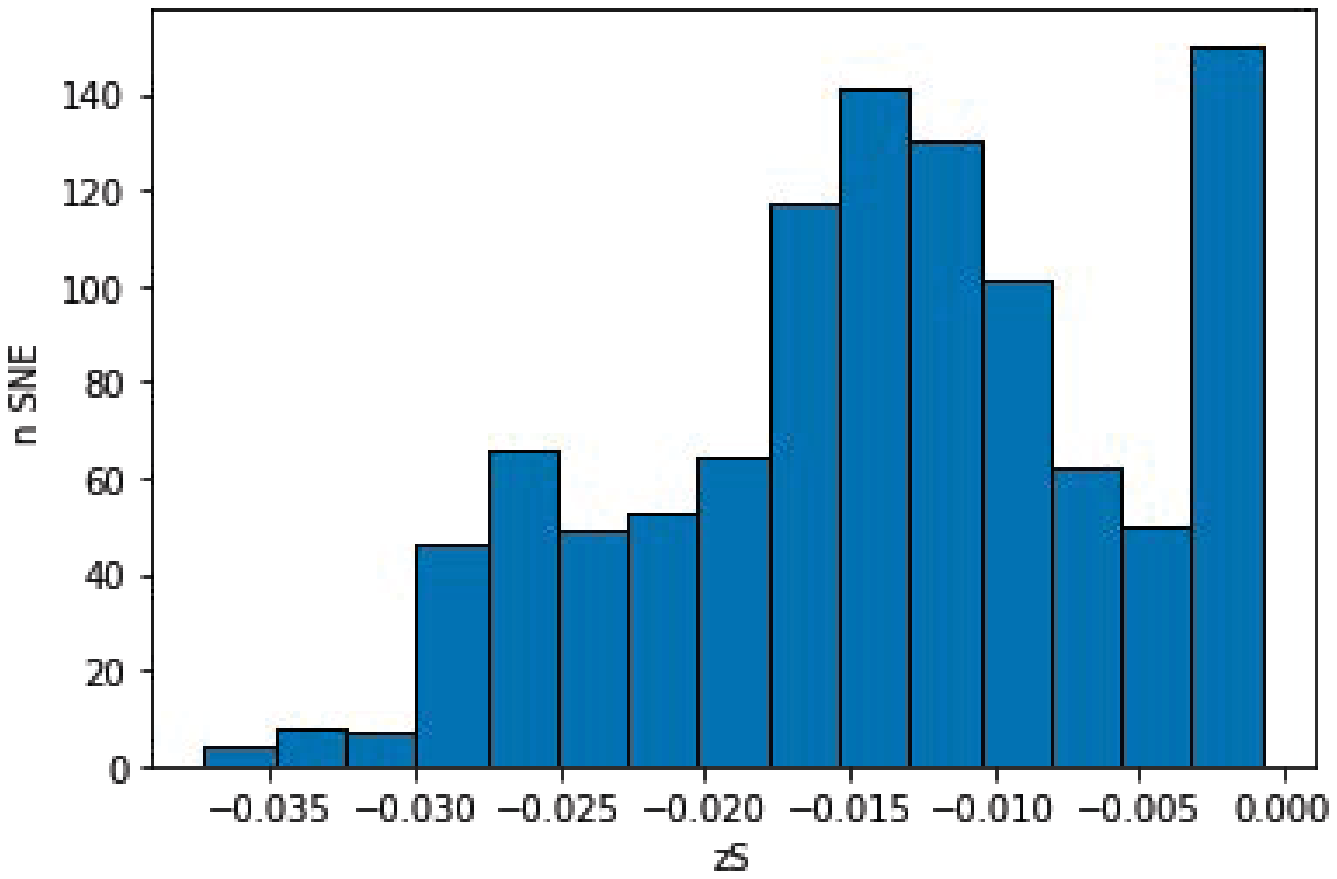}
    \includegraphics[width=0.33\hsize,height=0.3\textwidth,angle=0,clip]{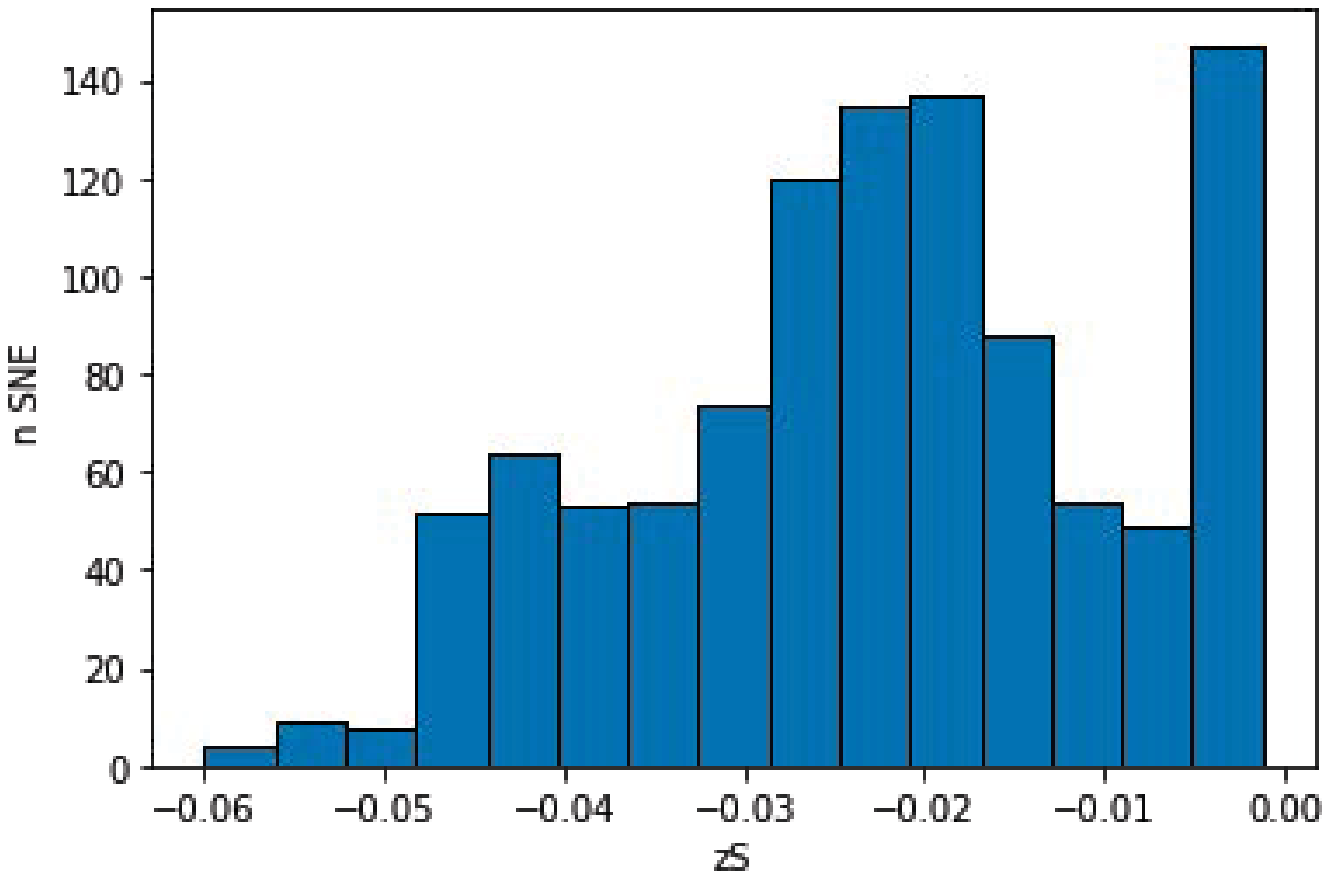}
    \includegraphics[width=0.33\hsize,height=0.3\textwidth,angle=0,clip]{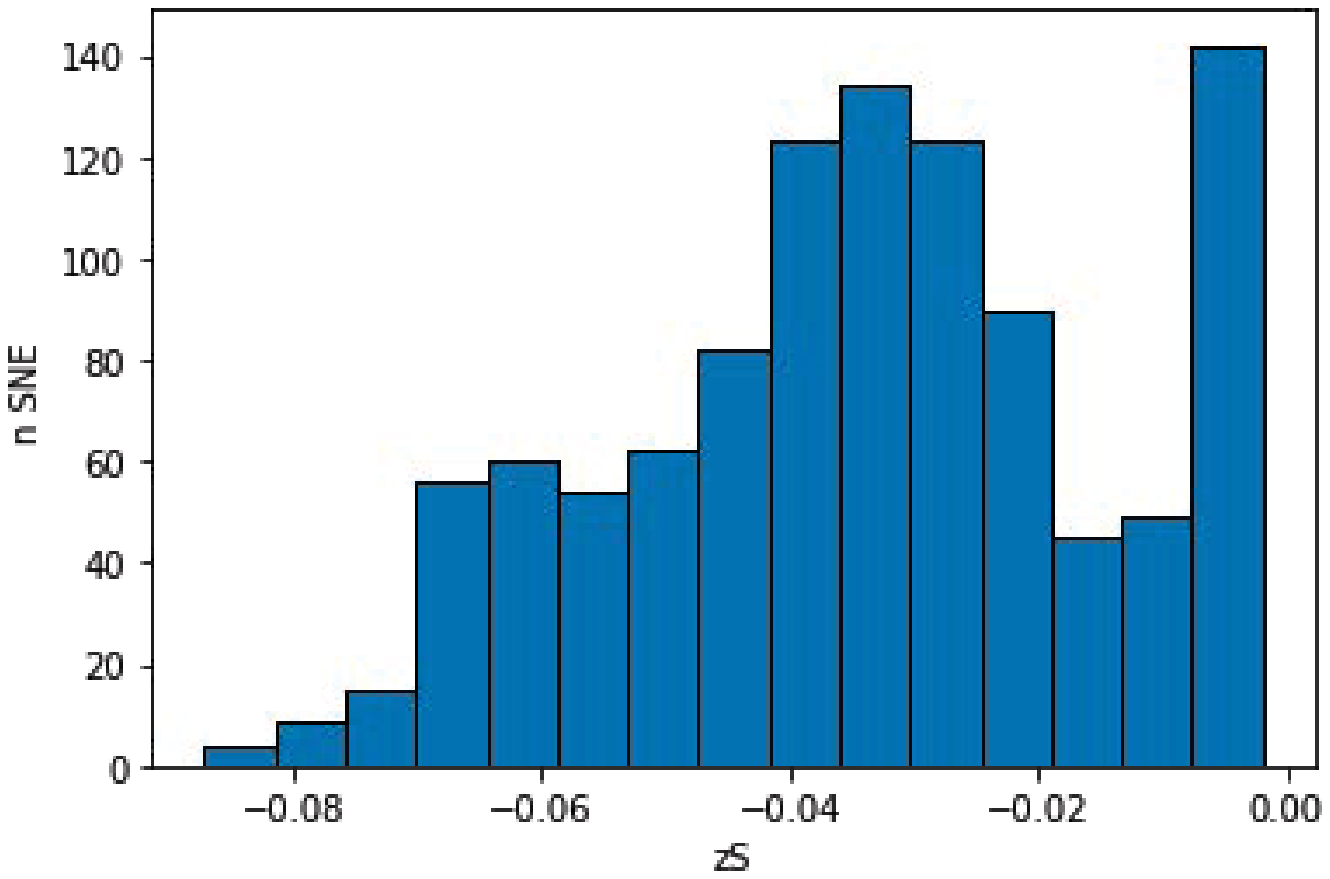}
    \caption{
    Histograms of the computed $z_{\rm S}$ from the $k_1$ parameter considering the Cosmology model C, where $\Omega_{M}=1$, $\Omega_{k}=\Omega_{\Lambda}=0$ ($H_0=67, 70, 74$ for the left, central and right panels. The values for $H_0$ are in km s$^{-1}$ per Mpc).}
    \label{fig:thirdthreecasesgeneralredshift}
\end{figure}

Given the computed best fit of $k_i$, we derive $z_{\rm S}$ using the expressions in Tab. \ref{tabdeltanu}. The histograms of $z_{\rm S}$ are shown in Figs. \ref{fig:firstthreecasesgeneralredshift}, \ref{fig:secondthreecasesgeneralredshift}, \ref{fig:thirdthreecasesgeneralredshift}. Having observed that different $k_i$ provide similar results for the same cosmological model and for the same values of $H_0$, we have decided to show only the nine cases for $k_1$, bearing in mind that similar arguments can be put forward for the other parameters.

In Fig. \ref{fig:firstthreecasesgeneralredshift}, the Cosmology model A is shown. The positive values for $z_{\rm S}$ reflect the negative sign for $k_1$. It is worth noticing that the magnitudes reached by $z_{\rm S}$ are similar to those computed in the individual cases. Again, these values decrease with the increasing of $H_0$, without changing the overall distribution, as we have seen in the correspondent previous cases. 

In Fig. \ref{fig:secondthreecasesgeneralredshift}, the Cosmology Model B is shown. Once more, the negative values for $z_{\rm S}$, are compliant to the sign of $k_1$. The main difference with the individual SNe Ia computation lies in the magnitude of $z_{\rm S}$, being its maximum absolute value smaller; further, we do not obtain positive $z_{\rm S}$ values, found in the corresponding cases beforehand. The methods do not lead to the same results, and we attribute this discrepancy to the higher precision of $k_1$ in the general computation, that accompanies a smaller spread on the results for $z_{\rm S}$. Finally, increasing $H_0$ determines a larger absolute mean value of $z_{\rm S}$.
 
In Fig. \ref{fig:thirdthreecasesgeneralredshift}, the Cosmology model C is shown.
We obtain only negative $z_{\rm S}$, with a larger mean absolute value that than in Fig. \ref{fig:secondthreecasesgeneralredshift}, but smaller than those computed for the individual cases, Fig. \ref{fig:thirdthreecasessingular}. Of course, the distributions of these values depend on the computed best fit for $k_1$, Tabs. \ref{tab:resultskiSingular}, \ref{tab:resultskigeneral}. Again, increasing $H_0$ increases the mean absolute value for $z_{\rm S}$.

\subsection{Hubble diagrams}

\begin{figure}
    \centering
    \includegraphics[width=0.33\hsize,height=0.3\textwidth,angle=0,clip]{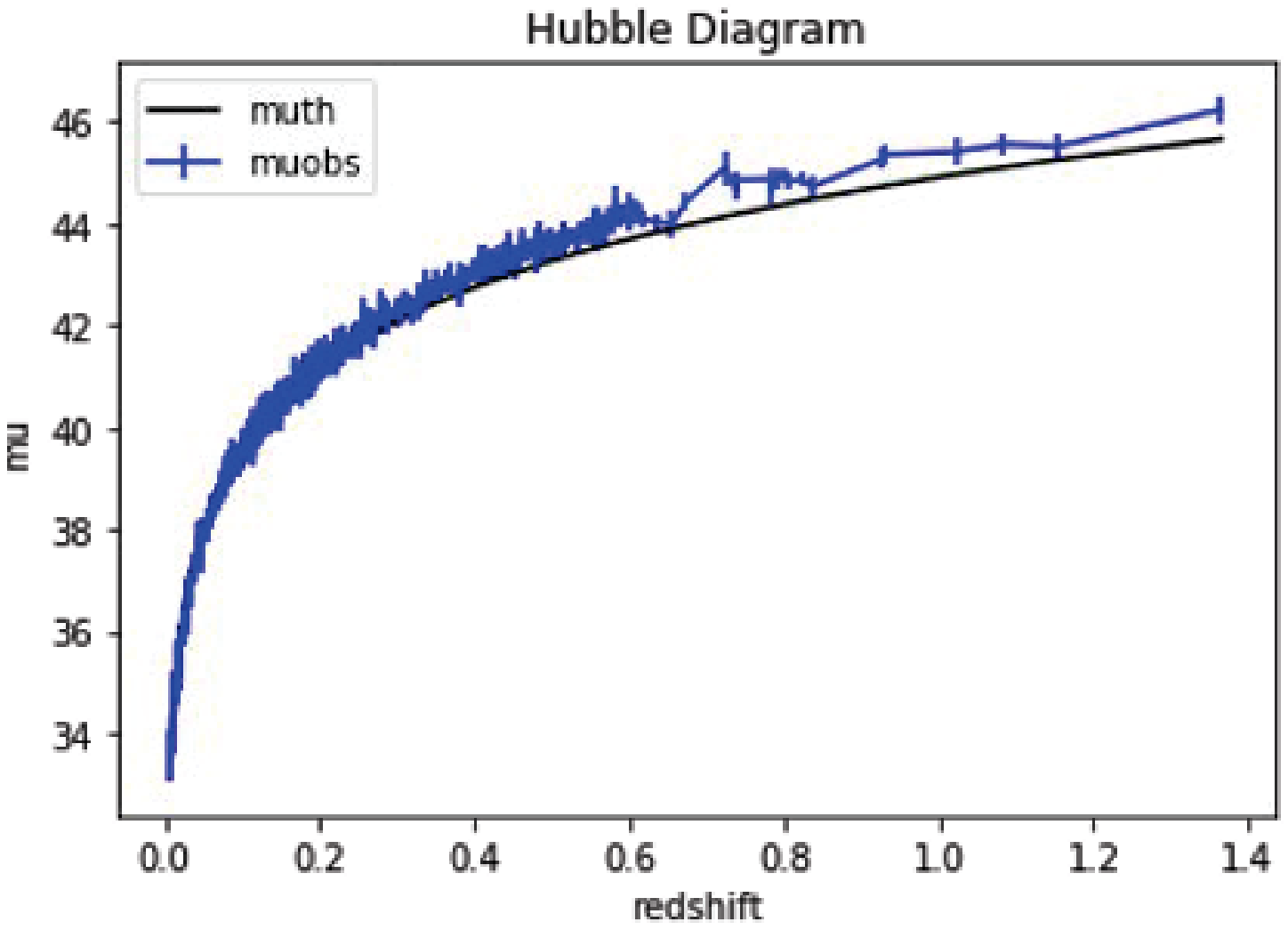}
    \includegraphics[width=0.33\hsize,height=0.3\textwidth,angle=0,clip]{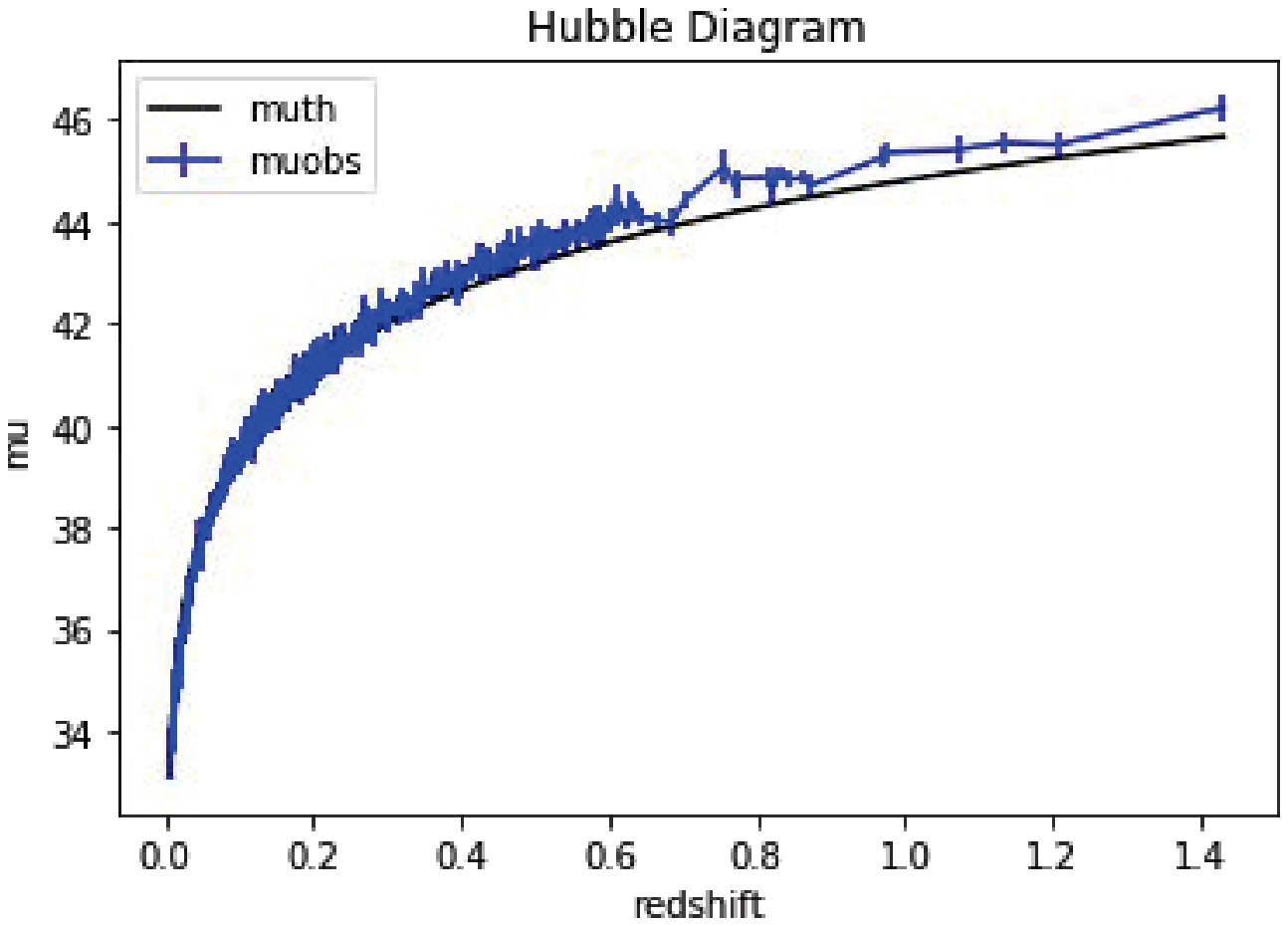}
    \includegraphics[width=0.33\hsize,height=0.3\textwidth,angle=0,clip]{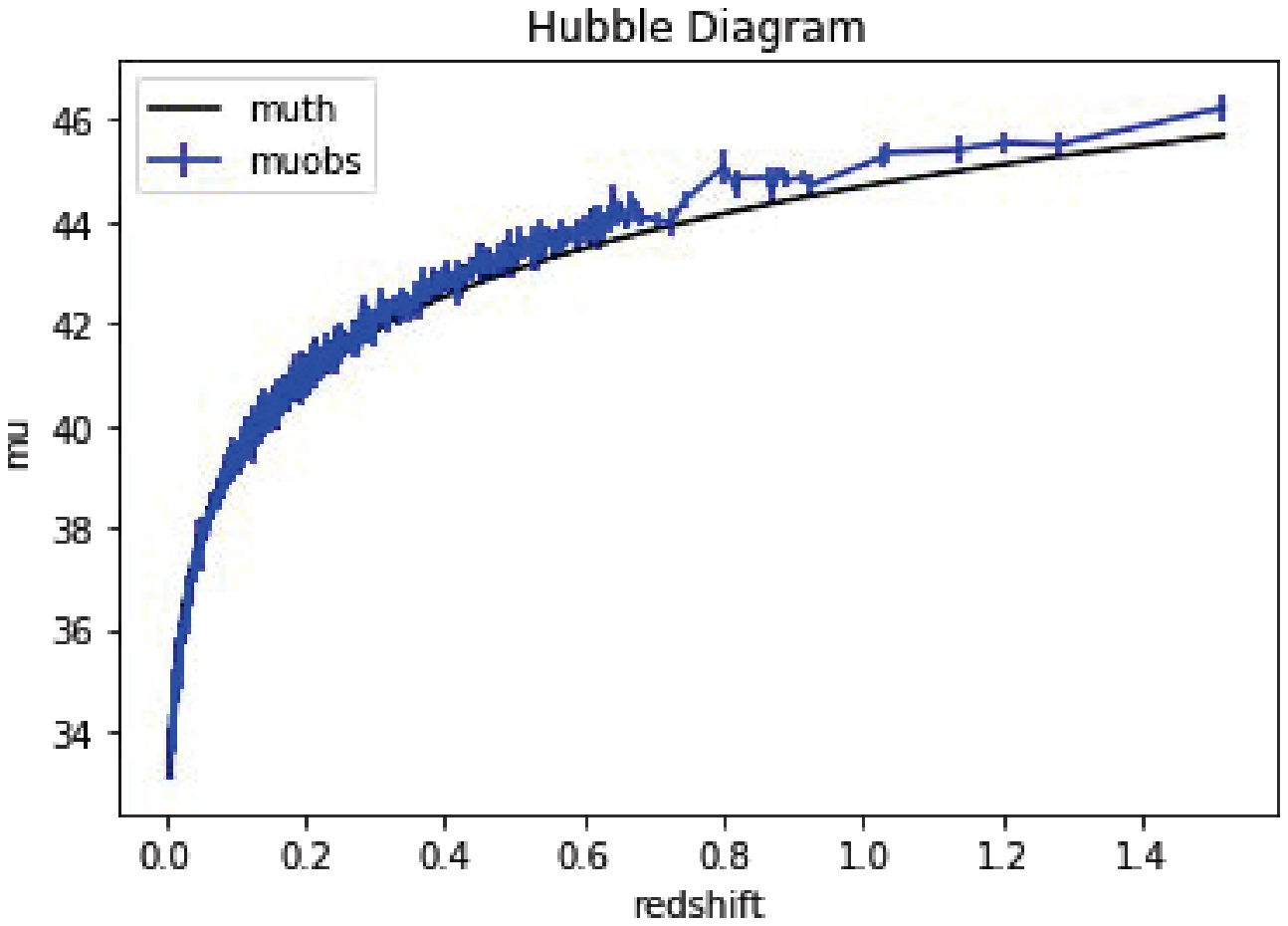}
    \includegraphics[width=0.33\hsize,height=0.3\textwidth,angle=0,clip]{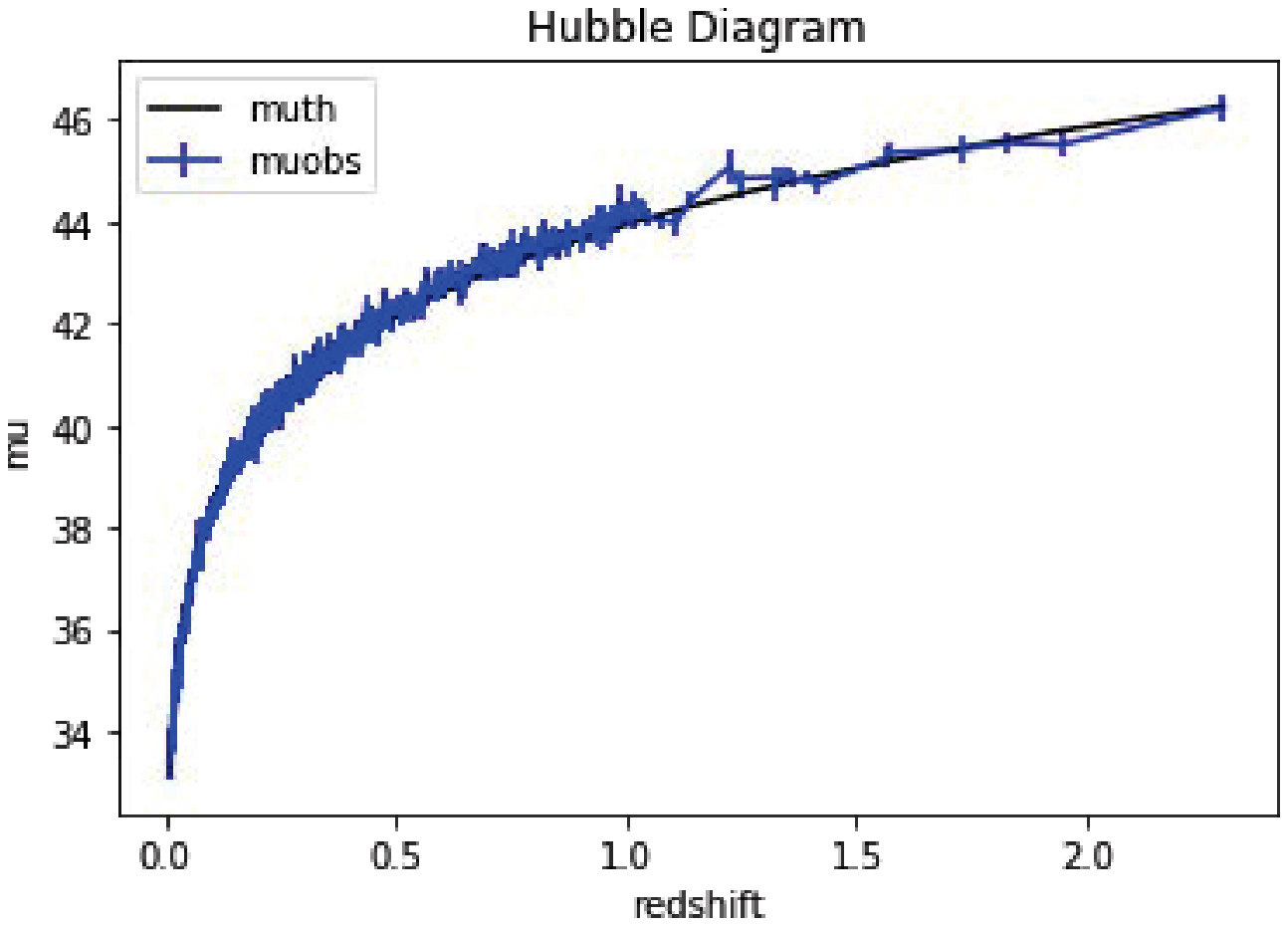}
    \includegraphics[width=0.33\hsize,height=0.3\textwidth,angle=0,clip]{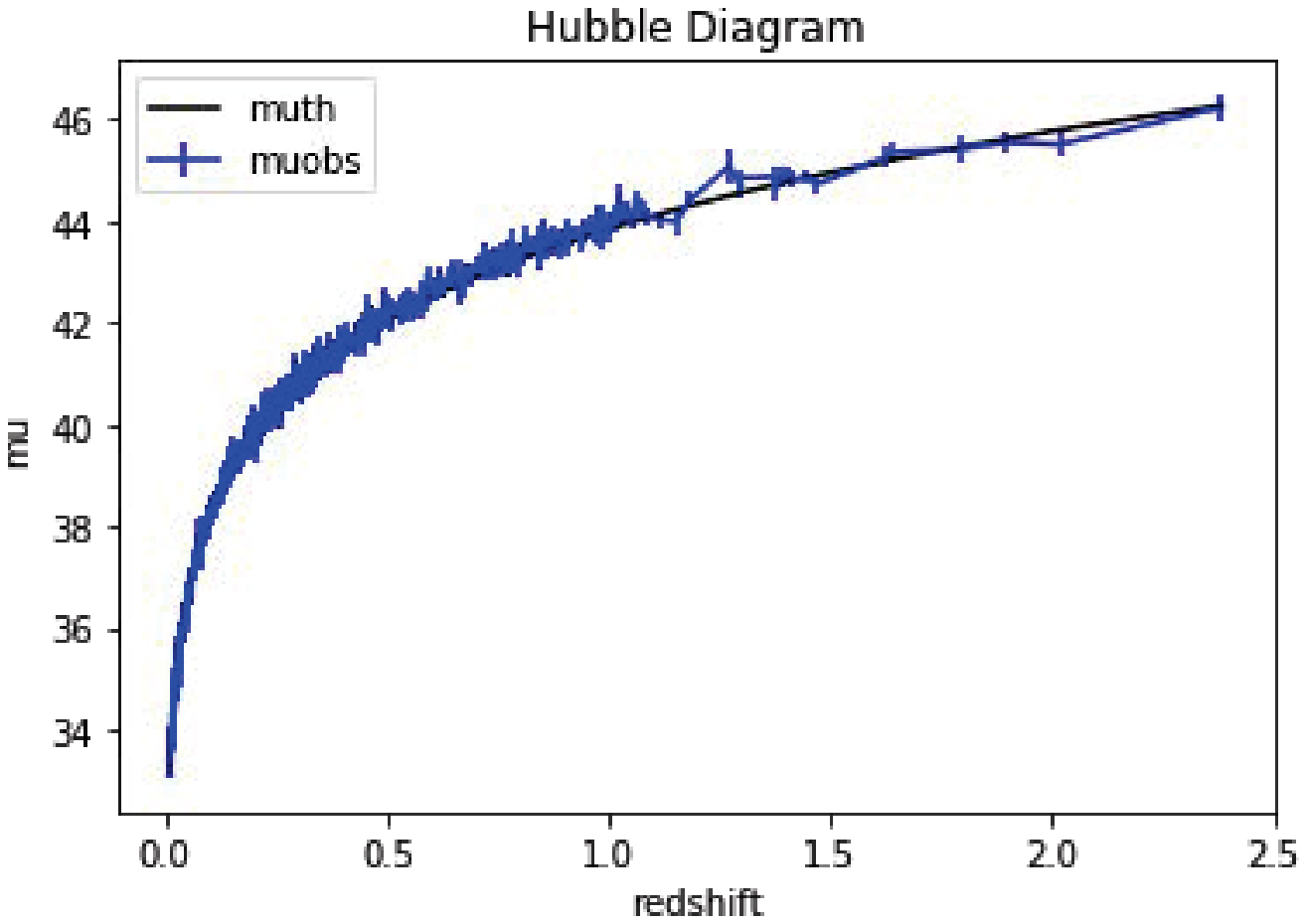}
    \includegraphics[width=0.33\hsize,height=0.3\textwidth,angle=0,clip]{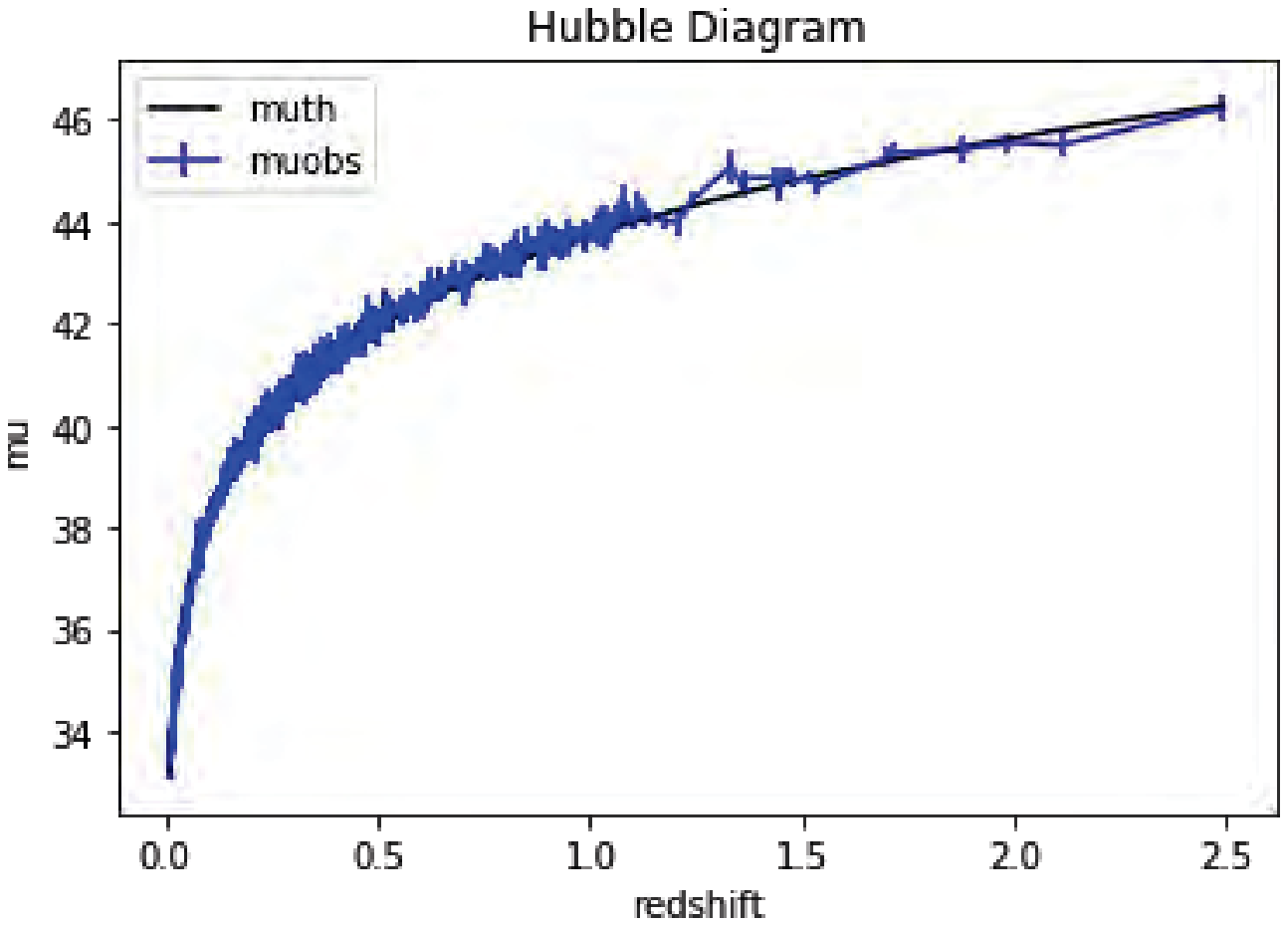}
    \includegraphics[width=0.33\hsize,height=0.3\textwidth,angle=0,clip]{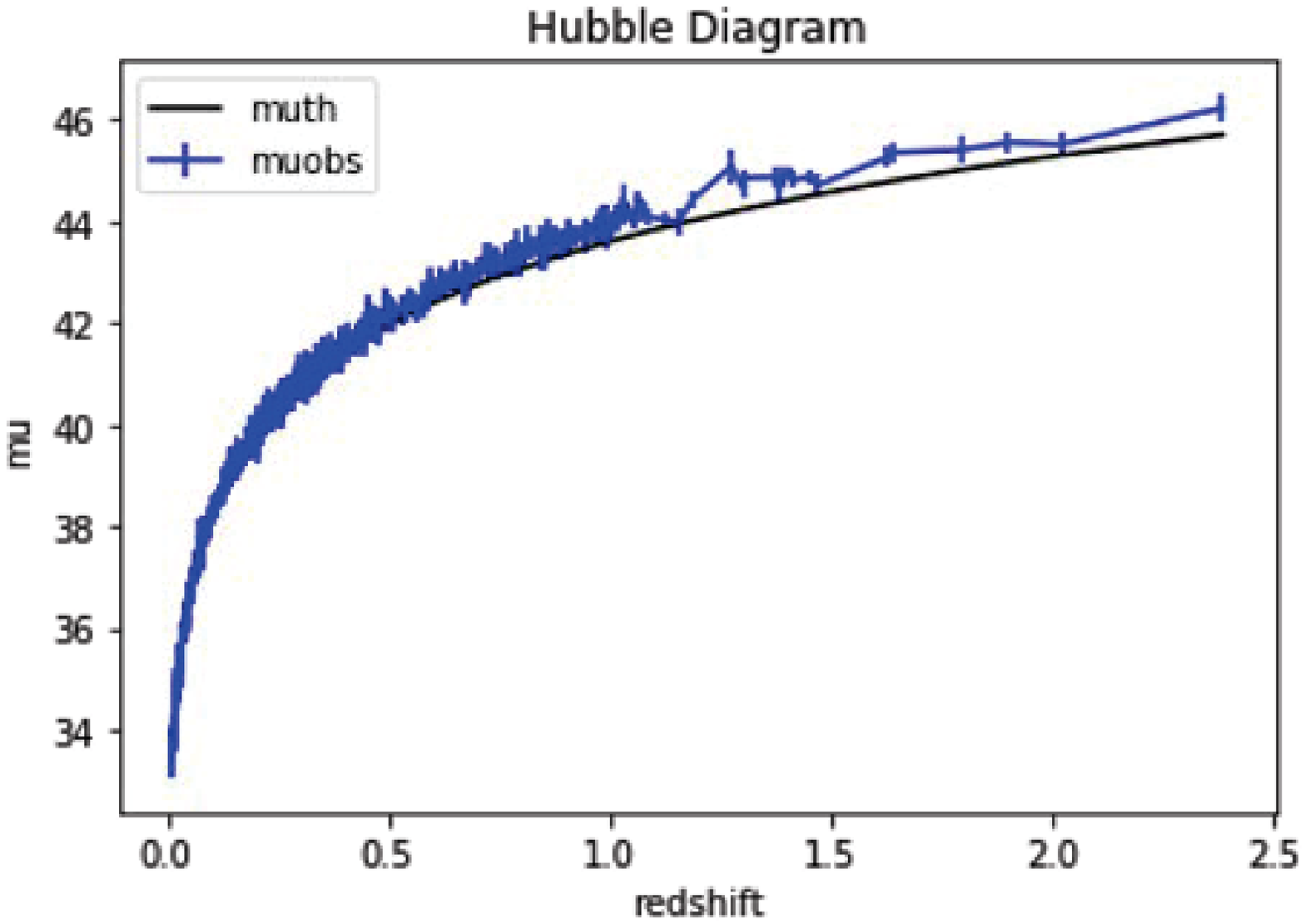}
    \includegraphics[width=0.33\hsize,height=0.3\textwidth,angle=0,clip]{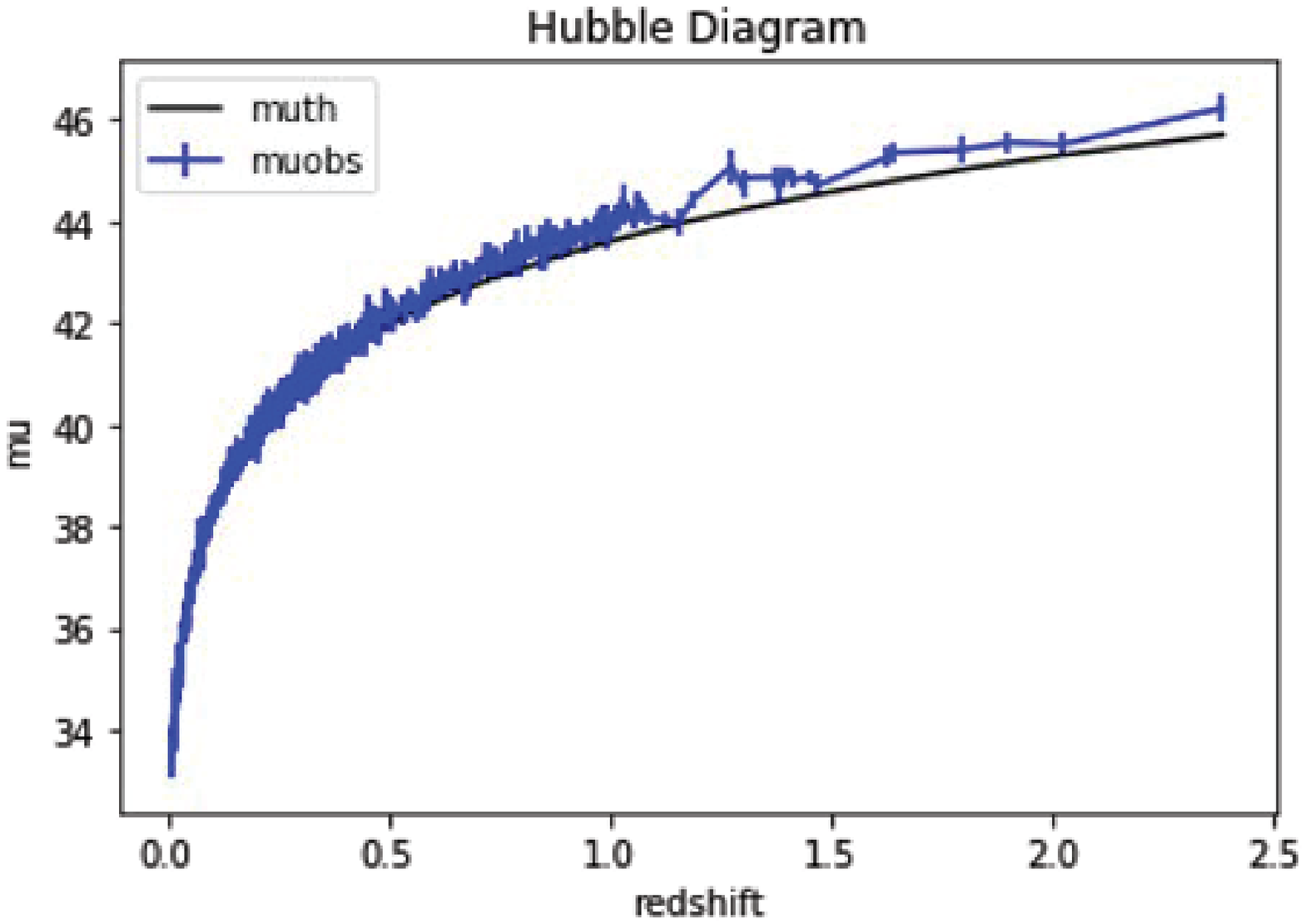}
    \includegraphics[width=0.33\hsize,height=0.3\textwidth,angle=0,clip]{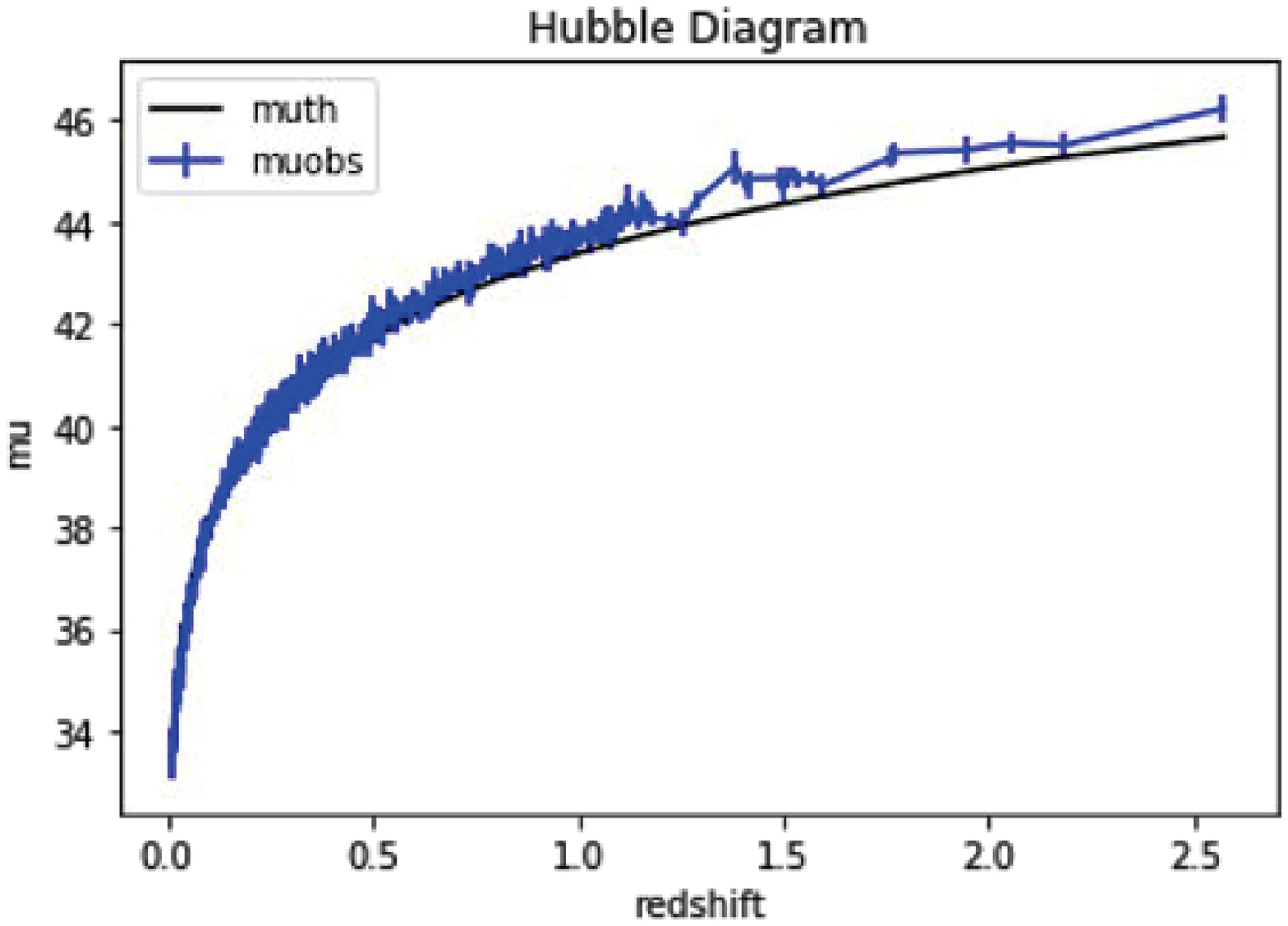}
    
    \caption{Through the Hubble diagrams, we compare the three cosmological models, each for row, based upon the best fit values of $k_1$ with data from the Pantheon Sample, with the usual three values of $H_0$ (67, 70, 74 km s$^{-1}$ per Mpc, each for column. The black lines represent the models, while the blue marks trace the SNe Ia data with their errors. We underline that the red shift on the x-axis is the computed expansion red shift $z_{\rm C}$. This explains why the scale changes between different cosmological models.}
    \label{fig:hubblediagrams}
\end{figure}

Once acquired the best fit values of $k_i$, and thus $z_{\rm S}$, we can draw the Hubble diagrams for the three cosmological models considering the Pantheon Sample. Thus, we will assess if our numerical results are in good agreement with the observations, specifically with the observed distance-modulus with its error. Again, we show the plots only for the $k_1$ parameter. The Hubble diagrams are shown in Fig. \ref{fig:hubblediagrams}.

  In the Hubble diagrams, the theoretical curve is very consistent with the majority of the data points, especially for the Cosmology model B. Instead, in the other models; the theoretical curve falls below the data line at high red shifts, for the three values of $H_0$. The reduced $\chi^{2}$, corresponding to the $k_{1}$ best fit value, the root mean squared deviation (RMSD) and the normalised root mean squared deviation (NRMSD) have been computed for all $k_{i}$, and they all support the same trend. For $k_{1}$, the results are shown in Tab. \ref{tab:statistics}, indicating that the lowest values are associated to Cosmology model B, a non-flat cosmology. Thus, these statistical indicators confirm our comments concerning Fig. \ref{fig:hubblediagrams}. The other Cosmology models show acceptable fits too, although worse than the results obtained by Cosmology model B. 
 
 \begin{table}
    \centering
    \begin{tabular}{c|c|c|c}
    \hline
       Cosmology  &  Reduced $\chi^2$ & RMSD & NRMSD \\\hline
         Cosmology 1, $H_0=67$ & 1.58 & 0.244 & 0.154 \\\hline
         Cosmology 1, $H_0=70$ & 1.62 & 0.247 & 0.155 \\\hline
         Cosmology 1, $H_0=74$ & 1.59 & 0.244 & 0.155\\\hline
        Cosmology 2, $H_0=67$ & 1.08 & 0.165 & 0.150\\\hline
         Cosmology 2, $H_0=70$ & 1.07 & 0.164 & 0.150\\\hline
         Cosmology 2, $H_0=74$ & 1.06 & 0.162 & 0.150 \\\hline
         Cosmology 3, $H_0=67$ & 1.46 & 0.228 & 0.151 \\\hline
        Cosmology 3, $H_0=70$ & 1.47 & 0.230 & 0.151 \\\hline
          Cosmology 3, $H_0=74$ & 1.47 & 0.230 & 0.151\\\hline
    \end{tabular}
    \caption{Best fit statistics for the $k_1$ parameter in all the cases considered in our work. The NRMSD has been derived from the RMSD divided by the difference between the maximum and minimum values of the difference between the theoretical and observed distance-modulus. The mean value of the uncertainty on the observed data is $\Delta_{\mu obs}=0.142$. The values for $H_0$ are in km s$^{-1}$ per Mpc.} 
    \label{tab:statistics}
\end{table}

\subsection{Adding BAO data}

 \begin{table}
    \centering
    \begin{tabular}{c|c|c|c|c}
    \hline
       Cosmology  &  $k_1$ &  $k_2$ &  $k_3$ &  $k_4$\\\hline
         Cosmology 1, $H_0=67$ & $(-8.72\pm 0.02)  \times 10^{-5}$ & $(-8.22\pm 0.02)  \times 10^{-5}$ & $(-5.49\pm 0.01)  \times 10^{10}$ & $(-9.19\pm 0.02)  \times 10^{-5}$\\\hline
         Cosmology 1, $H_0=70$ & $(-8.37\pm 0.02)  \times 10^{-5}$ & $(-7.92\pm 0.02)  \times 10^{-5}$ & $(-5.30\pm 0.01)  \times 10^{10}$ & $(-8.85\pm 0.02)  \times 10^{-5}$ \\\hline
         Cosmology 1, $H_0=74$ & $(-7.88\pm 0.02)  \times 10^{-5}$ & $(-7.49\pm 0.02)  \times 10^{-5}$ & $(-5.02\pm 0.01)  \times 10^{10}$ & $(-8.28\pm 0.02)  \times 10^{-5}$\\\hline
         Cosmology 2, $H_0=67$ & $(1.23\pm 0.03)  \times 10^{-5}$  & $(1.23\pm 0.03)  \times 10^{-5}$ & $(0.82\pm 0.02)  \times 10^{10}$ & $(1.21\pm 0.03)  \times 10^{-5}$ \\\hline
        Cosmology 2, $H_0=70$ & $(1.23\pm 0.04)  \times 10^{-5}$  & $(1.24\pm 0.04)  \times 10^{-5}$ & $(1.40\pm 0.02)  \times 10^{10}$ & $(2.07\pm 0.03)  \times 10^{-5}$\\\hline
          Cosmology 2, $H_0=74$ & $(3.24\pm 0.03)  \times 10^{-5}$ & $(3.30\pm 0.03)  \times 10^{-5}$ & $(2.21\pm 0.02)  \times 10^{10}$ & $(3.20\pm 0.03)  \times 10^{-5}$  \\\hline
          Cosmology 3, $H_0=67$ & $(-0.91\pm 0.03)  \times 10^{-5}$ & $(-0.91\pm 0.03)  \times 10^{-5}$ & $(-0.61\pm 0.02)  \times 10^{10}$ & $(-0.91\pm 0.04)  \times 10^{-5}$ \\\hline
         Cosmology 3, $H_0=70$ & $(0.31\pm 3.43)  \times 10^{-7}$ & $(-0.77\pm 3.77)  \times 10^{-7}$ & $(-0.81\pm 2.31)  \times 10^{8}$ & $(-0.29\pm 3.81)  \times 10^{-7}$ \\\hline
         Cosmology 3, $H_0=74$ & $(1.23\pm 0.03)  \times 10^{-5}$  & $(1.24\pm 0.04)  \times 10^{-5}$  & $(0.84\pm 0.03)  \times 10^{10}$  & $(1.22\pm 0.04)  \times 10^{-5}$ \\\hline
    \end{tabular}
    \caption{Results for the $k_i$ parameter considering the general best fit for SNe Ia of the Pantheon Sample together with the BAO constraints, for the three cosmological models. The values for $H_0$ are in km s$^{-1}$ per Mpc.} 
    \label{tab:resultskiBAO}
\end{table}
 
Due to the different nature of the BAO data, we just display the values obtained for the $k_{i}$ parameters, without the Hubble diagrams. The results are gathered in Tab. \ref{tab:resultskiBAO}, which we compare with those in Tab. \ref{tab:resultskigeneral}. For all $k_{i}$ for the Cosmology model A, the BAOs contribute with a negative value, and thus a positive contribution to $z_{\rm S}$, strengthening previous findings. Instead, for the Cosmology models B and C, the impact of considering BAOs may strengthen or weaken previous findings even changing the sign of the shift. For the Cosmology model C, we remark that for $H_0=70$ km s$^{-1}$ per Mpc, we find the values of the $k_i$ consistent with 0, despite the large errors. This might be of interest, because it corresponds to $z_{\rm S}=0$. Recalling that the Cosmology model C considers $\Omega_M=1$, this implies that the combination of SNe Ia+BAOs leads to the same results of the $\Lambda$CDM model. 

In \cite{spallicci-etal-2021}, it was argued that considering BAO date would have not falsified our proposition of recasting the observed $z$, according to Eq. (\ref{newz}). This is indeed the case, as it has been shown herein. 

\section{Time dilation, CMB and gravitational lensing}

Time dilation, Cosmic Microwave Background (CMB) and gravitational lensing data have been extensively discussed in \cite{spallicci-etal-2021} in the context of recasting the observed $z$ following Eq. (\ref{newz}). Therein, we observed that the error on time dilation  measurements \cite{goldhaber-etal-2001,blondin-etal-2008} is compatible with the ratio $z_{\rm S}/z \leq 10\%$. This consideration leads us to validate the Cosmology model B, and to some extent model C. The largest discrepancies occur for the Cosmology model A, at very low $z$, where the above ratio may rise up to 45\%. It would be interesting to run again time dilation measurements and verify with actual data the size of this discrepancy. 

Time dilation analysis has allowed somewhat ruling out tired light cosmology, first proposed by Zwicky \cite{zwicky-1929}, since it would not exist in a static Universe (more precisely, massive photons - including the effective mass photon of the SME - would have a negligible contribution to time dilation in a static universe at these optical frequencies). It represents a full alternative to red shift as manifestation of expansion and nowadays it suffers of several shortcomings \cite{lopezcorredoira-2017}. We take a different stand, and while we don't plead for a tired light {\it cosmology}, we question whether a tired light {\it mechanism} might have an impact on an expanding universe. Furthermore, while the tired light mechanism is always dissipative, the massive photon, the SME and the NLEM theories allow blue shifts too. This could be expected given that in this latter model the cosmic triangle relation is not fulfilled - a priori but a posteriori - and the contribution normally attributed to $\Omega_{\Lambda}$ in the $\Lambda$CDM model is now turned to $z_{\rm S}$. 

Regarding the CMB, referring to the discussion in \cite{spallicci-etal-2021}, the data can be interpreted to fit different cosmological models \cite{lopezcorredoira-2013}. Furthermore,  the CMB on its own is limited in constraining the dark energy hypothesis, since it is an indirect quantity derived from secondary assumptions \cite{Astier-Pain-2012}. Indeed, information related to the dark energy content in the Universe depends also on the distance between the CMB and us, given that in the early times for the $\Lambda$CDM the effects of dark energy were negligible. A reinterpretation of the red shift does not change the intrinsic physics of the CMB.

Finally for the weak lensing as a tool to derive cosmological parameters and specifically those concerning dark energy, the deformation linked to the shear  \cite{heavens-2009,Huterer-2010} 
also depends on the value of the red shift at which the weak lensing has been detected. Thus, we expect that the recasting of the red shift will imply moderate departures, similarly to cases analysed for the other cosmological probes following a methodology similar to that adopted for SNe Ia and BAOs, and supported by the  the proper formulas for the weak lensing. 

\section{Discussion, conclusions, and perspectives}

We have shown that the observed red shift $z$ might be composed by the expansion red shift $z_{\rm C}$ and an additional frequency shift $z_{\rm S}$, towards the red or the blue, derived from Extended Theories of Electromagnetism (ETE). They consist of: massive photon, Standard-Model Extension and Non-Linear Electro-Magnetism theories, and induce a static, {\it i.e.} expansion independent, frequency shift in presence of background (inter-) galactic electromagnetic fields and where applicable LSV fields, even when both fields are constant. The shift has been formulated for four different types, supposing its proportionality to: 1) the instantaneous frequency and the distance; 2) the emitted frequency and the distance; 3) only the distance; 4) the observed frequency and the distance. 

We have tested this prediction against the Pantheon Catalogue, composed by 1048 SNe Ia, with the addition of 15 BAO data, for different cosmological models characterised by the absence of a cosmological constant. 
The Cosmology models are: A) $\Omega_{M}=0.3$, $\Omega_{K}=0$, implying a flat universe where the "cosmic triangle" relation $\Omega_{M}+\Omega_{K}+\Omega_{\Lambda}=1$, is not satisfied {\it ab initio}, but {\it a posteriori}, through the effect of $z_{\rm S}$, which would act as an effective "Dark Energy" component; B) an open universe model, where $\Omega_{M}=0.3$ and $\Omega_{K}=0.7$, and $\Omega_{K}+\Omega_{M}=1$; C) the Einstein-de Sitter model of a flat, matter dominated universe with $\Omega_{M}=1$. 
The BAO data are largely in agreement with SN Ia data in our findings.

For all models, we span three values of the Hubble-Lema\^itre constant $H_0$:  $67, 70, 74$ km s$^{-1}$ per Mpc.  

After a preliminary analysis with mock red shifts, we have drawn from the data which values of $z_{\rm S}$ match better the observations of SNe Ia and BAOs. Further, we have determined the values of the parameters for each SN Ia as well as a best fit applicable to every SN Ia and BAO. In the former case, fitting each single SN Ia allows to accommodate specific distances, light-path dependency and, ultimately, all sorts of anisotropies. In the latter case, we identify universal parameters applicable to all SNe Ia and BAOs.
 
The static shift $z_{\rm S}$ for the Cosmology model A is always positive, thus corresponding to a dissipative effect for the photons. It implies, that the astrophysical objects are actually closer than what we deduce from the observed $z$. 

For the Cosmology model B with mock red shifts, from the individual computations we deduce the presence of a first zone at low red shift where $z_{\rm S}$ is negative, followed by a zone where there is change of sign of the shift around $z \sim 4$. The shift towards the red turns into blue, and the it increases with $z$. But the real data cover only up to $z=2.24$, where the furthest SN Ia in the Pantheon Sample is located. Thus, only negative values for $z_{\rm S}$ should be observed. Once more there is concordance between our assumptions and real data. A negative $z_{\rm S}$ implies the photons gaining energy when crossing (inter-)galactic fields and that the astrophysical objects are actually farther than what we deduce from the observed $z$. 

For the Cosmology model C, we note through the mock red shift tests, that $z_{\rm S}$ should be always negative, which is confirmed when using real data. Nevertheless, when including BAOs, for $H_0=67$ km s$^{-1}$ per Mpc, we get positive values for $z_{\rm S}$, while for $H_0=70$ km s$^{-1}$ per Mpc we obtain $z_{\rm S}$, consistent with 0. 

The BAO contribution strengthens the findings based on SNe IA for the Cosmology models A and B, but opposes the findings for the Cosmology model C based on SNe Ia only.

We conclude that the frequency shift  $z_{\rm S}$ can support  an alternative to accelerated expansion, naturally accommodating each SN Ia position in the distance-modulus versus red shift diagram, due to the light-path dependency of $z_{\rm S}$. 

For the perspectives, the general approach for the $k_i$ parameters renders possible to enlarge our computations by varying also the cosmological parameters like $H_0$, or $\Omega_M$, together with the ETE parameters, to both further improve our fits in this new framework as well as to see how the cosmological quantities will diverge with respect to the $\Lambda$CDM values.

It is mandatory considering the statistical errors on the red shift of the studied objects: for instance, even if in the Pantheon catalogue the errors on the red shifts have not been stated, they are still present, as for every physical observation. This means that, when possible, we can compare the magnitude of $z_{\rm S}$ with the uncertainty on the observed red shift. Indeed, discussions on the errors can have an influence on the cosmological results \cite{palanquedelabrouille-etal-2010,calcino-davis-2017,davis-hinton-howlett-calcino-2019,steinhardt-sneppen-bidisha-2020}.

The $z_{\rm S}$ shift provides a physical explanation of red shift remapping \cite{bassett-etal-2013,wojtak-prada-2016,wojtak-prada-2017,tian-2017}
and is not limited to the SN Ia case. It is naturally suited to explaining recently discovered expansion anisotropies  \cite{morenoraya-etal-2016,colin-etal-2019,migkas-etal-2020,salehi-etal-2020}. 

In future explorations, we will have to deal with the comparison of $z_{\rm S}$
with the error on $z$. The analysis of the error on spectroscopic and photometric measurements is destined to become a pivotal issue for cosmology 
\cite{palanquedelabrouille-etal-2010,calcino-davis-2017,davis-hinton-howlett-calcino-2019}. 

Finally, the $H_0$ parameter corresponds to $2.3 \times 10^{-18}$ m/s per meter in SI units. This value interpreted as static shift corresponds to $H_0/c = 7.7 \times 10^{-27} \Delta \nu/\nu$ per meter. Herein, we have made the assumption that $z_{\rm S}$ is just a (minor) part of the observed $z$. Thereby, the relative frequency shift per meter has $H_0/c$ as upper limit. 
For the Earth-Moon distance, the relative frequency shift $\Delta \nu/\nu$ is below $3 \times 10^{-18}$. It is worthwhile to dig whether this effect could be detected with interferometry and delay lines. Moreover, this experiment would verify the absence of expansion at small scale \cite{wiensnevskyschiller2016}.     

\section{Acknowledgements}
.
Acknowledgements are due to J.A. Helay\"el-Neto (Rio de Janeiro) for our common work on the Extended Theories of Electromagnetism, to S. Savastano (Potsdam) for the work on the Cobaya routine, M. G. Dainotti (Tokyo) and B. De Simone (Salerno) for the work on the SNe Ia data, in particular on the correlation matrix, and to M. Lopez Corredoira (La Laguna), C. L\"ammerzahl and V. Perlick (Bremen), A.D.V. di Virgilio (Pisa) for general comments and inputs. GS is
grateful to the LPC2E laboratory for its hospitability during the work
period on this manuscript. GS and SC acknowledge the support of Istituto Nazionale di Fisica Nucleare, Sez. di Napoli, Iniziative
Specifiche MOONLIGHT-2 and QGSKY.

\bibliography{references_spallicci_220129}

\end{document}